\documentclass[twocolumn,aps,revtex,groupedaddress,nofootinbib,final,superscriptaddress,floatfix]{revtex4-2}
\usepackage{graphicx}
\usepackage{caption}
\usepackage{subcaption}
\usepackage{amssymb}
\usepackage{stmaryrd}
\usepackage{amsmath}
\usepackage[hidelinks]{hyperref}
\usepackage[capitalize]{cleveref}
\usepackage{pgfplots}
\usepackage{pgfplotstable}
\usepgfplotslibrary{units}
\usepackage{mathtools}
\usepackage{bm}
\usepackage[toc,page]{appendix}
\usepackage{etoolbox}
\usepackage[utf8]{inputenc}
\usepackage{boondox-cal}
\AtBeginEnvironment{appendices}{\crefalias{section}{appendix}}
\pgfplotsset{
        /tikz/>=stealth,
        compat=1.12,
        /pgf/declare function={
            f(\x) =  -0.0000000059501342740085*\x+0.0000000228909948704699  ;
            lb = 8;
            ub = 20;
            a = 0.4;
            Y(\x) = exp(a*\x);
            X(\x) = (Y(\x) - Y(lb))/(Y(ub) - Y(lb)) * (ub - lb) + lb;
        },
        }
\usepackage{pgfplotstable}
\usepackage{xcolor}
\usepackage{enumitem}
\pgfplotsset{compat=1.8}% <-- moves axis labels near ticklabels (respects tick label widths)
\usepgfplotslibrary{external} 
\usepgfplotslibrary{units}
\usetikzlibrary{external}
\tikzset{external/mode=graphics if exists}
\makeatletter
\pretocmd{\@sect}{\def\@currentlabel{#8}}{}{}% Store title of \section
\pretocmd{\@ssect}{\def\@currentlabel{#5}}{}{}% Store title of \section*
\makeatother

\makeatletter
\newcommand{\setlabel}[1]{\edef\@currentlabel{#1}\label}
\makeatother

\begin{document}

%\preprint{Caltech MAP-300}
%%%%%%%%%%%%%%%%%%%%%%%%%%%%%%%%%%%%%%%%%%%%%
\maxdeadcycles=200
\makeatletter
\newcommand\begi
{%
  \begingroup
  \parindent\z@
  \par\advance\@itemdepth\@ne
  \advance\@totalleftmargin\csname leftmargin\romannumeral\the\@itemdepth\endcsname
  \leftskip\@totalleftmargin
  \everypar
  {%
    \llap{%
      \makebox[\labelsep][l]
      {%
        \csname labelitem\romannumeral\the\@itemdepth\endcsname
      }
    }%
  }%
  \obeylines
}
\newcommand\ei
{%
  \bottom
  \par\endgroup
}

\newcommand\up
{%
  \par
  \begingroup
    \advance\@itemdepth\@ne
    \advance\@totalleftmargin\csname leftmargin\romannumeral\the\@itemdepth\endcsname
    \leftskip\@totalleftmargin
}

\newcommand\down{\par\endgroup}

\newcommand\bottom
{%
  \ifnum\@itemdepth>\@ne
    \down\bottom
  \fi
}
\makeatother

\newcommand{\EL}{{\cal{E}}_L}
\newcommand{\nMAX}{10} 
\newcommand{\kMAX}{20}
\newcommand{\vrprime}[0]{\vec{r}^{~\!\prime}}
\newcommand{\hatrprime}[0]{\hat{r}^{\prime}}
\newcommand{\rprime}[0]{r^{\prime}}
\newcommand{\mus}[0]{\mu_{\text{s}}}
\newcommand{\st}[0]{\slash{T}}
\newcommand{\mun}[0]{\mu_\text{N}}
\newcommand{\vmun}[0]{\bm{\mu}_\text{\bf{N}}}
\newcommand{\lambdal}[0]{\lambda_{\text{L}}}
\newcommand{\omegad}[0]{\omega_{\text{D}}}
\newcommand{\bc}[0]{B_{\text{c}}}
\newcommand{\nbb}[0]{{0\nu}\beta\beta}
\newcommand{\den}[0]{d_{\text{e}}^{\text{N}}}
\newcommand{\vden}[0]{\mathbf{d}_\text{\bf{e}}^\text{\bf{N}}}
\newcommand{\denbar}[0]{\bar{d}_{\text{e}}^{\text{N}}}
\newcommand{\tqcd}[0]{\bar{\theta}_\text{QCD}}
\newcommand{\vlr}[0]{\bm{\lambda}}
\newcommand{\deneut}[0]{d_{\text{e}}^{\text{n}}}
\newcommand{\fp}[0]{f_\pi}
\newcommand{\eq}[1]{eq.~(\ref{#1})}
\newcommand{\eqs}[2]{eqs.~(\ref{#1,#2})}
\newcommand{\Eq}[1]{Eq.~(\ref{#1})}
\newcommand{\Eqs}[2]{Eqs.~(\ref{#1},\ref{#2})}
\newcommand{\fig}[1]{fig.~\ref{#1}}
\newcommand{\figs}[2]{figs.~\ref{#1},\ref{#2}}
\newcommand{\Fig}[1]{Fig.~\ref{#1}}
\newcommand{\tab}[1]{tab.~\ref{#1}}
\newcommand{\Tab}[1]{Tab.~\ref{#1}}
\newcommand{\qbb}[0]{Q_{\beta\beta}}
\newcommand{\tl}[0]{\text{L}}
\newcommand{\tr}[0]{\text{R}}
\newcommand{\nc}[0]{N_{\text{c}}}
\newcommand{\mw}[0]{M_{\text{W}}}
\newcommand{\mz}[0]{M_{\text{Z}}}
\newcommand{\mr}[0]{M_{\text{R}}}
\newcommand{\md}[0]{m_{\text{D}}}
\newcommand{\mn}[0]{m_\nu}
\newcommand{\lh}[0]{\Lambda_{\text{H}}}
\newcommand{\aif}[0]{a^{ff^\prime}_{i;ll^\prime}}
\newcommand{\gs}[0]{\Gamma_{\text{S}}=1}
\newcommand{\gps}[0]{\Gamma_{\text{PS}}=\gamma^5}
\newcommand{\lr}[0]{\Lambda_{\text{R}}}
\newcommand{\wrt}[0]{W_{\text{R}}}
\newcommand{\wl}[0]{W_{\text{L}}}
\newcommand{\ls}[0]{\Lambda_{\text{S}}}
\newcommand{\gf}[0]{G_{\text{F}}}
\newcommand{\mm}[0]{M_{\text{M}}}
\newcommand{\sst}[1]{{\scriptscriptstyle #1}}
\newcommand{\beeq}{\begin{equation}}
\newcommand{\eeeq}{\end{equation}}
\newcommand{\beqa}{\begin{eqnarray}}
\newcommand{\eeqa}{\end{eqnarray}}
\newcommand{\dida}[1]{/ \!\!\! #1}
\renewcommand{\Im}{\mbox{\sl{Im}}}
\renewcommand{\Re}{\mbox{\sl{Re}}}
\def\simge{\hspace*{0.2em}\raisebox{0.5ex}{$>$}
     \hspace{-0.8em}\raisebox{-0.3em}{$\sim$}\hspace*{0.2em}}
\def\simle{\hspace*{0.2em}\raisebox{0.5ex}{$<$}
     \hspace{-0.8em}\raisebox{-0.3em}{$\sim$}\hspace*{0.2em}}
\def\dn{{d_n}}
\def\de{{d_e}}
\def\datom{{d_{\sst{A}}}}
\def\grhobar{{{\bar g}_\rho}}
\def\gpibar{{{\bar g}_\pi^{(I) \prime}}}
\def\gpibarz{{{\bar g}_\pi^{(0) \prime}}}
\def\gpibaro{{{\bar g}_\pi^{(1) \prime}}}
\def\gpibart{{{\bar g}_\pi^{(2) \prime}}}
\def\mx{{M_X}}
\def\mrho{{m_\rho}}
\def\qpv{{Q_{\sst{W}}}}
\def\lamtv{{\Lambda_{\sst{TVPC}}}}
\def\lamtvs{{\Lambda_{\sst{TVPC}}^2}}
\def\lamtvc{{\Lambda_{\sst{TVPC}}^3}}

%     \hspace{-0.8em}\raisebox{-0.3em}{$\sim$}\hspace*{0.2em}}
%DEFINITIONS OF CONSTANTS
\def\currVal{1e-11}
\def\nanocurrVal{1e-9}
\def\thicknesst{1e-3}
\def\day{8.64e4}
\def\Rc{0.1}
\def\Bfield{1e-2}
\def\Rw{1e-3}
\def\Ll{5.e-8}
\def\ms{1e-3}
\def\mN{5.050783699e-27}
\def\mzero{1.2566370614e-6}
\def\Nspin{0.5}
\def\ncar{1e28} 
\def\kb{1.38064852e-23} 
\def\echarge{1.6021766208e-19}
\def\cspeed{2.99792458e8}
\def\pival{3.1415926535}
\def\omegadFac{\cspeed^2*\mzero*\ms/\Ll*(\Nspin+1)/\Nspin*\ncar/3*\mN/\kb}
%END DEFINITIONS OF CONSTANTS
\def\bra#1{{\langle#1\vert}}
\def\ket#1{{\vert#1\rangle}}
\def\coeff#1#2{{\scriptstyle{#1\over #2}}}
\def\undertext#1{{$\underline{\hbox{#1}}$}}
\def\hcal#1{{\hbox{\cal #1}}}
\def\sst#1{{\scriptscriptstyle #1}}
\def\eexp#1{{\hbox{e}^{#1}}}
\def\rbra#1{{\langle #1 \vert\!\vert}}
\def\rket#1{{\vert\!\vert #1\rangle}}

\def\lsim{{ <\atop\sim}}
\def\gsim{{ >\atop\sim}}
\def\nubar{{\bar\nu}}
\def\psibar{{\bar\psi}}
\def\Gmu{{G_\mu}}
\def\alr{{A_\sst{LR}}}
\def\wpv{{W^\sst{PV}}}
\def\evec{{\vec e}}
\def\notq{{\not\! q}}
\def\notl{{\not\! \ell}}
\def\notk{{\not\! k}}
\def\notp{{\not\! p}}
\def\notpp{{\not\! p'}}
\def\notder{{\not\! \partial}}
\def\notcder{{\not\!\! D}}
\def\notA{{\not\!\! A}}
\def\notv{{\not\!\! v}}
\def\Jem{{J_\mu^{em}}}
\def\Jana{{J_{\mu 5}^{anapole}}}
\def\nue{{\nu_e}}
\def\mns{{m^2_{\sst{N}}}}
\def\me{{m_e}}
\def\mes{{m^2_e}}
\def\mq{{m_q}}
\def\mqs{{m_q^2}}
\def\mw{{M_{\sst{W}}}}
\def\mz{{M_{\sst{Z}}}}
\def\mzs{{M^2_{\sst{Z}}}}
\def\ubar{{\bar u}}
\def\dbar{{\bar d}}
\def\sbar{{\bar s}}
\def\qbar{{\bar q}}
\def\sstw{{\sin^2\theta_{\sst{W}}}}
\def\gv{{g_{\sst{V}}}}
\def\ga{{g_{\sst{A}}}}
\def\pv{{\vec p}}
\def\pvs{{{\vec p}^{\>2}}}
\def\ppv{{{\vec p}^{\>\prime}}}
\def\ppvs{{{\vec p}^{\>\prime\>2}}}
\def\qv{{\vec q}}
\def\qvs{{{\vec q}^{\>2}}}
\def\xv{{\vec x}}
\def\xpv{{{\vec x}^{\>\prime}}}
\def\yv{{\vec y}}
\def\tauv{{\vec\tau}}
\def\sigv{{\vec\sigma}}
\def\istc{$\mathbf{_{I_\text{a}}\!S_T^C}$}
\def\ittc{$\mathbf{_{I_\text{a}}\!t_T^C}$}
\def\imtc{$\mathbf{_{I_\text{a}}\!M_T^C}$}
\def\iitc{$\mathbf{_{I_\text{a}}\!I_{\text{\bf ps},T}^C}$}
\def\ips{$I_\text{ps}$}
\def\Tmin{$T_\text{min}$}
\def\Tmax{$T_\text{max}$}
\def\Is{$I_\text{s}$}
\def\Isol{$I_\text{sol}$}
\def\Lsol{$L_\text{sol}$}
\def\Lloop{$L_\text{loop}$}
\def\dIs{$\Delta I_\text{s}$}
\def\Ib{$I_\text{B}$}
\def\Ia{$I_\text{a}$}
\def\Tset#1{$\mathbf{T}^\text{\bf{#1}}$}
\def\Iset#1{$\mathbf{I}_\text{\bf{a}}^\text{\bf{#1}}$}
\def\Fedm#1{F_\text{edm}^\text{#1}}
\def\Sedm{S_\text{edm}}
\def\distc{$\mathbf{ _{I_\text{a1}}^{I_\text{a2}}\!S_T^{C,o,n} }$}
\def\diitc{$\mathbf{ _{I_\text{a1}}^{I_\text{a2}}\!I_{\text{\bf ps},T}^{C,o,n} }$}
\newcommand{\stc}[3]{$\mathbf{_{#1}\!S_{#2}^{\text{\bf#3}}}$}
\newcommand{\stcbar}[3]{$\mathbf{_{#1}\!\bar{S}_{#2}^{\text{\bf#3}}}$}
\newcommand{\dstc}[5]{$\mathbf{ \prescript{\bf#2}{\bf#1}\!S_{#3}^{\text{\bf#4}#5}}$}
\newcommand{\dsstc}[5]{$\mathbf{ \prescript{\bf#2}{\bf#1}\!s_{#3}^{\text{\bf#4}#5}}$}
\newcommand{\itc}[3]{$\mathbf{_{#1}\!I_{\text{\bf ps},#2}^{\text{\bf#3}}}$}
\newcommand{\sitc}[3]{$\mathbf{_{#1}\!i_{\text{\bf ps},#2}^{\text{\bf#3}}}$}
\newcommand{\ditc}[5]{$\mathbf{ \prescript{\bf#2}{\bf#1}\!I_{\text{\bf ps},#3}^{\text{\bf#4}#5}}$}
\newcommand{\dgtc}[5]{$ \prescript{#2}{#1}\!\gamma_{#3}^{\text{#4},#5}$}
\newcommand{\mtc}[3]{$\mathbf{_{#1}\!M_{#2}^{\text{\bf#3}}}$}
\newcommand{\mtcbar}[3]{$\mathbf{_{#1}\!\bar{M}_{#2}^{\text{\bf#3}}}$}
\newcommand{\gtc}[3]{$_{#1}\!\gamma_{#2}^{\text{#3}}$}
\newcommand{\gbar}[3]{$_{#1}\!\bar{\gamma}_{#2}^{\text{#3}}$}
\newcommand{\ttc}[3]{$\mathbf{_{#1}\!t_{#2}^{\text{\bf#3}}}$}
\newcommand{\IStc}[3]{$\mathbf{_{#1}\!{IS}_{#2}^{\text{\bf#3}}}$}

\def\sst#1{{\scriptscriptstyle #1}}
\def\gpnn{{g_{\sst{NN}\pi}}}
\def\grnn{{g_{\sst{NN}\rho}}}
\def\gnnm{{g_{\sst{NNM}}}}
\def\hnnm{{h_{\sst{NNM}}}}
\def\xivz{{\xi_\sst{V}^{(0)}}}
\def\xivt{{\xi_\sst{V}^{(3)}}}
\def\xive{{\xi_\sst{V}^{(8)}}}
\def\xiaz{{\xi_\sst{A}^{(0)}}}
\def\xiat{{\xi_\sst{A}^{(3)}}}
\def\xiae{{\xi_\sst{A}^{(8)}}}
\def\xivtez{{\xi_\sst{V}^{T=0}}}
\def\xivteo{{\xi_\sst{V}^{T=1}}}
\def\xiatez{{\xi_\sst{A}^{T=0}}}
\def\xiateo{{\xi_\sst{A}^{T=1}}}
\def\xiva{{\xi_\sst{V,A}}}
\def\rvz{{R_{\sst{V}}^{(0)}}}
\def\rvt{{R_{\sst{V}}^{(3)}}}
\def\rve{{R_{\sst{V}}^{(8)}}}
\def\raz{{R_{\sst{A}}^{(0)}}}
\def\rat{{R_{\sst{A}}^{(3)}}}
\def\rae{{R_{\sst{A}}^{(8)}}}
\def\rvtez{{R_{\sst{V}}^{T=0}}}
\def\rvteo{{R_{\sst{V}}^{T=1}}}
\def\ratez{{R_{\sst{A}}^{T=0}}}
\def\rateo{{R_{\sst{A}}^{T=1}}}
\def\mro{{m_\rho}}
\def\mks{{m_{\sst{K}}^2}}
\def\mpi{{m_\pi}}
\def\mpis{{m_\pi^2}}
\def\mom{{m_\omega}}
\def\mphi{{m_\phi}}
\def\Qhat{{\hat Q}}
\def\FOS{{F_1^{(s)}}}
\def\FTS{{F_2^{(s)}}}
\def\GAS{{G_{\sst{A}}^{(s)}}}
\def\GES{{G_{\sst{E}}^{(s)}}}
\def\GMS{{G_{\sst{M}}^{(s)}}}
\def\GATEZ{{G_{\sst{A}}^{\sst{T}=0}}}
\def\GATEO{{G_{\sst{A}}^{\sst{T}=1}}}
\def\mdax{{M_{\sst{A}}}}
\def\mustr{{\mu_s}}
\def\rsstr{{r^2_s}}
\def\rhostr{{\rho_s}}
\def\GEG{{G_{\sst{E}}^\gamma}}
\def\GEZ{{G_{\sst{E}}^\sst{Z}}}
\def\GMG{{G_{\sst{M}}^\gamma}}
\def\GMZ{{G_{\sst{M}}^\sst{Z}}}
\def\GEn{{G_{\sst{E}}^n}}
\def\GEp{{G_{\sst{E}}^p}}
\def\GMn{{G_{\sst{M}}^n}}
\def\GMp{{G_{\sst{M}}^p}}
\def\GAp{{G_{\sst{A}}^p}}
\def\GAn{{G_{\sst{A}}^n}}
\def\GA{{G_{\sst{A}}}}
\def\GETEZ{{G_{\sst{E}}^{\sst{T}=0}}}
\def\GETEO{{G_{\sst{E}}^{\sst{T}=1}}}
\def\GMTEZ{{G_{\sst{M}}^{\sst{T}=0}}}
\def\GMTEO{{G_{\sst{M}}^{\sst{T}=1}}}
\def\lamd{{\lambda_{\sst{D}}^\sst{V}}}
\def\lamn{{\lambda_n}}
\def\lams{{\lambda_{\sst{E}}^{(s)}}}
\def\bvz{{\beta_{\sst{V}}^0}}
\def\bvo{{\beta_{\sst{V}}^1}}
\def\Gdip{{G_{\sst{D}}^\sst{V}}}
\def\GdipA{{G_{\sst{D}}^\sst{A}}}
\def\fks{{F_{\sst{K}}^{(s)}}}
\def\FIS{{F_i^{(s)}}}
\def\fpi{{F_\pi}}
\def\fk{{F_{\sst{K}}}}
\def\RAp{{R_{\sst{A}}^p}}
\def\RAn{{R_{\sst{A}}^n}}
\def\RVp{{R_{\sst{V}}^p}}
\def\RVn{{R_{\sst{V}}^n}}
\def\rva{{R_{\sst{V,A}}}}
\def\xbb{{x_B}}
\def\mlq{{M_{\sst{LQ}}}}
\def\mlqs{{M_{\sst{LQ}}^2}}
\def\lscal{{\lambda_{\sst{S}}}}
\def\lvect{{\lambda_{\sst{V}}}}
\def\PR#1{{{\em   Phys. Rev.} {\bf #1} }}
\def\PRC#1{{{\em   Phys. Rev.} {\bf C#1} }}
\def\PRD#1{{{\em   Phys. Rev.} {\bf D#1} }}
\def\PRL#1{{{\em   Phys. Rev. Lett.} {\bf #1} }}
\def\NPA#1{{{\em   Nucl. Phys.} {\bf A#1} }}
\def\NPB#1{{{\em   Nucl. Phys.} {\bf B#1} }}
\def\AoP#1{{{\em   Ann. of Phys.} {\bf #1} }}
\def\PRp#1{{{\em   Phys. Reports} {\bf #1} }}
\def\PLB#1{{{\em   Phys. Lett.} {\bf B#1} }}
\def\ZPA#1{{{\em   Z. f\"ur Phys.} {\bf A#1} }}
\def\ZPC#1{{{\em   Z. f\"ur Phys.} {\bf C#1} }}
\def\etal{{{\em   et al.}}}
\def\delalr{{{delta\alr\over\alr}}}
\def\pbar{{\bar{p}}}
\def\lamchi{{\Lambda_\chi}}
\def\qw0{{Q_{\sst{W}}^0}}
\def\qwp{{Q_{\sst{W}}^P}}
\def\qwn{{Q_{\sst{W}}^N}}
\def\qwe{{Q_{\sst{W}}^e}}
\def\qem{{Q_{\sst{EM}}}}
\def\gae{{g_{\sst{A}}^e}}
\def\gve{{g_{\sst{V}}^e}}
\def\gvf{{g_{\sst{V}}^f}}
\def\gaf{{g_{\sst{A}}^f}}
\def\gvu{{g_{\sst{V}}^u}}
\def\gau{{g_{\sst{A}}^u}}
\def\gvd{{g_{\sst{V}}^d}}
\def\gad{{g_{\sst{A}}^d}}
\def\gvftil{{\tilde g_{\sst{V}}^f}}
\def\gaftil{{\tilde g_{\sst{A}}^f}}
\def\gvetil{{\tilde g_{\sst{V}}^e}}
\def\gaetil{{\tilde g_{\sst{A}}^e}}
\def\gvqtil{{\tilde g_{\sst{V}}^e}}
\def\gaqtil{{\tilde g_{\sst{A}}^e}}
\def\gvutil{{\tilde g_{\sst{V}}^e}}
\def\gautil{{\tilde g_{\sst{A}}^e}}
\def\gvdtil{{\tilde g_{\sst{V}}^e}}
\def\gadtil{{\tilde g_{\sst{A}}^e}}
\def\delp{{\delta_P}}
\def\delzp{{\delta_{00}}}
\def\deld{{\delta_\Delta}}
\def\dele{{\delta_e}}
\def\lnew{{{\cal L}_{\sst{NEW}}}}
\def\osffp{{{\cal O}_{7a}^{ff'}}}
\def\oszg{{{\cal O}_{7c}^{Z\gamma}}}
\def\osgg{{{\cal O}_{7b}^{g\gamma}}}

%%%%%%%%%%%%%%%%%%%%%%%%%%%%%%%%%%%%%%%%%%%%%

\def\slash#1{#1\!\!\!{/}}
\def\beq{\begin{eqnarray}}
\def\eeq{\end{eqnarray}}
\def\bea{\begin{eqnarray*}}
\def\eea{\end{eqnarray*}}
\def\NCA{\em Nuovo~Cimento}
\def\IJMP{\em Intl.~J.~Mod.~Phys.}
\def\NP{\em Nucl.~Phys.}
\def\PLB{{\em Phys.~Lett.}~B}
\def\JETPLett{{\em JETP Lett.}}
\def\PRL{\em Phys.~Rev.~Lett.}
\def\MPL{\em Mod.~Phys.~Lett.}
\def\PRD{{\em Phys.~Rev.}~D}
\def\PR{\em Phys.~Rev.}
\def\PRP{\em Phys.~Rep.}
\def\ZPC{{\em Z.~Phys.}~C}
\def\PTP{{\em Prog.~Theor.~Phys.}}
% Some other macros used in the sample text
\def\Baryon{{\rm B}}
\def\Lepton{{\rm L}}
\def\sbar{\overline}
\def\stilde{\widetilde}
\def\sst{\scriptscriptstyle}
\def\vac{|0\rangle}
\def\argh{{{\rm arg}}}
\def\G{\stilde G}
\def\Wmess{W_{\rm mess}}
\def\NI{\stilde N_1}
\def\antivac{\langle 0|}
\def\infinity{\infty}
\def\mco{\multicolumn}
\def\epp{\epsilon^{\prime}}
\def\psibar{\overline\psi}
\def\nmess{N_5}
\def\chibar{\overline\chi}
\def\lagr{{\cal L}}
\def\drbar{\overline{\rm DR}}
\def\msbar{\overline{\rm MS}}
\def\conj{{{\rm c.c.}}}
\def\Et{{\slashchar{E}_T}}
\def\Etot{{\slashchar{E}}}
\def\mZ{m_Z}
\def\MPlanck{M_{\rm P}}
\def\mW{m_W}
\def\cbeta{c_{\beta}}
\def\sbeta{s_{\beta}}
\def\cW{c_{W}}
\def\sW{s_{W}}
\def\deltaeps{\delta}
\def\sigmabar{\overline\sigma}
\def\epsilonbar{\overline\epsilon}
\def\vep{\varepsilon}
\def\ra{\rightarrow}
\def\half{{1\over 2}}
\def\ko{K^0}
\def\alr{A_{\sst{LR}}}

%  \gsim and \lsim provide >= and <= signs.
\def\centeron#1#2{{\setbox0=\hbox{#1}\setbox1=\hbox{#2}\ifdim
\wd1>\wd0\kern.5\wd1\kern-.5\wd0\fi
\copy0\kern-.5\wd0\kern-.5\wd1\copy1\ifdim\wd0>\wd1
\kern.5\wd0\kern-.5\wd1\fi}}
\def\ltap{\;\centeron{\raise.35ex\hbox{$<$}}{\lower.65ex\hbox{$\sim$}}\;}
\def\gtap{\;\centeron{\raise.35ex\hbox{$>$}}{\lower.65ex\hbox{$\sim$}}\;}
\def\gsim{\mathrel{\gtap}}
\def\lsim{\mathrel{\ltap}}
%%%%%%%%%%%%%%%%%%%%%%%%%%%%%%%%%%%%%%%
%  Slash character...
\def\slashchar#1{\setbox0=\hbox{$#1$}           % set a box for #1
   \dimen0=\wd0                                 % and get its size
   \setbox1=\hbox{/} \dimen1=\wd1               % get size of /
   \ifdim\dimen0>\dimen1                        % #1 is bigger
      \rlap{\hbox to \dimen0{\hfil/\hfil}}      % so center / in box
      #1                                        % and print #1
   \else                                        % / is bigger
      \rlap{\hbox to \dimen1{\hfil$#1$\hfil}}   % so center #1
      /                                         % and print /
   \fi}                                        %

%%EXAMPLE:  $\slashchar{E}$ or $\slashchar{E}_{t}$
\setcounter{tocdepth}{2}

%%%%%%%%%%%%%%%%%%%%%%%%%%%%%%%%%%%%%%%%%%%%%

%%\begin{tabular}{r}
%%{\normalsize Caltech MAP-300}
%%\end{tabular}

%%\preprint{Caltech MAP-300}
%
%
%\hfill

%Caltech MAP-300

%\vskip

{
%\date{\today}

\title{First non-zero measurement of a nuclear electric dipole moment}
\author{Gary Pr\'{e}zeau}
\email{email: gary.prezeau@loisat.org}
\affiliation{League of Independent Scientists and Teachers (LOISAT), Torrance, 90501, CA, USA}
\begin{abstract}
This paper reports the first non-zero measurement of a nuclear electric dipole moment using a novel method based on the rate of change of a supercurrent first proposed in 2016~\cite{https://doi.org/10.48550/arxiv.1604.02152} and fleshed out in this current paper. The theory, experimental concept and implementation are described in detail. The non-zero nuclear electric dipole moment measured with over 1000 hours of data was that of $^{181}$Ta producing a best value $|d_\text{e}^\text{Ta}|=(3.39\pm0.31_\text{stat})\cdot10^{-32}e\cdot\text{cm}$ and $|d_\text{e}^\text{Ta}|=(3.39\pm3.18)\cdot10^{-32}e\cdot\text{cm}>0$ at 99.985\%CL. There is an uncertainty on the value of overall multiplicative parameters such as the self-inductance of the superconducting circuit ($\pm4\%$), the mutual inductance between the SQUID pickup coil and the sample wire ($\pm15\%$), and the magnitude of the solenoid current ($\pm5\%$). An upper-limit was estimated for the control element, $^{207}$Pb, $|d_\text{e}^\text{Pb}|\lesssim1.2\cdot10^{-31}e\cdot\text{cm}$ at 95\%CL.
\end{abstract}
\maketitle
\tableofcontents
%\pagebreak

%
\section{Introduction}\label{sec:intro}
The amount of baryonic matter present in our universe today is a small portion of the total amount of matter and energy in our universe, according to the standard model of cosmology and experimental data. The existing baryonic matter is all that remained when, shortly after the Big Bang, all the anti-baryons annihilated with nearly all the baryons, leaving approximately one billionth of the original number of baryons behind. This excess of baryons is unexplained and indicates that our universe is not invariant under time reversal (a T-transformation). Equivalently, the baryon excess indicates that a mirror universe (the resulting universe under a global parity transformation or P-transformation) with anti-particles (the resulting universe under a global charge conjugation or C-transformation) would look different from ours. Succinctly, it is said that our universe is CP-odd and T-odd.\footnote{Other terminology includes CP-violating and T-violating.}

Of the three fundamental forces or sectors currently known,\footnote{Gravitational, strong and electroweak forces.} two of them are theoretically capable of producing CP-odd interactions between particles: The first one is the electroweak sector where CP-violation was discovered in 1964 in kaon meson decay~\cite{Christenson1964}, and, more recently, in beauty baryon $\Lambda_b^0$ decay~\cite{2025}. Another potential source of CP violation is in the strong sector. Strong interactions are described by Quantum Chromodynamics (QCD), a field theory that contains a CP-odd contribution whose strength is set by a dimensionless, multiplicative parameter denoted $\tqcd$.\footnote{The actual term in QCD is simply $\theta$, but it is shifted by a phase to $\tqcd$ when the quarks gain mass through spontaneous symmetry breaking.} The $\tqcd$ term contributes to the electric dipole moments (EDMs) of nucleons, as does the electroweak sector. A $\tqcd\sim1$ implies a nucleon EDM dominated by the $\tqcd$ term. The first attempt to measure the neutron EDM ($\deneut$) was in 1951~\cite{Smith:1957ht} and they found $|\deneut|<5\cdot10^{-20}e\cdot\text{cm}$. That upper limit on $\deneut$ suggested that $\tqcd<10^{-4}$. The unexpectedly small value of $\tqcd$ is known as the strong CP problem. The current limit on the neutron EDM using ultracold neutrons~\cite{Abel2020} is $\deneut<1.8\cdot10^{-26}e\cdot\text{cm}$ at the 90\% CL suggesting that $\tqcd<10^{-10}$.

Other experiments have sought to find atomic EDMs that arise from electron and nucleon EDMs, as well as CP-odd interactions between subatomic particles. The most stringent upper limit for an atomic EDM comes from $^{199}$Hg~\cite{Graner2016} with $|d_\text{e}^\text{Hg}|<7.4\cdot10^{-30}e\cdot\text{cm}$, a value determined using electric fields of the order of 10~kV/cm that were parallel or anti-parallel to an external magnetic field. The spin precession frequency (the Larmor frequency) of the atom depends on the orientation of the electric field relative to the magnetic field. That dependence is proportional to $|d_\text{e}^\text{Hg}|$.

Attempts to measure an atomic EDM through the application of an external electric field have to contend with Schiff screening~\cite{Schiff1963}. Schiff screening occurs because regardless of the form and shape of the external electric field, any atomic EDM results in a translational shift of the position of all atomic particles in such a way that the eigenstates and eigenvalues of the system do not depend linearly on the atomic EDM, assuming a pointlike nucleus and no relativistic corrections. Since real nuclei have finite extent and are typically not spherical, while high-$Z$ atoms have significant relativistic corrections, $Z$ being the atomic number, the screening of external electric fields is not complete resulting instead in a large suppression of any net EDM~\cite{Flambaum2002,Flambaum2012} rather than a cancellation. Hence, a considerable amount of theoretical work has gone into identifying circumstances under which EDMs are enhanced~\cite{Khrip1984,Auerbach1996,Spevak1997,Engel2000,Engel2003,Dobaczewski2018,Flambaum2020} prompting a number of experimental attempts to detect EDMs in atoms susceptible to these enhancements~\cite{Parker2015,Bishof2016,Zheng2022}.

Paramagnetic atoms and molecules are also laboratories to measure the electron EDM thanks to the fact that an unpaired electron is exposed to internal electric fields, and, in large atoms, relativistic effects can magnify the impact of an external electric field. Attempts at measuring the electron EDM in atoms and molecules have been going on for decades~\cite{Regan2002,Hudson2002,2018,Panda2019} with a current upper-bound of $|d_\text{e}|<4.1\cdot10^{-30}~e\cdot\text{cm}$ measured using HfF$^+$~\cite{Roussy2023}.

The nearly infinitesimal value of $\tqcd$ led to the hypothesis of an axial U(1) symmetry and a corresponding axial field that could undergo spontaneous symmetry breaking~\cite{Peccei1977}. In order to minimize the spontaneously broken vacuum, the $\tqcd$-term and the vacuum expectation value of the axial U(1) field must cancel, thus explaining the infinitesimal value of the neutron EDM. It was quickly realized that the spontaneous symmetry breaking would lead to the existence a pseudo-scalar particle (the axion)~\cite{Weinberg1978,Wilczek1978} that would have been produced copiously at the beginning of the universe~\cite{Abbott1983,Preskill1983,Dine1983} making it an excellent dark matter candidate (see \cite{ChadhaDay2022} for a non-technical review of the axion). Recent experimental searches for axion-like particles have focused on detecting oscillations in the precession of neutron/atomic magnetic moments due to an EDM induced by an axion background field~\cite{Abel2017,DiLuzio2024}.

In this paper, a new method for measuring nuclear EDMs is presented. This novel approach does not use external electric fields. A corresponding experimental concept was developed and implemented, as described in detail below. This work brings together different physics subfields including solid state physics, astronomy, astrophysics, and nuclear/particle physics. \Cref{sec:theory} details the theory behind the method by first deriving the Maxwell equations in the presence of a neutral pointlike ``nucleus''\footnote{The non-zero electric charge of a nucleus is not relevant to that initial discussion.} that has both a nuclear magnetic dipole moment and a nuclear EDM. The solutions for the electromagnetic fields produced by that pointlike nucleus are split into contact and long range contributions and generalized to a bulk material; expressions for the nuclear polarization and magnetization are given. The Wigner-Eckart theorem is then used to re-express the electric field stemming from the nuclear EDMs in terms of the magnetic field stemming from the nuclear magnetic moments. This leads to an expression for the long range electric field induced by the nuclear EDMs in terms of the nuclear magnetization, \Eq{ELR}. This long range electric field is referred to as the electrization field to distinguish it from the total electric field which is null. An expression for a net atomic EDM is next derived that includes the possibility that the valence electrons may no longer be bound to the atom, as is the case for atoms in a conductor or a superconductor. When valence electrons are in the conduction band, a pointlike nuclear EDM is no longer completely screened by the remaining core electrons. The total electric field which is the sum of the contact electric field plus the electrization field in the conductor is still zero, but in a superconductor, the contact electric field has decoupled from the supercurrent as the electrons forming the Cooper pairs can no longer scatter off the lattice sites. In a superconductor, the Cooper pairs in the penetration depth can only be affected by the bulk long range electric field of the nuclear EDMs, i.e., the electrization field.

In the presence of an external magnetic field, the electrization field in the penetration depth is shown to follow Curie's law as it is proportional to the nuclear magnetization (in the experiment, $\mun B/(k_\text{B}T)\lesssim10^{-3}$). Since supercurrents are located in the penetration depth of a superconductor, the linear differential equation for a supercurrent in a circuit is given in \Eq{didt}: it has a term for the impedance in the circuit due to the self-inductance, and an electromotive force equal to the line integral of the electrization field induced by an external magnetic field generated by a solenoid (see \Fig{sampAssembly}): that external magnetic field creates a non-zero nuclear magnetization in the penetration depth which in turn creates an electrization field proportional to the nuclear magnetization. The electrization field can drive an increasing supercurrent in the sample wire. The free magnetization, $\Fedm{ele}$ is then defined in \Eq{bele} and quantifies the ability of an element to generate an electrization field. Energy conservation is discussed and it is shown that the energy for the increasing supercurrent ultimately comes from the current in the solenoid that is producing the external magnetic field.

The experimental objectives are listed and rewritten here for convenience:
\begin{enumerate}
\item Use two elements for the sample wire with the principal element, tantalum ($^{181}_{73}$Ta), having a much larger free magnetization than the control element, lead. The ratio of free magnetizations are $\Fedm{Ta}/\Fedm{Pb}=18.8$, see \Tab{nucData}.
\item Determine the existence of an increasing current in the principal element sample wire and verify that it follows \Eq{betaSlope}, in other words, that it follows Curie's law
\begi
The steepness of the slope is inversely proportional to the temperature.
The steepness of the slope is proportional to the magnetic field, i.e., to the current in the sample solenoid.
\ei
\item Determine that in the control element sample wire, the effect is much smaller or undetectable.
\item Consider the correlation between the lead and tantalum data, and show that removing a non-zero contribution of the electrization field from the tantalum data improves the correlation.
\end{enumerate}

The experimental sample assembly is described in \cref{sec:data}. It consists of the sample wire which is positioned against a SQUID\footnote{Superconducting QUantum Interference Device.} that detects minute changes in the sample supercurrent, \Is. A segment of the sample wire is inside the sample solenoid, itself part of a single superconducting circuit that includes a booster solenoid that is wound around an aluminum core which boosts the solenoid wire current, \Isol, by a factor of approximately 10 when the temperature drops below the critical temperature of aluminum. The booster solenoid is itself inside the main solenoid that induces the initial $\text{\Isol}$ above the critical temperature of aluminum. The sample assembly also has various layers of shielding. The procedures for welding the sample wire leads and joining the sample/booster solenoid leads are also described, and data is provided showing that the welds and joint are superconducting.

In \cref{sec:squid}, the electronics of the SQUID system is described, as well as the calibration that allows the conversion of the SQUID voltage to a sample current value. The sample assembly is located in an adiabatic demagnetization refrigeration (ADR) system that can reach temperatures below 40~mK. In \cref{sec:regen}, the process of inducing $\text{\Isol}$ from an externally applied current $\text{\Ia}$ is described, as well as the REGEN process of the ADR. Going forward, $\text{\Ia}$ and $\text{\Isol}$ are used in an interchangeable fashion since they are proportional to each other through a ratio of inductances. It is during REGEN that the FAA salt pill\footnote{Ferric Ammonium Alum salt pill can reach a minimum practical temperature of about 35~mK~\cite{Shirron2014}.} has its spin entropy minimized using a powerful magnetic field generated by the pill solenoid carrying a current \ips. The REGEN process also provides experimental measurements of $\text{\Isol}$ that are compared with their theoretical values calculated from numerically determined inductances of all the solenoids.

The next sections describe over 1000 hours of SQUID voltage data that are split into two categories: the fixed temperature data and the zero-field data. The notation for the fixed temperature datasets is provided in \cref{sec:fixedT} as well as the definitions of the parameters appearing throughout this paper. The fixed temperature and fixed sample current datasets are used to verify \Eq{didt}. In an ADR, a constant temperature can be maintained by manipulating the spin entropy of the salt pill with \ips~ creating a potential systematic for fixed temperature data. The level of that systematic was assessed with \Isol=0 data, and controlled for by looking at ``differenced datasets''. These are defined as the resulting dataset when subtracting one dataset composed of data collected at a temperature $T$ and current \Isol, from another dataset composed of data collected at the same temperature $T$ but at a different current \Isol. The main figures for tantalum are in \Fig{fixedTempCurr-T} where the datasets collected at five different temperatures and four different currents are plotted. \Fig{gammaTTaMod} plots the slopes from the middle column of \Fig{fixedTempCurr-T} (the differenced datasets that minimize systematics) as a function of inverse temperature. \Fig{SpropIa} was used to determine the dependence of the fixed temperature tantalum data on \Ia. \Fig{RevB} plots the tantalum data at a fixed temperature of 0.065~K with the direction of the solenoid magnetic field reversed; this data is referred to as the reverse field data and was used in both the determination of the inverse temperature dependence and the linear dependence on \Ia. The fixed temperature data for lead is given in \Fig{fixedTempCurrPb} with the middle column showing the plots of the differenced datasets that minimize systematics.

The zero-field datasets in \cref{sec:zfd} are each collected at a fixed current $\text{\Ia}$ over a continuous range of temperatures, from the minimum, base temperature shortly after REGEN, to a maximum temperature reached after the sample assembly has warmed up over days from ambient heat absorption. Zero-field data, where the pill solenoid current is zero, is used to test the time integral of \Eq{didt}. The zero-field datasets are plotted in \cref{TaPbsubKall,TaPbPlots}.

\Cref{sec:systematics} discusses systematics and it is shown that the data is inconsistent with either external systematics or an internal systematic in the form of a finite resistance in the sample solenoid. Errors are calculated and the final value $d_\text{e}^\text{Ta}$ is given.

\Cref{sec:disc} discusses potential research applications of the electrization field:
\begi
The same method used to measure $d_\text{e}^\text{Ta}$ can be used to measure the nuclear EDM of a large number of elements, creating a vast dataset that over-constrain leading order CP-violating interactions in atoms in order to experimentally determine their coupling constants.
The electrization field can also be generated by $p$-wave superconductor Cooper pairs that have spin 1 and can be used for the measurement of the electron EDM with a sensitivity many orders of magnitude above current and projected limits of future experiments.
The electrization field could detect Schiff moment enhancements of nuclear EDMs.
The great sensitivity of the electrization field to nuclear EDMs suggests that it could detect or strongly constrain EDMs induced by the background axion field including local variations in the density of the axion background~\cite{Sikivie2002,Przeau2015}.
\ei

This paper has four appendices: \cref{app:potential} discusses the electrization field from the point of view of a potential and provides the solutions to the Shr\"{o}dinger equation; \cref{app:osc} shows that the electrization field leads to oscillations of the supercurrent even in the absence of a magnetic field; \cref{app:Tstddev} describes ``jumps'' in the fixed temperature data and how they were handled in the data analysis; \cref{app:avrMagMom} calculates the average magnetization of $p$-wave superconductor Cooper pairs.

\section{Theory}\label{sec:theory}
Consider an  effective Lagrangian that includes only photons and a pointlike, neutral ``nucleus'' with only magnetic and electric dipole moments,  $\mun$ and $\den$ respectively:
\beq\label{efflag}
\mathcal{L}_{\text{eff}}= -\bar{N} \frac{\sigma^{\mu\nu}}{2}\! \left[ \mu_\text{N} \!-\! ic\den\gamma_5 \right]\!N F_{\mu\nu} -\frac{F^{\mu\nu}F_{\mu\nu}}{4\mu_0}
\eeq
where $N$ is the nuclear field centered at the origin, $\langle N,\mathbf{r}|N,\mathbf{r}^\prime\rangle=\delta(\mathbf{r}-\mathbf{r}^\prime)$ and $F_{\mu\nu}=\partial_\mu A_\nu-\partial_\nu A_\mu$ is the electromagnetic tensor. In the static, non-relativistic limit, the Gauss and Ampère equations are found using the Euler-Lagrange equations applied to \Eq{efflag}
\beq
\bm{\nabla}\cdot\mathbf{E}&=& \frac{1}{\epsilon_0}  \vden\cdot\bm{\nabla}\delta(\mathbf{r}) \\
\bm{\nabla}\times\mathbf{B}&=& \mu_0 \vmun\times\bm{\nabla}\delta(\mathbf{r})~.
\eeq
The solutions to the electrostatic and vector potentials that  vanish at infinity are the usual results for pointlike dipoles
\beq
V(\bm{r})&=&-\vden\cdot\bm{\nabla}\frac{1}{4\pi\epsilon_0 r} \\
\mathbf{A}(\bm{r})&=&-\vmun\times\bm{\nabla}\frac{\mu_0}{4\pi r}~,
\eeq
with electric and magnetic fields given by
\beq\label{LRcontact}
\mathbf{E}(\bm{r})&=&\mathbf{E}_{\vlr}(\bm{r})+\mathbf{E}_{\bm{\delta}}(\bm{r}) \\
\mathbf{E}_{\vlr}(\bm{r})&\equiv& \frac{3(\vden \cdot \hat{\mathbf{r}}) \hat{\mathbf{r}}-\vden }{4\pi \epsilon_0 r^3} \\ \label{contactE}
\mathbf{E}_{\bm{\delta}}(\bm{r})&\equiv&-\frac{1 }{3 \epsilon_0}\vden \delta(\mathbf{r}) \\
\mathbf{B}(\bm{r})&=&\mathbf{B}_{\vlr}(\bm{r})+\mathbf{B}_{\bm{\delta}}(\bm{r}) \\
\mathbf{B}_{\vlr}(\bm{r})&\equiv& \mu_0\frac{3(\vmun \cdot \hat{\mathbf{r}}) \hat{\mathbf{r}}-\vmun }{4\pi r^3} \\
\mathbf{B}_{\bm{\delta}}(\bm{r})&\equiv&\mu_0\frac{2 }{3}\vmun \delta(\mathbf{r})
\eeq
where the contributions to the electromagnetic fields that are long range, $\mathbf{E}_{\vlr}, ~\mathbf{B}_{\vlr}$ are kept separate from the short range, contact terms proportional to Dirac delta function, $\mathbf{E}_{\bm{\delta}},~\mathbf{B}_{\bm{\delta}}$.  Given the field solutions for a single pointlike electric/magnetic dipole, the field solutions in a bulk material with the same boundary condition can be found by superposition of the single dipole solutions
\begin{align}\label{gauss}
&\bm{\nabla}\cdot[\epsilon_0\mathbf{E}(\bm{r}) \!+\! \mathbf{P}(\bm{r})  ] =0
\\ \label{ampere}
&\bm{\nabla}\!\times\!\left[\frac{\mathbf{B}(\bm{r})}{\mu_0}\!-\! \mathbf{M}(\bm{r})\right] \! =0
\\ \label{polEq}
&\frac{\mathbf{P}(\bm{\bar{r}})}{\epsilon_0} = -\frac{1}{\mathcal{v}} \!\!\int_{\mathcal{v}}\! \sum_{i\in\text{Lat}}\! \!  \left[\mathbf{E}_{\vlr}(\bm{r}\! -\! \bm{r}_i)\!+\!\mathbf{E}_{\bm{\delta}}(\bm{r}\! -\! \bm{r}_i)\right]\text{d}^3r   \\ \label{magEqB}
&\mu_0\mathbf{M}(\bm{\bar{r}}) = \frac{1}{\mathcal{v}} \!\!\int_{\mathcal{v}}\! \sum_{i\in\text{Lat}} \! \! \left[\mathbf{B}_{\vlr}(\bm{r}\! -\! \bm{r}_i)\!+\!\mathbf{B}_{\bm{\delta}}(\bm{r}\! -\! \bm{r}_i)\right]\text{d}^3r
\end{align}
where sums are over all lattice points and the integrals are over an infinitesimal volume ${\mathcal{v}}$ centered at $\bm{\bar{r}}$ that is much larger than the volume of a single atom; these integrals are necessary as magnetization and polarization are bulk properties representing the number of magnetic and electric dipoles per unit volume respectively. Applying the Wigner-Eckart theorem provides the relationship
\beq\label{WE}
& &\vden = \frac{\langle N| \mathbf{\hat{J}}\cdot\mathbf{\hat{d}}_\text{\bf{e}}^\text{\bf{N}} | N \rangle}{\langle N|\mathbf{\hat{J}}\cdot\bm{\hat{\mu}}_\text{\bf{N}} | N\rangle }\vmun \\ \label{WEdef}
& &\langle N|\mathbf{\hat{J}}\cdot\bm{\hat{\mu}}_\text{\bf{N}} | N\rangle=g\mu_\text{N}J(J+1)
\eeq
where $|N\rangle$ is the nuclear wave function, $g$ is the g-factor, $J$ is the nuclear spin and $\mu_\text{N}$ is the nuclear magnetic moment. Hence a suppressed nuclear magnetic moment leads to a suppressed nuclear EDM (NEDM). The polarization $\mathbf{P}(\bm{r})$ can be rewritten in terms of $\mathbf{B}_{\vlr},~\mathbf{B}_{\bm{\delta}}$ using \Eq{WE}, \Eq{WEdef} and the fact that $\mun=g\mu_\text{N}J$
\begin{align}\label{newP}
&\frac{\mathbf{P}(\bm{r})}{\epsilon_0} = -c^2\frac{\denbar}{\mun\mathcal{v}}\int_{\mathcal{v}}\! \sum_{i\in\text{Lat}} \![\mathbf{B}_{\vlr}(\bm{r}\! -\! \bm{r}_i)\!-\!\frac{1}{2}\!\mathbf{B}_{\bm{\delta}}(\bm{r}\! -\! \bm{r}_i)]\text{d}^3r\\ \label{denbar}
&\denbar\equiv\frac{\langle N| \mathbf{\hat{J}}\cdot\mathbf{\hat{d}}_\text{\bf{e}}^\text{\bf{N}} | N \rangle}{(J+1)}
\end{align}
where $\denbar$ is the average NEDM of the a nucleus with spin $J$ aligned with the axis defined by the direction of the uniform magnetic field which  sets the direction of the magnetization $\mathbf{M}(\bm{r})$. \Eq{newP} shows that  a non-zero bulk electrical polarization can arise by applying a magnetic field to produce a non-zero nuclear magnetization. To take a concrete case, if the bulk material is cylindrical and a  co-axial  magnetic field polarizes the nuclear magnetic moments, it can simultaneously create a nuclear electric polarization with one important difference: the magnetic field lines due to the magnetization in the cylinder will be continuous across the end surfaces of the cylinder. In contrast, only the surface charge densities of the cylinder will contribute to the electric field and the field lines will be discontinuous across the surfaces because they change direction. This important difference between the magnetization and polarization field lines is entirely due to the short range $\bm{\delta}$ interactions: the long and short range fields add in the case of magnetization while they cancel for polarization. Hence, in a torus where there are no end surfaces co-axial with the axis of the torus, there can be no electric polarization, unlike magnetization that can exist in a torus or, more specifically, a wire whose ends have been welded. Setting $\mathbf{P}(\bm{r})=0$ in \Eq{newP} and using \Eq{magEqB} we obtain\footnote{\Eq{ELR} could also have been obtained from the fact that $1/\mathcal{v}\!\int_{\mathcal{v}} \sum_{i}\delta(\bm{r}\! -\! \bm{r}_i)\text{d}^3r=N/V$, the number density of nuclei in the bulk material, and using \Eq{polEq}, \Eq{magEqB} and \Eq{denbar}.}
\beq\label{ELR}
 \frac{1}{\mathcal{v}} \!\!\int_{\mathcal{v}}\! \sum_{i\in\text{Lat}} \!\! \mathbf{E}_{\vlr}(\bm{r}\! -\! \bm{r}_i)\text{d}^3r=\frac{\denbar}{3\epsilon_0\mun}\mathbf{M}(\bm{\bar{r}})
\eeq
Including the possibility that electrons may have an EDM as well, a general criterion for a non-zero atomic electric dipole moment, $d_\text{A}$, is derived for an atom located in a bulk material immersed in a uniform magnetic field. The bulk material is described by an ideal crystalline lattice where the nuclei occupy the lattice sites. $d_\text{A}$ can be considered as the sum of all the atomic particle EDMs, including EDMs induced by CP-odd interactions between the atomic particles. We want to see if an atomic EDM $d_\text{A}$ located at a specific lattice site changes that atom's electron energy levels and, relatedly, its Fermi surface, ignoring sub-leading ${\cal{O}}(d_{\text{A}}^2)$, ${\cal{O}}(m_e/m_\text{N})$ corrections where $m_e$ and $m_\text{N}$ are the electron and nuclear masses respectively. Taking the static uniform magnetic field direction to be +$\bf{\hat{z}}$, the Hamiltonian acquires a potential energy term due to $d_\text{A}$
\beq\label{bulkV}
H = -k\sum_{i=1}^{N_\text{b}}\frac{Ze^2}{r_i} -k\sum_{i=1}^{N_\text{b}} \frac{e d_\text{A}\cos\theta_i}{r_i^2} + H_\text{r} ~, 
\eeq
where $Z$ is the atomic number of the material atoms, $N_\text{b}\le Z$ is the number electrons bound to the atom, $r_i$ is the distance of the $i^\text{th}$ bound electron from the nucleus and $\theta_i$ is the polar angle between the $i^\text{th}$ electron coordinate vector and the $z$-axis. The first sum adds up the potential energy of each  electron in the nuclear electric field while the second sum adds up the potential energy of each electron in the atomic EDM field. Since the atomic magnetic dipole points in the $\bm{\hat{z}}$ direction, the atomic EDM is either parallel or anti-parallel with the $z$-axis; it will be taken as aligned below.  $H_\text{r}$ contains the electron kinetic terms, the interactions between the uniform magnetic field and the atomic particles, and the potential energy of electron-electron interactions. Note that in \Eq{bulkV}, there is no term for the interaction energy between the pointlike nucleus at the lattice site and the atomic EDM as the potential energy of a pointlike charged particle equidistant to two opposite, pointlike charges of equal magnitude is zero. To leading order in $d_{\text{A}}$
\beq
H &\cong& -k\sum_{i=1}^{N_\text{b}}\frac{Ze^2}{|\vec{r}_i-\vec{r}_0|} +{\cal{O}} ( d_{\text{A}}^2)\label{effH} + H_\text{r} \\
\vec{r}_0&\equiv& \frac{d_{\text{A}}}{Ze}\bm{\hat{z}}~,
\eeq
Taking $\hat{p}_i$ as the momentum operator of the $i^\text{th}$ electron and writing the Hamiltonian as an operator, we have
\beq
\hat{H}=e^{\sum_i\hat{p}_i\cdot\vec{r}_0/\hbar}\hat{H}_0e^{-\sum_i\hat{p}_i\cdot\vec{r}_0/\hbar}~,
\eeq
where $\hat{H}_0$ is the unperturbed Hamiltonian with $d_\text{A}=0$.  Since $\vec{r}_0$ is a classical number, the translation operator $e^{\sum_i\hat{p}_i\cdot\vec{r}_0/\hbar}$ does not affect the electron kinetic energy terms. Furthermore, an overall translation of the electron cloud does not affect the electron-electron potential energy nor the uniform magnetic field interactions terms. Taking the unperturbed Hamiltonian eigenvectors as $|n\rangle$ with $\hat{H}_0|n\rangle=E_n|n\rangle$ yields
\beq
\hat{H}e^{\sum_i\hat{p}_i\cdot\vec{r}_0/\hbar}|n\rangle=E_ne^{\sum_i\hat{p}_i\cdot\vec{r}_0/\hbar}|n\rangle
\eeq
It is seen that the eigenvalues are identical to the eigenvalues of the unperturbed system. Therefore, the Fermi surface is unaffected by the presence of the EDM at the lattice site, as are any bulk material properties that depend on the structure of the Fermi surface.

The net EDM of the atom, $\vec{d}_\text{net}$, is $d_\text{A}\bm{\hat{z}}$ plus the dipole due to the displacement of the electrons:
\begin{align}\label{nedEDM}
&\vec{d}_\text{net}= d_{\text{A}}\bm{\hat{z}} +\sum_i^{N_\text{b}}q_j\vec{r}_i= d_{\text{A}}\bm{\hat{z}}-N_\text{b}e\vec{r}_0 =\sigma_\text{sf} d_{\text{A}} \bm{\hat{z}}\\ \label{sigmasf}
&\sigma_\text{sf}\equiv \frac{N_\text{c}}{Z}\ne0 ~ \text{for metals.}
\end{align}
where $N_\text{c}=Z-N_\text{b}$ is the number of electrons in the conduction band and $\sigma_\text{sf}$ is a suppression factor of the atomic dipole. When $N_\text{b}=Z$ as in an insulator, $d_\text{net}=0$ similar to the Schiff theorem for neutral atoms. For a pure metal composed of a single element, $N_\text{c}$ is the number of valence electrons in the conduction band and $d_\text{net} \ne 0$.\footnote{Since the metal atom at a lattice site has a net charge, the dipole moment depends on the location of the origin. However, this is an artifact since adding the potential for each atomic constituent would result in a dipole of charge $N_\text{b}$ plus a Coulomb contribution from the excess charge, the latter's electric field always being canceled by the conduction band electrons.} However, $d_\text{net} \ne 0$ for $N_\text{c}<0$ also. $N_\text{c}<0$ can occur for elements in a compound where the target nuclei for measuring the EDM has a higher electronegativity than its neighbors in which case $N_\text{c}$ may not be an integer (see also \Eq{didtgen}).

If the valence electrons are all in the conduction band as is the case for Pb and Ta, the inner electrons will all be in closed shells and an applied magnetic field cannot induce a significant atomic magnetic dipole moment from the electron cloud\footnote{The magnetic shielding constant is defined as $B_\text{eff}=(1-\sigma_\text{shielding})B_\text{ext}$ and is due to diamagnetism. For Xe and Rn, the noble gases nearest Pb and Ta, the shielding constants are 0.007 and 0.02 respectively~\cite{Fukuda2003}.  For the largest magnetic field applied in this experiment, $\sim0.01$~T, this  corresponds to an induced atomic magnetic moment $\sim\mu_\text{N}$. If one imagines a process by which atomic diamagnetism can induce an electric dipole moment, it will be seen experimentally that this was not the case in the present work since it, presumably, would have affected Ta and Pb at a similar order of magnitude.}. Furthermore, any nuclear EDM induced by $s$-wave electrons undergo CP-violating interactions inside the nucleus effectively changes the nucleus EDM. As for the conduction electrons, they will form Cooper pairs with total spin 0 in the materials used in this current experiment. Hence, for lead and tantalum, the atomic electric dipole moment stems entirely from the nuclear EDM. Including the screening factor, \Eq{ELR} becomes
\beq\label{ELRs}
\mathbf{\mathcal{E}}_{\bm{\lambda}}(\bm{\bar{r}})= \frac{1}{\mathcal{v}} \!\!\int_{\mathcal{v}}\! \sum_{i\in\text{Lat}} \!\! \mathbf{E}_{\vlr}(\bm{r}\! -\! \bm{r}_i)\text{d}^3r=\denbar\frac{\sigma_\text{sf}}{3\epsilon_0\mun}\mathbf{M}(\bm{\bar{r}})
\eeq
where $\mathbf{\mathcal{E}}_{\bm{\lambda}}(\bm{\bar{r}})$ will be referred to as the electrical magnetization or electrization for short.

Hence, an atom  with a pointlike nucleus and a non-zero nuclear magnetic moment can exhibit a partially unscreened atomic electric dipole if it is in a bulk metal. Furthermore, when the metal is immersed in a uniform magnetic field $\mathbf{B}$, a fraction of the magnetic and electric dipole moments align with $\mathbf{B}$, leading to a magnetization $\mathbf{M}(\bm{\bar{r}})$ and an electrization $\mathbf{\mathcal{E}}_{\bm{\lambda}}(\bm{\bar{r}})$. However, in a normal metal, the overall electric field can only come from the polarized surfaces of the bulk material as $\mathbf{\mathcal{E}}_{\bm{\lambda}}(\bm{\bar{r}})$ is precisely canceled by the contact electric field $\frac{1}{\mathcal{v}} \!\!\int_{\mathcal{v}}\!\sum_i\mathbf{E}_{\bm{\delta}}(\bm{r}\! -\! \bm{r}_i)\text{d}^3r$.

In the BCS theory of superconductivity however, the Cooper pairs can no longer scatter off the lattice sites which are effectively integrated out leaving only long range electrical interactions whose quantized form is the phonon. If the probability that a Cooper pair scatters off a lattice site is zero, its wavefunction must be zero there. Formally, electrons, photons and phonons constitute the complete particle content of the BCS theory and electrons form Cooper pairs.  The absence of the lattice sites from BCS theory implies the absence of the contact electric field which is non-zero only at the location of the nucleus; the Cooper pairs thus interact only with the non-zero electrization field.

A magnetic field can penetrate a superconductor within the penetration depth which is also where the supercurrents are located. Hence, if the magnetic field induces a $\mathbf{M}(\bm{\bar{r}})\ne0$ in the penetration depth, the supercurrents cannot perceive the contact electric field and will accelerate under the influence of the long range electrization field.

\subsection{Experimental model, supercurrent and energy conservation}\label{expModel}

A sample loop wire of radius $R_\text{w}$ is located on the axis of a tightly wound superconducting solenoid with $N_\text{sol}$ loops per meter. That superconducting solenoid will be referred to as the sample solenoid with a supercurrent $\text{\Isol}$ induced at the start of the experiment. The magnitude of the magnetic field strength in the wire is given by
\beq\label{Hfield}
H=e^{(\rho-R_\text{w})/\lambda_\text{L}}N_\text{sol}I_\text{sol}
\eeq
where $\rho$ is the distance from the wire axis. The magnetization $M$ is related to the magnetic field strength by
\beq
{M}=\chi_{\text{N}}{H}~.
\eeq
Curie's law for the magnetic susceptibility of a nucleus with spin $I_\text{N}$ is
\beq
\chi_{\text{N}}=\frac{ \mu_0a_\text{b}N_\text{N}  }{  3k_\text{B}T  }  \mun^2  \frac{I_\text{N}+1}{I_\text{N}}
\eeq
yielding for the electrization field
\begin{align}\label{curieSol}
{\mathcal{E}}_{\lambda}(\rho)=-e^{\frac{r-R_\text{w}}{\lambda_\text{L}}}\denbar \mu_0^2c^2    \frac{N_\text{c}}{Z} \frac{I_\text{N}+1}{I_\text{N}} \frac{a_\text{b}N_\text{N}}{9} \frac{\mun}{k_{\text{B}}T}N_\text{sol}I_\text{sol}
\end{align}
where $a_\text{b}$ is the natural abundance factor of the isotope, $N_\text{N}$ is the nuclear number density of the superconductor, $\mun$ is the nuclear magnetic moment, and $T$ is the temperature.

The electrization field is used to calculate the electromotive force and write down the circuit equation for the sample wire where a segment is inside the solenoid sample of length $I_\text{sol}$. This is calculated by averaging the electrization field over the cross-sectional area of the wire where the supercurrents exist. That area is the penetration depth defined by the exponential factor $e^{(r-R_\text{w})/\lambda_\text{L}}$ yielding  
\beeq
\begin{aligned}
{\mathcal{\bar{E}}}_{\lambda}&=\frac{\int_0^{R_\text{w}}{\mathcal{E}}_{\lambda}(\rho)\rho\text{d}\rho}{\int_0^{R_\text{w}}e^{\frac{r-R_\text{w}}{\lambda_\text{L}}}\rho\text{d}\rho} \\ \label{Ebar}
&=\denbar \mu_0^2c^2    \frac{N_\text{c}}{Z} \frac{I_\text{N}+1}{I_\text{N}} \frac{a_\text{b}N_\text{N}}{9} \frac{\mun}{k_{\text{B}}T}N_\text{sol}I_\text{sol}
\end{aligned}
\eeeq
The electromotive force is then ${\mathcal{\bar{E}}}_{\lambda}l_\text{sol}$. Other elements of the circuit are the impedances: resistive, capacitive and inductive:
\begi
Resistive: considering limits on the resistance in a superconducting loop from persistent current experiments~\cite{File1963}, we will assume that the potential drop across a possibly non-zero resistance in the circuit is negligible to be verified experimentally with the data.
Capacitive: the sample wire has parallel segments that are about 1~mm apart, that could lead to a parasitic capacitance. The expected signal will have a near infinitesimal frequency and therefore any parasitic capacitance impedance will be immensely larger than the self-inductance impedance and can be ignored as a current path.
Inductive: There are three contributions to this impedance:
\up
The self-inductance of the sample wire, \Lloop, which multiplies the time derivative of the sample wire current, \Is~ (see \Fig{sampAssembly}). This measurement of this time derivative is the experimental objective (see below).
The mutual inductance with a free current external to the sample assembly, $M_\text{ps}$, namely the current of the pill solenoid, \ips\footnote{No other external current was identified as a source of noise.}.
A thermo-electric effect that depends only on the temperature and stems from the booster solenoid since the magnitude of temperature-dependent variations of the SQUID data decreases with \Isol; that inductive effect will be written as \Isol$P$(\Isol)$\text{d}f(T)/\text{d}t$ where $P$(\Isol) is taken to be a polynomial of \Isol.
\down
\ei
Hence, the circuit equation from Faraday's law only has the electromotive force and inductive elements. Defining the electromotive force per unit inductance, ${\cal{E}}_L={\mathcal{\bar{E}}}_{\lambda}l_\text{sol}/\text{\Lloop}$ gives\footnote{We emphasize that it is the nuclear magnetic moments that are aligned with the magnetic field, and that the nuclear electric dipole moment must be either parallel or anti-parallel to the magnetic moments.}
%\begin{equation}
\begin{align}\label{didt} 
\frac{\text{d}\text{\Is}}{\text{d}t}&={\cal{E}}_L + M_\text{ps}\frac{\text{d}\text{\ips}}{\text{d}t} + I_\text{sol}P(I_\text{sol})\frac{\text{d}f(T)}{\text{d}t}  \\ \label{betaSlope}
{\cal{E}}_L&\equiv \denbar \mu_0^2c^2  \frac{N_\text{c}}{Z} \frac{I_\text{N}+1}{I_\text{N}} \frac{a_\text{b}N_\text{N}}{9} \frac{\mun}{k_{\text{B}}T} \frac{l_\text{sol} N_\text{sol}I_\text{sol}  }{\text{\Lloop}}~. 
\end{align}
%\end{equation}
${\cal{E}}_L$ is constant at fixed temperature $T$ and current \Isol. The factors that are element-dependent in \Eq{betaSlope} can be separated from those that are depend on the experiment-specific parameters
\beq
{\cal{E}}_L&\equiv& \Fedm{ele} S_\text{edm}   \frac{ \denbar }{ e }   \\ \label{bele}
\Fedm{ele} &\equiv&  \frac{N_\text{c}}{Z} \frac{I_\text{N}+1}{I_\text{N}}a_\text{b}N_\text{N}\mun \\
\Sedm&\equiv&  \frac{ ec^2 \mu_0^2  l_\text{sol} N_\text{sol} I_\text{sol} }{9\text{\Lloop}k_{\text{B}}T  } 
\eeq
where $e$ is the electron charge magnitude. $\Fedm{ele}$ has units of magnetization and it is defined as the multiplication of the atomic element parameters that can boost or suppress the magnetization. In this work, it will be referred to as the {\it free magnetization} and refers to the amount of magnetization available in a given element for producing an electrization field. The relevant parameters for Ta and Pb used in this experiment are given in \Tab{nucData}. $\Sedm$ has units of frequency and is the experimental sensitivity to the EDM; it is a large number that boosts the detectability of the EDM. For this experiment, the greatest sensitivity we reached was $\Sedm=4.4\cdot10^{19}$~Hz.

Before turning to energy conservation, the experimental objectives are listed and the experimental strategy to measure $\EL$ summarized. The experimental objectives are
\begin{enumerate}
\item Use two elements for the sample wire with the principal element, tantalum, having a much larger free magnetization than the control element, lead. The ratio of free magnetizations are $\Fedm{Ta}/\Fedm{Pb}=18.8$, see \Tab{nucData}.
\item Determine the existence $\EL$ in the principal element sample wire and verify that it follows \Eq{betaSlope}, in other words, that it follows Curie's law
\begi
The steepness of the slope is inversely proportional to the temperature.
The steepness of the slope is proportional to the magnetic field, i.e., to the current in the sample solenoid.
\ei
\item Determine that in the control element sample wire, the effect is much smaller or undetectable.
\item Consider the correlation between the lead and tantalum data, and show that removing a non-zero contribution of the electrization field from the tantalum data improves the correlation. The correlation between the two datasets stems from the fact that the same equipment and solenoids were used for collecting the data for both lead and tantalum.
\end{enumerate}
$\EL$ can be obtained from \Eq{didt} by using the fact that at fixed temperature, the last term disappears while the second term does not depend on \Isol: Two datasets collected at the same fixed temperature but with different $\text{\Isol}$ can be subtracted from each other in order to drastically minimize the systematic of the mutual inductance between \ips~ and \Is.

500 hours of data was collected to implement this strategy in tantalum to verify the $1/T$ dependence of the electrization field leading to \Fig{gammaTTaMod} showing the inverse temperature dependence and rate of change of \Is~ in \Eq{gammabarM}. In addition, the data was used to show that $\EL\propto$$\text{\Isol}$ in \Fig{SpropIa} remembering that \Isol$\propto$\Ia; also see \cref{eqratioa,eqratiob,eqratioc} and accompanying text. Hence with tantalum, we were able to show that the tantalum data is~$\propto$\Ia/$T$.

Furthermore, no such $\propto$\Ia/$T$ dependence could be confirmed in the control metal, lead, which has a free magnetization 19 times smaller than that of tantalum.

The zero-field case where \ips=0 could also be tested. When \ips=0, the second term of the RHS of \Eq{didt} disappears, and  \Eq{didt} can be integrated to a time $t$. Setting the integration constant to 0 yields
\begin{align}\label{intdidt}
\text{\Is}(t)&=\int_0^t \frac{{\cal{E}}_L}{T} \text{d}t + I_\text{sol}P(I_\text{sol})f(T)
\end{align}
where now $T=T(t)$ (see \Fig{tempTa}).

The systematic depends on both temperature and $\text{\Isol}$ so that subtracting datasets collected at different $\text{\Isol}$ will no longer cancel it. Instead, the fact that the electrization field of lead is far weaker than that of tantalum was used: the integral on the RHS of \Eq{intdidt} was subtracted from the zero-field tantalum datasets to show that the result correlated with the lead zero-field datasets far better than the original tantalum datasets (see \Fig{TaPbPlots} and text). This is particularly remarkable as it involved subtracting time integrals of the inverse temperature over 50+~hours.

\begin{figure}
\resizebox{8.75 cm}{!}{
\begin{tabular}{|l|l|l|l|l|l|l|l|l|}
\hline
Isotope   &   $N_\text{c}$   &    $N_\text{N}$    &    $\mun~(\mu_{\text{N}})$   &   $I_\text{N}$  &  $a_\text{b}$~\%  & P \%& $\Fedm{ele}$~(A/m) &  $H_\text{c}$~(T)\\
\hline
$_{73}^{181}\text{Ta}$ & 5 & 5.54 & 2.37 & 7/2& 99.988 & 99.98 & 58.4 & 0.09\\
\hline
$_{82}^{207}\text{Pb}$ & 4 & 3.30 & 0.58 & 1/2& 22.1 & 99.99 & 3.1 & 0.08\\
\hline
$_{13}^{27}\text{Al}$ & 3 &  6.05 & 3.64 & 5/2& 100 & -- & 358.11 & 0.01\\
\hline
$_{42}^{95,97}\text{Mo}$ & 6 &  6.45     & -0.9 & 5/2 & 25.47 & -- & 15.18 &0.0096\\
\hline
$_{23}^{51}\text{Va}$ & 5 &  7.22 & 5.1514 & 7/2& 99.75 & -- & 523.99 & 1\\
\hline
$_{3}^{7}\text{Li}$ & 1 &  4.64 & 3.26 & 3/2& 92.4 & -- & 425($a_\text{b}$=100) & -- \\
\hline
$_{3}^{6}\text{Li}$ & 1 &  4.64 & 0.822 & 1& 7.6 & -- & 128($a_\text{b}$=100) & -- \\
\hline
$_{41}^{93}\text{Nb}$ & 5 &   5.57  & 6.17 & 9/2 & 100 & -- & 259 &0.82\\
\hline
$_{6}^{13}\text{C}$ & $1$ &  11.3     & 0.702 & 1/2 & 1.07 & -- & 2.15 & -- \\
\hline
$_{26}^{57}\text{Fe}$ & 3 &    8.49    & 0.091 & 1/2 & 2.12 & -- &   0.29 & -- \\
\hline
$_{19}^{39}\text{K}$ & 1 &  1.31      & 0.391 & 3/2 & 93.26 & -- & 2.13   & -- \\
\hline
\end{tabular}
}
\caption{Bulk and atomic parameters for sample elements where $a_\text{b}$ is the natural abundance of the isotope and ``P'' refers to the purity of the metal purchased for the experiment, Ta and Pb. $N_\text{N}$ is given in units of $10^{28} \text{m}^{-3}$. Aluminum was used to build the booster solenoid, but molybdenum may be a good alternative with a much smaller nuclear magnetization but with similar critical magnetic field and temperature; note that $_{42}^{95}\text{Mo}$ and $_{42}^{97}\text{Mo}$ have nearly identical parameters and were combined in a single entry. The critical field of the superconductor is given as it sets an upper-limit on \Isol. Vanadium has the largest free magnetization of all elements that are superconducting in their pure form while lithium has the smallest number of nucleons\footnote{For $_{3}^{6}\text{Li}$ and $_{3}^{7}\text{Li}$ the quoted values of $\Fedm{Li}$ are assuming $a_\text{b}$=100}. Niobium is a widely used in superconductivity applications and research. Pure carbon is not superconducting but graphite intercalated compounds (GICs) are and provide an experimental pathway to measuring the nuclear EDM of other elements that are not superconducting in their pure form like potassium. In a GIC, only 1 out of 4 valence ``carbon electrons'' are in the conduction band and $_{6}^{13}\text{C}$ is the only stable isotope with a non-zero spin. Similarly, Iron is not superconducting, but in iron-based $p$-wave superconductors, it can provide a pathway to measure the electron EDM.}\label{nucData}
\end{figure}

\subsubsection{Energy conservation}

Consider a volume ${\cal{V}}$ that contains a circular superconducting wire carrying a current \Is~  and co-axial with a superconducting circular solenoid that has a persistent current \Isol. The radii of the wire and  solenoid are much smaller than the radius of the circular loop, the  solenoid self-inductance is \Lsol, the sample wire loop has self-inductance \Lloop, and the mutual inductance between the solenoid and the wire is $M_\text{s}$. The sample wire nuclei have a nuclear EDM $\den$ with a corresponding electrization field $\mathbf{\mathcal{E}}_{\bm{\lambda}}$ To calculate the total rate of work done on the free charges by the electrization field, we start with
\beq
\text{d}W&=&q\mathbf{\mathcal{E}}_{\bm{\lambda}}\cdot\mathbf{v}\text{d}t \\
\frac{\text{d}W}{\text{d}t} &=& \int_{\cal{V}}{\mathcal{E}}_{\bm{\lambda}}\cdot\mathbf{J}\text{d}V
\eeq
The entire sample assembly is in the volume ${\cal{V}}$ including the superconducting wire that forms the sample solenoid. $\mathbf{\mathcal{E}}_{\bm{\lambda}}$ includes any field, internal or external to the sample assembly that can perform work on the supercurrents; this necessarily excludes the contact electric field $\mathbf{E}_{\bm{\delta}}(\bm{r})$ proportional to Dirac delta functions located on the lattice sites (see \Eq{contactE}). Using Maxwell's equations yields
\beq
\frac{\text{d}W}{\text{d}t} = -\frac{\text{d}~}{\text{d}t}\int_{\cal{V}}\frac{1}{2}\mathbf{B}\cdot\mathbf{H}\text{d}V -\frac{\text{d}~}{\text{d}t}\int_{\cal{V}}\frac{1}{2}\mathbf{\mathcal{E}}_{\bm{\lambda}}\cdot\mathbf{D}\text{d}V - \int_{\cal{S}} (\mathbf{\mathcal{E}}_{\bm{\lambda}}\times\mathbf{H})\cdot\text{d}\mathbf{a}
\eeq
In the absence of external fields, the net work on the system is zero. Taking the volume to be all space, the Poynting vector contribution vanishes. Keeping only terms linear in $\den$, and noting that in the absence of external fields $\mathbf{\mathcal{E}}_{\bm{\lambda}}\propto\den$, we are left with
\beq
\frac{\text{d}~}{\text{d}t}\int_{\cal{V}}\frac{1}{2}\mathbf{B}\cdot\mathbf{H}\text{d}V=\frac{\text{d}~}{\text{d}t}\left(\frac{1}{2}\text{\Lloop}I_\text{s}^2+ \frac{1}{2}\text{\Lsol}I_\text{sol}^2 + M_\text{s}\text{\Is\Isol} \right)=0~,
\eeq
where \Lloop is the self-inductance of the sample wire circuit and \Lsol that of the sample solenoid. Constructing a solenoids/wire loop assembly with vanishingly small mutual inductance $M_\text{s}$ leads to the result
\beq\label{energyCons}
\frac{\text{d}~}{\text{d}t}\left(\frac{1}{2}\text{\Lloop}I_\text{s}^2\right)=-\frac{\text{d}~}{\text{d}t}\left( \frac{1}{2}\text{\Lsol}I_\text{sol}^2 \right)
\eeq
proving that the rate of energy  gained by the \Is~ magnetic field is precisely equal to the rate of energy lost by the $\text{\Isol}$ magnetic field up to  vanishingly small corrections of ${\cal{O}}({\den}^2)$. \Eq{energyCons} combined with \Eq{didt} setting $M_\text{ps}=f(T)=0$ suggest the existence of oscillatory solutions. Solving  \Eq{betaSlope} for \Isol, substituting in \Eq{energyCons}  and defining ${\cal{E}}_L^\prime\equiv{\cal{E}}_L$\Lloop/\Isol, we obtain
\beq
\frac{\text{\Lloop\Lsol}}{{{\cal{E}}_L^\prime}^2}\frac{\text{d}^2\text{\Is}}{\text{d}t^2}&=&-I_\text{s}\\
\text{\Is}(t)&=&A\text{sin}(\omega_\text{EDM} t+\phi)\\
\text{\Isol}(t)&=&\frac{\text{\Lloop}}{{\cal{E}}_L^\prime}A\omega_\text{EDM}\text{cos}(\omega_\text{EDM} t+\phi)\\
\omega_\text{EDM}&=&\frac{{{\cal{E}}_L^\prime}}{\sqrt{\text{\Lloop\Lsol}}}~,
\eeq
with boundary conditions at $t$=0, \Isol(0)=\Isol, \Is=0, we have $A$=$\text{\Isol}\sqrt{\text{\Lsol/\Lloop}}$ and $\phi=0$. Expanding \Is~ and $\text{\Isol}$ leads to
\beq\label{sinwt}
\text{\Is}(t)&\cong&{\cal{E}}_Lt + {\cal{O}}({\den}^3) \\ \label{coswt}
\text{\Isol}(t)&\cong&\text{\Isol}-\frac{1}{2}(\omega_\text{EDM}t)^2=\text{\Isol}+{\cal{O}}({\den}^2)
\eeq
Hence, it is justified to neglect induced electric field due to the variations of $\text{\Isol}$ in \Eq{betaSlope}. This can also be seen directly from \Eq{energyCons} where we have
\beq\label{dIsoldt}
\frac{\text{d}\text{\Isol}}{\text{d}t}=-\frac{\text{\Lloop}}{\text{\Lsol}}\frac{\text{\Is}}{\text{\Isol}}\frac{\text{d}\text{\Is}}{\text{d}t}~.
\eeq
However, \Lsol$\gg$\Lloop and \Isol$\gg$\Is. Indeed, \Is~ can be estimated from the fact that the supercurrent  through the Josephson Junctions is suppressed for magnetic fields $>10^{-5}$~T. For our setup this corresponds to  $I_\text{d}\sim10~\mu$A which is much less than the current through the sample solenoid $\text{\Isol}$=~3.5~A showing that  the electromotive force stemming from the variation of  $\text{\Isol}$ can be neglected. 

The solutions \Eq{sinwt} and \Eq{coswt} are not applicable at all times as they suggest that at some point the current will transfer back to $\text{\Isol}$ which is not necessarily true. The issue is that \Eq{energyCons} does not include the azimuthal current component of \Is~  that can also be driven by an electrization field as discussed in \cref{app:osc}.

%In addition an electric current cannot be generated from terms such as $\nabla(\vec{m}\cdot\vec{B})$ since there would be opposite gradients at the ends of the sample solenoid and the forces would cancel.

\section{Sample assembly and data analysis}\label{sec:data}

\subsection{Sample assembly}

The sample assembly is depicted in \Fig{sampAssembly} and it has 5 main components: The main solenoid, the booster solenoid, the sample solenoid, the sample wire and the SQUID. The solenoid parameters are given in \Tab{solParam}. The booster and sample solenoids are built from a single continuous wire whose ends are joined in a superconducting joint. The booster solenoid is located inside the main solenoid. All the solenoids are constructed with a single filament superconducting Cu Clad Nb47\%Ti (NbTi for short) wire with formvar insulation and a NbTi core diameter of 0.31~mm.

The main solenoid has leads that go to the exterior of the ADR to a DC power source that feeds it a current \Ia. The booster solenoid is located inside the main solenoid and has two layers of windings: the first lead of the booster solenoid superconducting wire is wound around an aluminum cylinder that has a layer of yellow tape for ease of winding; that first solenoid layer is held in place with yellow tape. The second lead forms the winding of the sample solenoid and then returns to wind around the first solenoid layer; that second layer is also held in place with yellow tape. The two leads are then joined with a Bi56Pb44 solder following the procedure outlined in~Ref.\cite{SiyuanLiu2013}. That joint was observed to preserve a persistent current for 9 days~(see \Tab{IsolVal}). The main and booster solenoids are encased in a Hy mu 80 shield. Not shown in \Fig{sampAssembly} is the absolute magnetic sensor ALT021 sold by NVE that was used to verify superconductivity as well as confirm data from the SQUID. Once it was no longer needed, it was removed from the sample assembly for the tantalum data collection to diminish heat transfers from the outside.
\begin{figure*}
\resizebox{15 cm}{!}{\includegraphics*{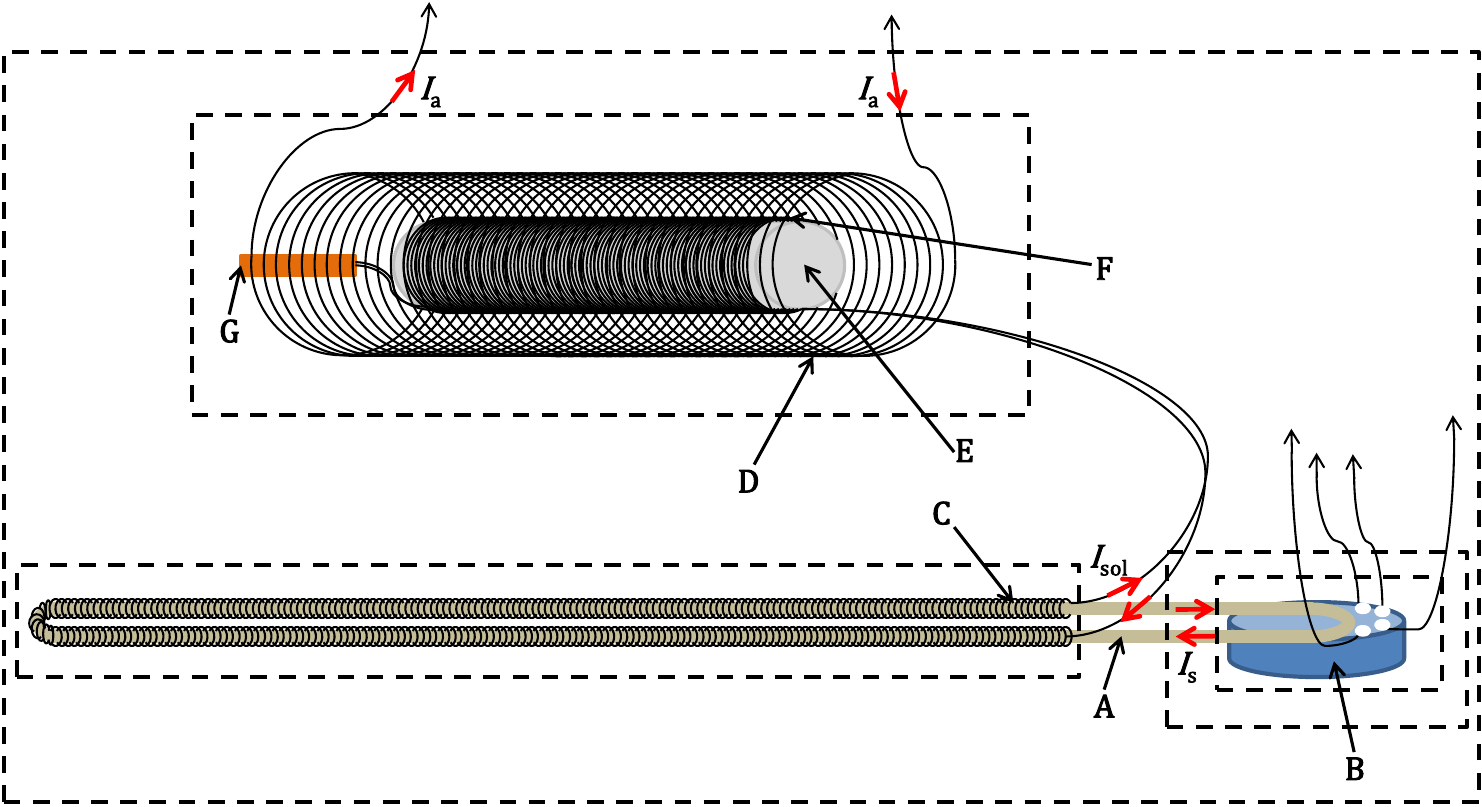}}
%\vspace{-2cm}
\caption{Sample assembly components: A) sample wire (Ta or Pb); B) SQUID; C) sample solenoid; D) main solenoid wound around copper tubing (not shown); E) aluminum booster solenoid core (length 5~cm radius 0.5~cm); F) booster solenoid with two layers of winding where one lead forms the first winding layer and the other lead proceeds to form the sample solenoid and returns to form the winding of the second layer; G) BiPb joint of the NbTi booster solenoid wire leads. The currents are indicated with the red arrows: The current applied by the DC power source to the main solenoid, \Ia; the induced booster solenoid current, \Isol; the sample current detected by the SQUID, \Is.  The dashed rectangles represent shielding: the inner and outer rectangles around the SQUID are made of lead and Hy mu 80 respectively; the shielding around the main and sample solenoids are made of  Hy mu 80; the overall shielding is made of lead and aluminum adhesive. The potential and current leads from the SQUID run to the exterior SQUID electronics while the main solenoid leads run to an exterior DC current source.}\label{sampAssembly}
\end{figure*}
\begin{center}
\begin{table}[t]
\begin{tabular}{|l|l|l|l|}
\hline
     &   \# loops  &    length (cm)   &   $\bar{r}$ (mm) \\
\hline
Main Sol.     &   109  &   6.1    &   7.545 \\
\hline
Booster Sol. layer 1    &   81  &   5    &   5.2 \\
\hline
Booster Sol. layer 2    &   80  &   5    &   5.7 \\
\hline
Sample Sol.     &   358  &   20.4    &  0.8 \\
\hline
\end{tabular}
\caption{Assembly solenoid loop number, length and mean radius. The mean radius of the sample solenoid was calculated using the total length of the wire used to make the solenoid, the solenoid length and the number of loops.}\label{solParam}
\end{table}
\end{center}
\begin{figure*}
\centering
\captionsetup[subfigure]{labelformat=empty}
\resizebox{16 cm}{!}{
\begin{tabular}{cc}
 \includegraphics*{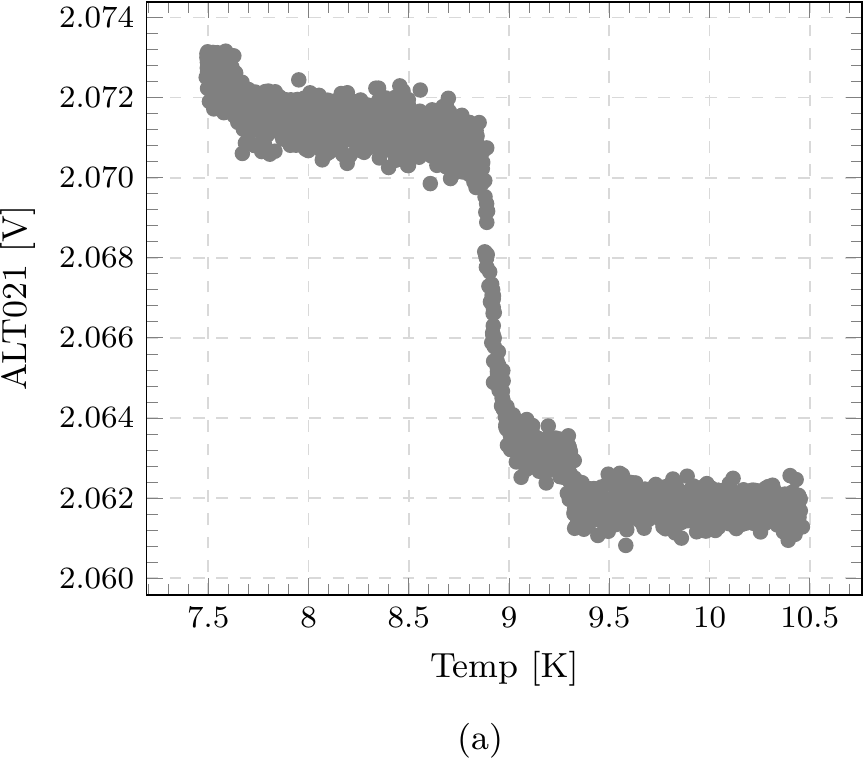}
&
  \includegraphics*{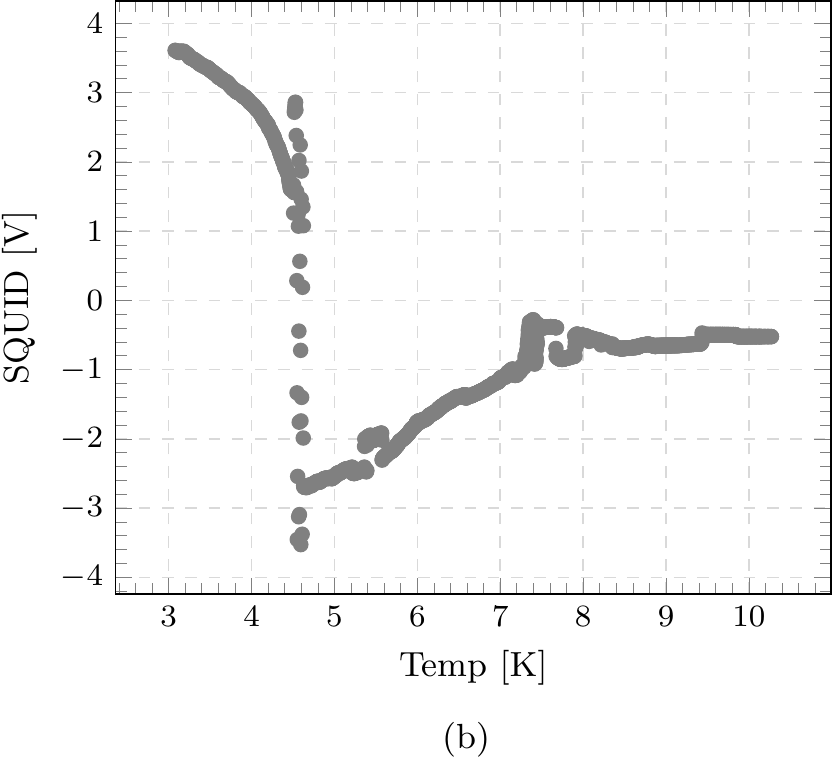}
\end{tabular}
}
\caption{(a) Bi56Pb44 joint ALT021 data  showing the transition from the superconducting state to the normal state at $T$=8.9~K. (b) SQUID data for the tantalum sample wire showing the transition from the superconducting state to the normal state at $T$=4.5~K.}\label{Bi56Pb44}
\end{figure*}
\begin{center}
\begin{table}[t]
\begin{tabular}{|c|c|c|}
\hline
     &   Ta  &    Pb \\
\hline
Peak current     &   140~A  &   50~A     \\
\hline
Welding time:    &   10~ms  &   10~ms    \\
\hline
Mode    &   Cold  &   Cold     \\
\hline
Pre-flow    &   5~s  &   5~s     \\
\hline
Post-flow    &   4~s  &   4~s     \\
\hline
Ignition current  &  10~A  &   10~A     \\
\hline
\end{tabular}
\caption{Cold weld parameters for Ta and Pb.}\label{weldParam}
\end{table}
\end{center}
The purpose of the main solenoid is to create a magnetic flux through the booster solenoid at a temperature above the critical temperature of the Bi56Pb44 weld, about 8.9~K according to the ALT021 data in \Fig{Bi56Pb44}. Once the sample assembly is below 8.9~K as monitored by the SQUID signal, $\text{\Ia}$ is turned off inducing an initial $\text{\Isol}$ in the booster/sample solenoid. Part of the sample wire is inside the sample solenoid which is also  encased in a Hy mu 80 shield. The segments of the sample wire exiting the sample solenoid are wrapped in a single layer of yellow tape and brought together with copper tape also attached to the copper plate to which the sample assembly is affixed. That copper plate was tightly affixed to the 50~mK rod of the ADR. The leads of the sample wire are ``cold'' welded with the settings for both Ta and Pb given in \Tab{weldParam}.

Cold welding procedure: the sample wire leads are cleaned with acetone and twisted together and grounded. During the 5~s pre-flow of argon gas, the twisted wire is cut before the generation of the ark. This welding process was verified to be superconducting in test loops, but to verify the superconductivity of the actual Ta sample wire weld inside the sample assembly, SQUID data was used to verify that a superconducting transition occurred at $T$=4.50~K, consistent with the tantalum transition temperature of 4.48~K.\footnote{Note that the RuOx thermometer installed in the ADR has an accuracy $\pm$50~mK above 1.4~K.}

After welding, the welded end of the sample wire was put against the back of the SQUID\footnote{\Fig{sampAssembly} shows the sample wire against the front of the SQUID, but it was drawn this way to show the SQUID leads leaving the SQUID pads on their way to the outside.} and both were installed in the Pb shielding, itself installed into Hy mu 80 shielding. The whole sample assembly was put into Pb shielding layered with Aluminum adhesive.

The ADR was held solidly  in place by three steel support jack rods (see \fig{mylar}) and a regular jack below the ADR to suppress vibrations stemming from the compressor. The vibration suppression was tested by verifying that the SQUID signal remained unchanged when the compressor was toggled off and on.

\subsection{SQUID electronics and voltage to current conversion}\label{sec:squid}

\begin{figure*}
\resizebox{15 cm}{!}{\includegraphics*{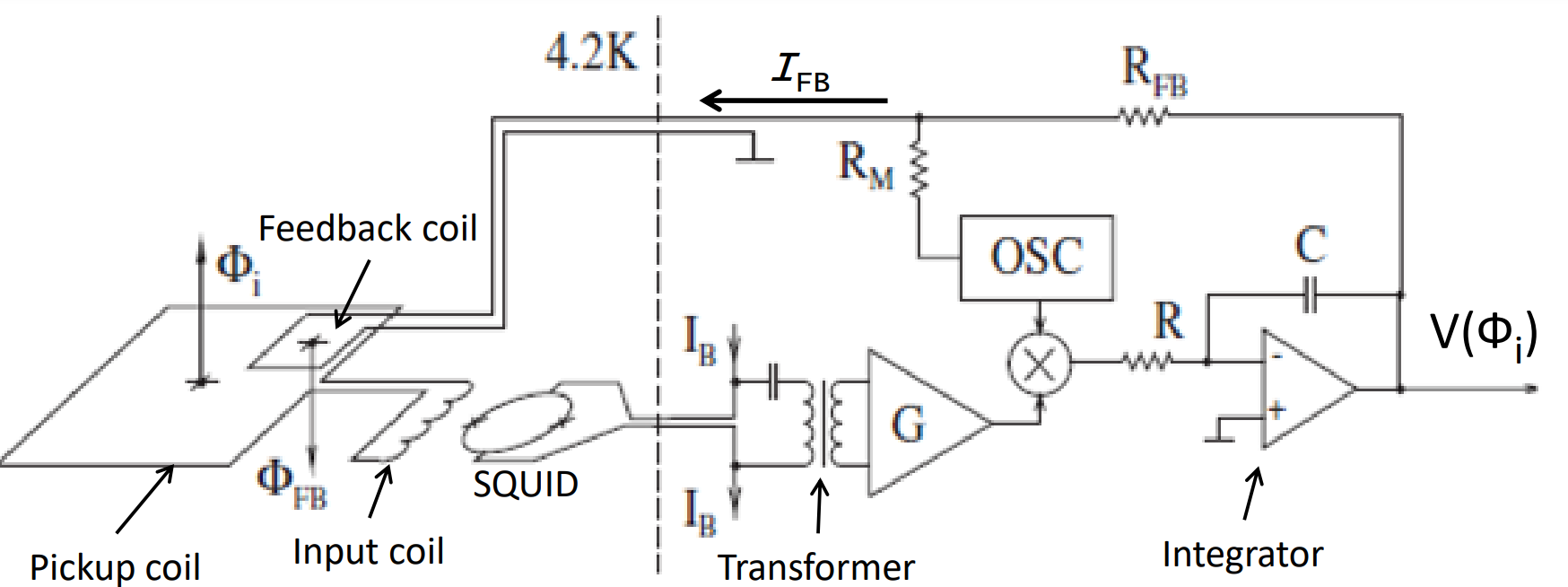}}
\caption{PFL circuit detailing the LOCKED mode components in the SQUID electronics, reproduced with the permission of the manufacturer, Starcryo. Labels were added by the author.}\label{squidLOCKcircuit}
\end{figure*}

The SQUID was operated in ``LOCKED'' mode which means that the total magnetic flux detected by the SQUID is kept constant. Operating a SQUID in LOCKED mode allows the detection of magnetic flux changes that are tiny fractions of a fluxon, a magnetic flux quantum. Referring to \Fig{squidLOCKcircuit}, the LOCKED circuit works as follows:
\begi
The superconducting pickup coil is exposed to an external magnetic flux  $\Phi_\text{i}$, the desired signal.
The change in flux due to $\Phi_\text{i}$ in the superconducting circuit of the pickup coil induces a counter-current to keep the flux constant.
The counter-current flows into the input coil which is transformer coupled to the SQUID inductance.
A potential difference appears in the SQUID which is the input signal to the warm transformer (to the right of the dashed line) where it is amplified by a gain 'G'.
The amplified signal is combined with an oscillating signal.
The combined signal is sent to an integrator that produces a current in the feedback coil\footnote{Note that the  feedback coil is not superconducting and cannot sustain persistent currents located inside the pickup coil.} and a SQUID output voltage $V(\Phi_\text{i})$ data point collected by the external data acquisition card.
The feedback flux $\Phi_\text{FB}$ of the feedback coil cancels the flux change in the pickup coil due to the original external signal.
The current of the pickup coil thus returns to its original value.
\ei 
The fact that the change in the flux through the pickup coil due to the external signal equals in magnitude the change in the flux through the pickup coil due to the current in the feedback coil implies the relation
\beq
\Delta I_\text{PC} = \frac{M_\text{FB-PC}}{L_\text{PC}}\Delta I_\text{FB}~,
\eeq
where $\Delta I_\text{PC}$ is the current variation in the pickup coil, $M_\text{FB-PC}$ is the mutual inductance between the pickup coil and the feedback coil, $L_\text{PC}$ is the self-inductance of the pick up coil and  $\Delta I_\text{FB}$ is the current variation in the feedback coil. Since the variation of the external flux through the pickup coil came from a variation of the current of the sample wire, $\Delta I_\text{S}$, it is also true that
\beq
\Delta I_\text{PC} = \frac{M_\text{S-PC}}{L_\text{PC}}\Delta I_\text{S}~,
\eeq
where $M_\text{S-PC}$ is the mutual inductance between the sample and the pickup coil and $\Delta I_\text{S}$ is the current variation in the sample. The output in LOCKED mode is a voltage that was calibrated to a single fluxon as 0.743~V/$\Phi_0$ with the parameters (bias, mode, phase, offset) used to configure the sensitivity of the SQUID given in \Tab{squidParam}. To convert a SQUID voltage variation to a feedback coil current variation, the current in the feedback coil was manually increased by 0.2442~$\mu$A every 10~s; that resulted in an average SQUID voltage variation of 0.214~V.  Hence the conversion factor $\Delta_\text{VI}$ between the SQUID voltage data and the feedback coil current is
\beq
\Delta_\text{VI} \equiv\frac{\Delta I_\text{FB}}{\Delta V}=1.14\mu\text{A}/\text{V}~.
\eeq
$\Delta_\text{VI}$ was found to be temperature independent from 0.05~K to 4~K which leads to a relation between the sample current and the voltage data
\beq
\Delta I_\text{S}= \frac{M_\text{FB-PC}}{M_\text{S-PC}}\Delta_\text{VI}\Delta V~.
\eeq
\begin{center}
\begin{table*}[t]
\begin{tabular}{|l|l|l|l|l|l|l|l|}
\hline
Bias($\mu$A)   &   Mode($\mu$A)  &    Phase(\%)    &    Offset ($\mu$A) & $M_\text{FB-PC}$(nH) & $L_\text{PC}$(nH) & $M_\text{Ta-PC}$ (nH) & $\Delta_{VI}~(\mu$A/V) \\
\hline
29.158 & 4.0904 & 3.137 & 1.7949 &  0.29 &  5.5 & 0.067$\pm0.01$& 1.14\\
\hline
\end{tabular}
\caption{Sensitivity configuration of the SQUID and inductance values. The bias, mode, phase and offset are chosen to maximize sensitivity of the SQUID to magnetic flux variations~\cite{STARCRYO}. The uncertainty on $M_\text{Ta-PC}$ comes from zeroing or doubling the distance between the SQUID surface and the nearest point on the sample loop.}\label{squidParam}
\end{table*}
\end{center}

\subsection{Inducing and measuring \Isol, REGEN, Shielding efficiency}\label{sec:regen}

The electrization field follows Curie's law and is therefore most intense at low temperatures. In order to maximize the odds of measuring a non-zero $\EL$, an adiabatic demagnetization refrigeration (ADR) system was used. The ADR has 4 temperature stages nominally called the 50~K, 3~K, 1~K and 50~mK stages. The 1~K and 50~mK stages are thermally connected to gadolinium gallium garnet (GGG) and ferric	ammonium	alum (FAA) salt pills respectively. During the REGEN process, the spin entropies of the salt pills are minimized with a 4~T magnetic field generated by a pill solenoid current, \ips. The minimal salt pill spin entropies are now heat reservoirs for the salt pill vibrational entropies, thus lowering the temperatures of the 1~K and 50~mK stages to sub-Kelvin levels. The REGEN process involves creating a strong thermal contact between the 50~mK, 1~K and 3~K stages by ``closing the heat switch'' which simply means pressing two gold-plated plates thermally connected to the 3~K stage against gold-plated sections of the the 1~K and 50~mK stages. During REGEN, the pill solenoid current is ramped up to 8.5~A and the closed heat switch means that the 3~K stage acts like a heat sink to the GGG and FAA pills. After absorbing most of the heat from the pills during the DWELL stage, \ips~ is brought back down to zero at which point the 1~K and 50~mK stages reach their base temperature of about 540~mK and 40~mK respectively.

In order to measure $\EL$, a  persistent and large superconducting $\text{\Isol}$ is also required by Curie's law. A persistent, minimum $\text{\Isol}$  is initially created by induction in the booster solenoid as follows (refer to \Fig{sampAssembly}):
\begi
with the heat switch closed, the compressor is turned off to warm the sample assembly to $T>$8.9~K, the critical temperature of the BiPb joint.
With the solenoid wires in the normal state, the exterior DC current source is turned on to produce a current $\text{\Ia}$  in the main solenoid.
The compressor is then reactivated to cool the sample assembly back down below 8.9~K. The transition of the booster solenoid to the superconducting state is determined   with both  thermometer readings and the monitoring of the SQUID signal for its  transition to the superconducting state\footnote{The SQUID has a critical temperature of 9.2~K.}.
After the transition to the superconducting state has been confirmed, the  current $\text{\Ia}$ is slowly turned off  inducing an initial, minimum current $\text{\Isol}$ in the booster solenoid.
Since the booster solenoid and the sample solenoid were built from a single continuous superconducting wire whose ends were fused together in a superconducting BiPb joint, $\text{\Isol}$ also flows in the sample solenoid.
\ei
The persistent, minimum $\text{\Isol}$ induced with  the above steps is approximately a tenth of the maximum $\text{\Isol}$ used for the measurement of ${\cal{E}}_L$ at sub-Kelvin temperatures. The boosting of $\text{\Isol}$ to its maximum value occurs during the REGEN procedure. REGEN data confirms that $\text{\Isol}$ is a persistent current, that it is boosted to its maximum value, and  that the shielding efficiently protects the sample assembly from magnetic fields. An example of SQUID data collected as a function of temperature during REGEN is given in \Fig{REGEN} to provide a visualization as the procedure is described in detail below:
\begi
As mentioned above, REGEN starts with the heat switch closed and the ``ramp up'' of the pill solenoid current \ips~ from 0~A to 8.5~A over 20 minutes. During ramp up, the temperature of the 50~mK stage increases steadily to about 3.06~K from 2.62~K.
During the ramp up to 8.5~A, the magnet is shielded with vanadium permendur magnetic shielding and at full \ips, the magnetic field should not go beyond 100-200~$\mu$T. During REGEN, the flux through the SQUID changed by a fraction of a fluxon as the \ips~ increased to 7~A during ramp up. This provides confidence that the shielding protects the sample assembly from external magnetic fields well.
At the end of ramp up, the ``dwell'' stage begins where heat accumulated in  the FAA pill during ramp up is dumped  into the 3~K stage which acts as a heat sink. The dwell stage lasts for 45 minutes.
At the end of the dwell stage, the heat switch is opened and ``ramp down'' begins during which \ips~ returns to 0~A over 45 minutes.
During ramp down, the temperature of the 50~mK stage monotonically decreases to sub-Kelvin temperatures. As Aluminum has a $T_\text{c}<$1.2~K in the presence of magnetic fields, aluminum components of the sample assembly such as the aluminum adhesive taped on the outer shield and the aluminum core of the booster solenoid transition to the superconducting state:
\up
At $T$=1.175~K the Al adhesive which is better thermally connected to the 50~mK rod becomes superconducting as seen in the sharp change in the SQUID signal slope in \Fig{REGEN} at that temperature. This feature occurs in all our REGEN data.
The booster solenoid Al core becomes superconducting and expels magnetic fields at lower temperatures for the following two reasons: 
\up
It is more thermally insulated from the 50~mK stage because yellow tape was used to help the superconducting wire stay in place around the aluminum cylinder.
It is exposed to the $\text{\Isol}$ magnetic field which lowers the critical temperature of Al.
\down
As the Meissner effect begins at the temperature \Tmax~ where the Al core begins the superconducting transition, the magnetic flux expulsion from the Al core reduces the self-inductance of the booster solenoid and boosts $\text{\Isol}$ since the total flux through both the booster and sample solenoids must remain constant\footnote{Flux through other components of the superconducting wire from which the solenoids are made is comparatively negligible.}. The relationship between temperature and $\text{\Isol}$ is given in \Eq{IB}.
The expulsion of  magnetic flux from the Al core is completed when the temperature of the Al core is below the new critical temperature of aluminum corresponding to exposure to the maximal \Isol-generated magnetic field. The maximal $\text{\Isol}$ is the current needed to maintain the flux inside the booster/sample solenoids when the booster solenoid self-inductance is at its minimal value.
$\text{\Isol}$ continuously increases until it reaches its maximal value. Correspondingly, the magnetic flux through the SQUID also increases either through a mutual inductance between the sample solenoid and the sample wire and/or the SQUID. A SQUID signal oscillates with increasing magnetic flux, and those oscillations are clearly visible in \Fig{REGEN} below the critical temperature of Al.  This is evidence that  $\text{\Isol}$ is  increasing.
The Meissner effect is completed at a temperature \Tmin~ where $\text{\Isol}$ is maximal.
\down
\ei
\begin{figure}
\centering
\resizebox{8 cm}{!}{
 \includegraphics*{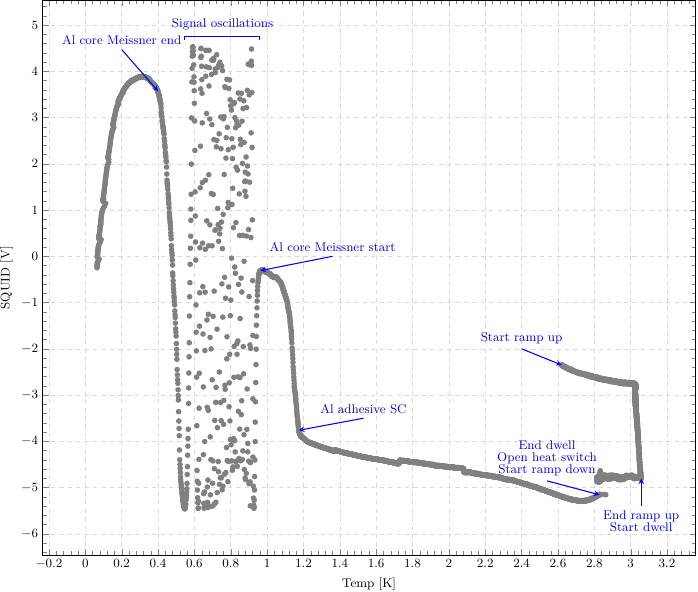}
}
\caption{Full REGEN data for Ta at \Ia=0.53~A. Each data point is a mean value of 1000 raw data points averaged over 1~second. $T_\text{max}$ is the temperature where the Meissner effect begins while $T_\text{min}$ is the temperature where it ends.}\label{REGEN}
\end{figure}
\begin{figure}
\centering
\resizebox{8 cm}{!}{
 \includegraphics*{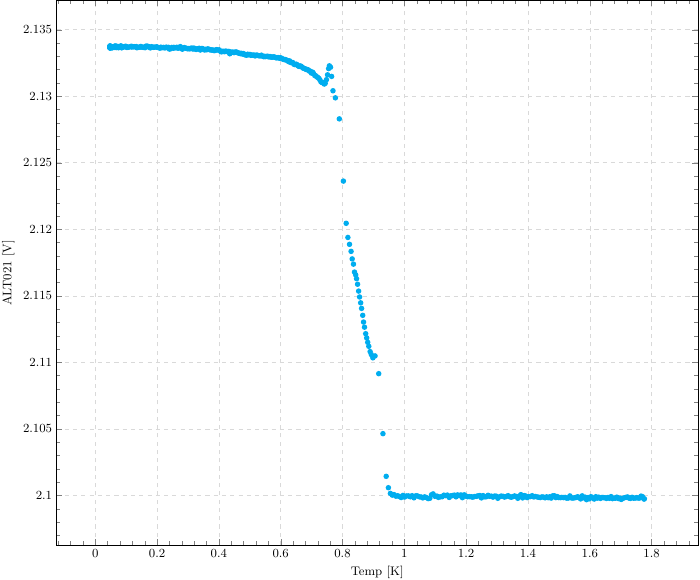}
 }
\caption{ALT021 data taken on September 17th 2024 starting at 8:41PM PDT showing the beginning of the Meissner effect around a temperature of 0.967~K followed by saturation of the sensor around a temperature of 0.761~K. Note that the temperature of the Aluminum booster core lags behind the thermometer stage temperature by about 0.194~K (see \Eq{IB}).}\label{ALT021meissner}
\end{figure}
In order to experimentally verify that the booster solenoid works as described, a first experimental test of the sample assembly was performed using a Pb sample wire with the ALT021. For that test,  4 loops of the booster/sample solenoid wire were wound around the ALT021 sensor and \Ia=0.5~A. That data is plotted in \Fig{ALT021meissner}: the rise of the current is clearly visible starting around 0.967~K. Around $T\cong$0.761, the ALT021 sensor saturates\footnote{The ALT021 saturates at magnetic fields around 0.5~mT.}  and could not track  $\text{\Isol}$ to its maximal value at $T$=\Tmin, but  it  did confirm that $\text{\Isol}$ sharply increases at a \Tmax~ nearly identical to those determined with the SQUID REGEN data for an applied current \Ia=0.53~A as seen in \Fig{regenPlots} and \Fig{PbregenPlots} where REGEN data is plotted against temperature for Ta and Pb respectively at different applied \Ia. Prior to a detailed analysis, the plots in \Fig{regenPlots} and \Fig{PbregenPlots} visually support the boosting of $\text{\Isol}$ as there are no oscillations for \Ia=0 (implying no boosted \Isol) and the oscillations occur over the widest temperature range (implying largest boosted \Isol) for the largest \Ia.
\begin{center}
\begin{table}[t]
\begin{tabular}{|l|l|}
\hline
              &   (H)       \\
\hline
Booster-Main sol. mutual Ind  ($M_\text{bm}$)&   $3.193\cdot10^{-5}$      \\
\hline
Booster Al self Ind  ($L_\text{bstr}$)&  $5.332\cdot10^{-5}$   \\
\hline
Booster Al self Ind  SC ($L_\text{bstr}$)&  $0.682\cdot10^{-5}$   \\
\hline
Sample solenoid (Ta) self Ind. ($L_\text{s}^\text{Ta}$)&  $0.096\cdot10^{-5}$  \\
\hline
Sample solenoid (Pb) self Ind. ($L_\text{s}^\text{Pb}$)&  $0.048\cdot10^{-5}$   \\
\hline
\Lloop &  $0.00769\cdot10^{-5}$   \\
\hline
\end{tabular}
\caption{Numerically calculated inductances using the  dimensions and number of loops of the solenoids. SC refers to the aluminum core of the booster solenoid being in the superconducting state. The sample solenoid into which the Ta and Pb sample wires were inserted has an average radius of 0.8~mm. The self-inductance of the sample solenoid depends on the Ta and Pb wire radii,  0.25~mm  and 0.5~mm respectively. The superconducting wires used to build the solenoids NbTi SC‐T48B‐M-0.5~mm were purchased from supercon.}\label{numInductance}
\end{table}
\end{center}
$\text{\Isol}$ was determined two ways: theoretically using the numerically calculated industances of the solenoid and experimentally from the REGEN data using the relationship between critical temperatures $T^\prime$ and applied magnetic fields
\beq\label{criticalT_B}
T^\prime=T_\text{c}\left[1-\left(\frac{B}{B_\text{c}}\right)\right]^{0.5}
\eeq
where $B_\text{c}$ is the critical magnetic field at $T=0$~K. Thus, knowing the critical temperatures \Tmin~ and \Tmax~ allows the minimal and maximal values of $\text{\Isol}$ to be determined by solving \Eq{criticalT_B} for $B=\mu_0n_\text{bstr}$$\text{\Isol}$ in terms of $T^\prime$ where $n_\text{bstr}$ is the number of loops per unit length of the booster solenoid (see \Tab{solParam})
\beq\label{IB}
I_\text{sol}^\text{\Ia}=\frac{B_\text{c}}{\mu_0n_\text{bstr}}\left[1-\left(\frac{T^\text{\Ia}_\text{adr}+\tau}{T_\text{c}}\right)^2\right]~,
\eeq
where $T^\text{\Ia}_\text{adr}$ is the temperature displayed by the ADR 50~mK thermometer and $\tau$ is the difference between $T^\text{\Ia}_\text{adr}$ and the actual temperature of the aluminum core. $\tau$ can be determined by using the fact that  data was collected at $\text{\Ia}$ and 2$\text{\Ia}$ (for example, at \Ia=0.265~A and \Ia=0.53~A in tantalum). Since \Isol$\propto$$\text{\Ia}$ (see \Eq{mbm}, \Eq{mbmsc}), use $I_\text{sol}^{2\text{\Ia}}$=2$I_\text{sol}^{\text{\Ia}}$ to solve for $\tau$ in terms of $T^{\text{\Ia}}_\text{adr},~T^{2\text{\Ia}}_\text{adr}$ known from the REGEN data
\beq\label{tau}
\tau=T^{2\text{\Ia}}_\text{adr}-2T^{\text{\Ia}}_\text{adr}+\sqrt{2(T^{\text{\Ia}}_\text{adr}-T^{2\text{\Ia}}_\text{adr})^2+T_\text{c}^2}~.
\eeq
The values for \Isol$_\text{,min}$ and \Isol$_\text{,max}$ and corresponding \Tmax~ and \Tmin~ are given in \Tab{IsolVal}. \Tmax~ and \Tmin~ were determined from \Fig{regenPlots} and \Fig{PbregenPlots} in the following fashion:
\begi
Define the REGEN dataset $\mathbf{R}$=\{$r_1,r_2,...,r_N$\} and the corresponding REGEN ADR temperature dataset $\mathbf{T}_\text{R}$=\{$T_1,T_2,...,T_N$\} where $r_i\in\mathbf{R}$ is a SQUID voltage data point taken at ADR temperature $T_i\in\mathbf{T}_\text{R}$.
From the REGEN {\bf R}=\{$r_1,r_2,...,r_N$\} dataset, create a differenced dataset $\Delta${\bf R}=\{$\delta r_i$  | $\delta r_i= r_{i+1}-r_{i-1} $\} in order to search for sudden, large changes in the slope of the REGEN plots.
Difference datasets to identify  \Tmax~ are plotted in the middle columns of \Fig{regenPlots} and \Fig{PbregenPlots}; the last column of those figures contain the differenced dataset plots to identify \Tmin.
Note that the differenced plots for \Tmax~ and \Tmin~ are  different due to the asymmetry in the plots of the first column:  after \Tmax~  the $r_i$ values   vary extremely rapidly. On the other hand,  before \Tmin, the $r_i$ values vary more slowly. The asymmetry is also visible in the fact that the oscillation peaks in the plots of the first column are less resolved at temperatures closer to \Tmax~ than \Tmin.
To determine \Tmax, we seek  the first of the $\delta r_i$ that does not decrease.
To determine \Tmin, we seek the last clearly identifiable peak in the $\delta r_i$ that is at a lower temperature than the last oscillation peak.
\ei
In addition to the experimentally determined \Isol, theoretical estimates of \Isol$_\text{,min}$ and \Isol$_\text{,max}$(columns $I_\text{min}^\text{num}$, $I_\text{max}^\text{num}$ of \Tab{IsolVal}) were calculated using the numerically evaluated inductances of \Tab{numInductance} inserted in \Eq{mbm}  and \Eq{mbmsc} 
\beq\label{mbm}
M_\text{bm}\text{\Ia}&=&\left(L_\text{s}^\text{ele}-L_\text{bstr}\right)\text{\Isol} \\ \label{mbmsc}
M_\text{bm}\text{\Ia}&=&\left(L_\text{s}^\text{ele}-L_\text{bstr}^\text{sc}\right)\text{\Isol} 
\eeq
where `ele' is Ta or Pb and noting that the mutual inductance between the main solenoid and sample solenoid is zero. The relative negative sign between the sample solenoid and booster solenoid stems from the fact that they were wound in opposite directions to minimize the inductance and increase the $\text{\Isol}$ boost. The experimentally and theoretically determined $\text{\Isol}$ agree well and the difference between them will be taken as the uncertainty on \Isol.

Having found $\text{\Isol}$ using \Tmax, \Tmin~ and numerical estimates, the persistence of $\text{\Isol}$ is also confirmed in the extracted \Tmax~ given  in \Tab{IsolVal}. Indeed,  \Tmax~ was extracted for two REGEN datasets taken 9 days apart using a current that was induced from $\text{\Ia}$ on the earlier date. The fact that $\text{\Isol}$ persisted unchanged over 9 days in addition to the limits on the resistance  of the PbBi joint presented in Ref.~\cite{SiyuanLiu2013} for currents order $\sim10^3$A provides confidence that resistivity in the joint is not a significant systematic that could have affected our results presented below.

\begin{figure*}
\centering
\pgfplotsset{
    small,
    legend style={
        at={(0.01,0.01)},
        anchor=south west,
    },
   }%
\resizebox{16 cm}{!}{
\begin{tabular}{ccc}
  \includegraphics*{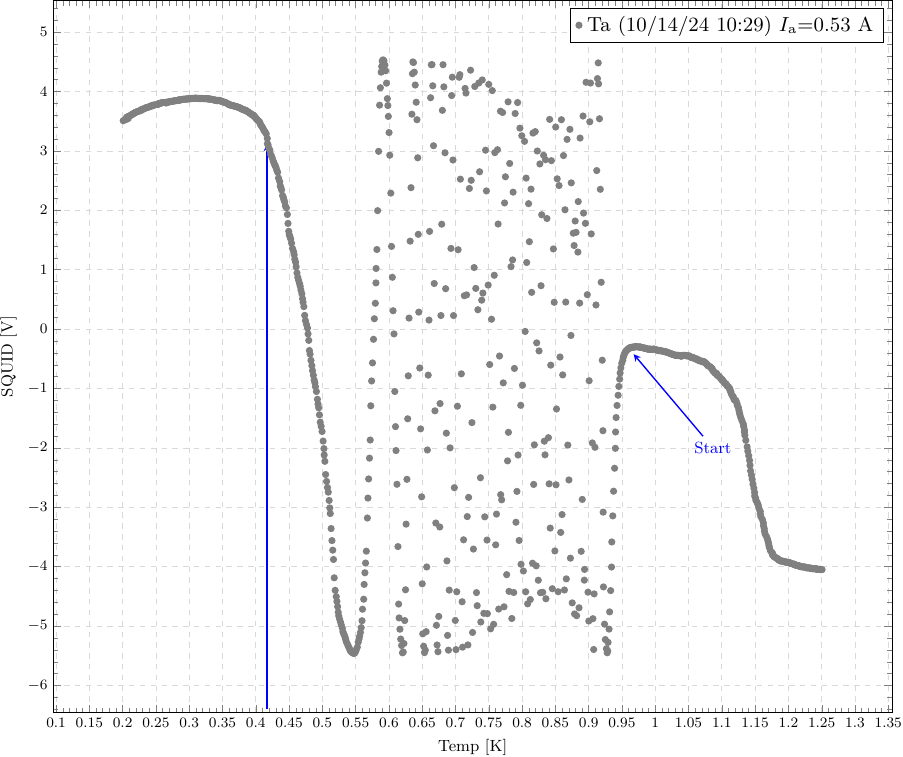}
&
  \includegraphics*{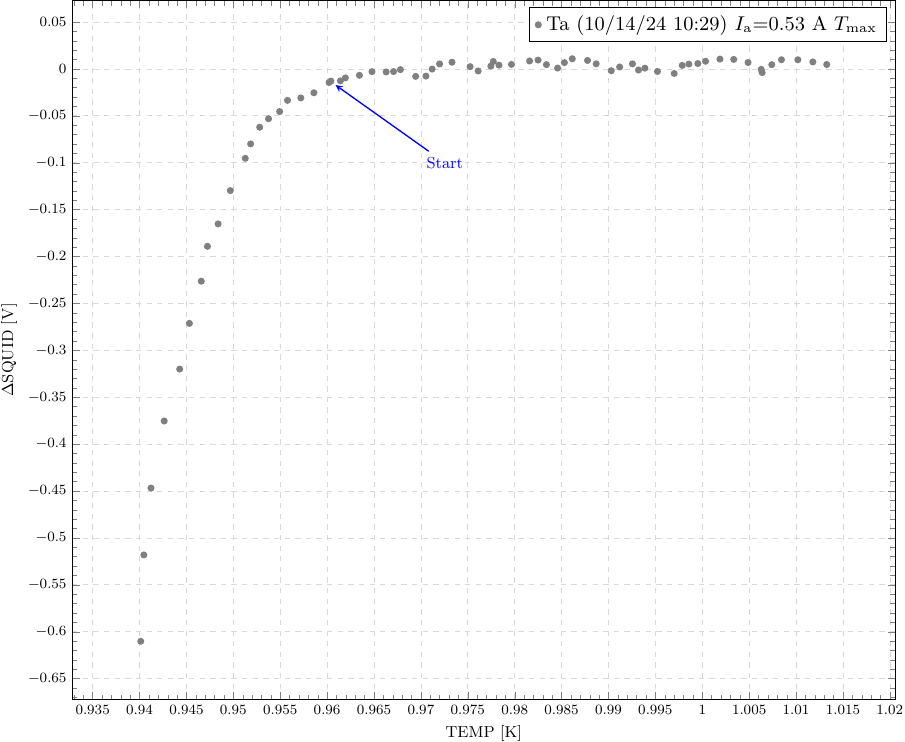}
&
  \includegraphics*{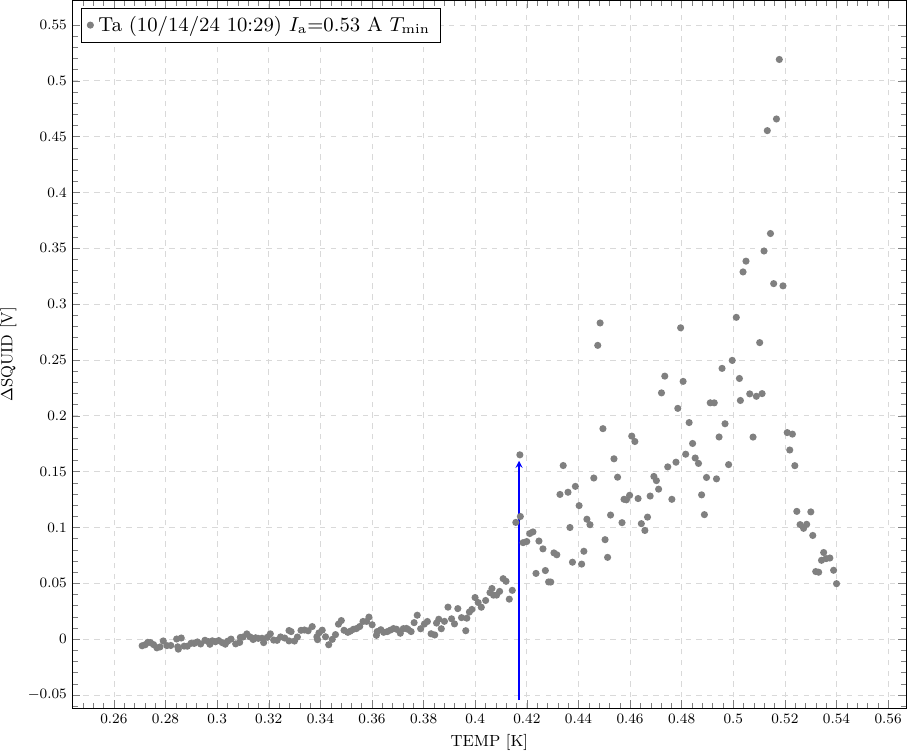}
\\
  \includegraphics*{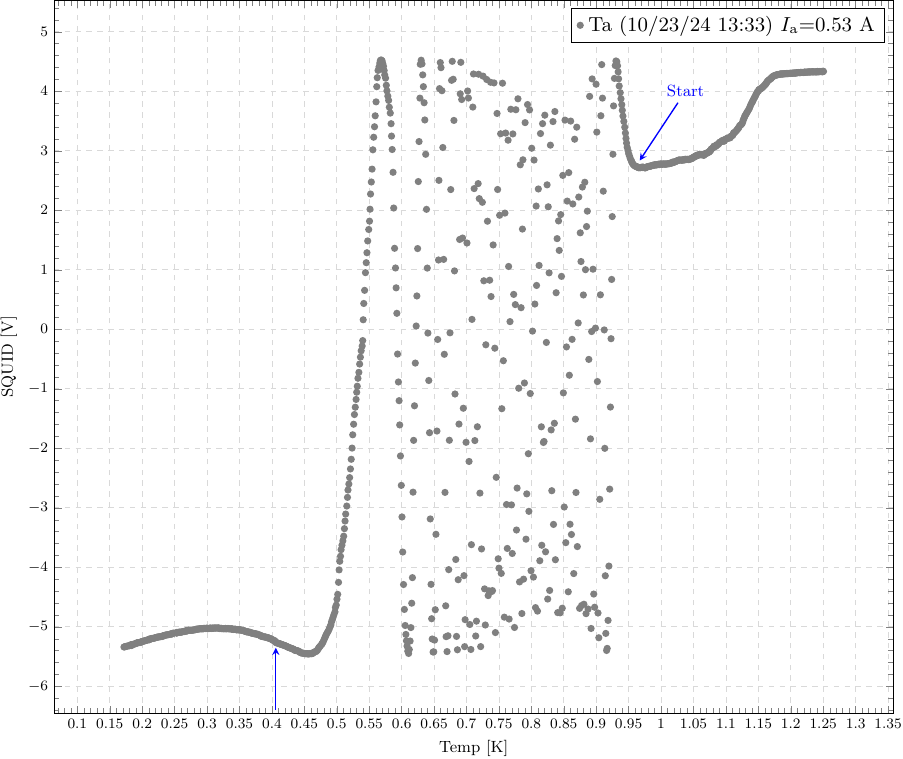}
&
  \includegraphics*{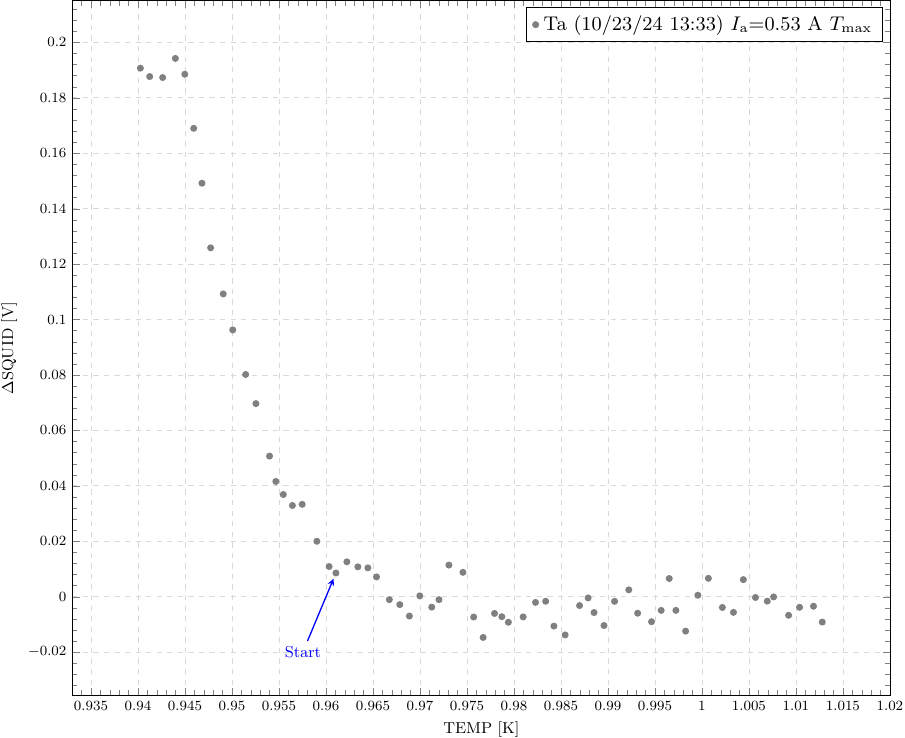}
&
  \includegraphics*{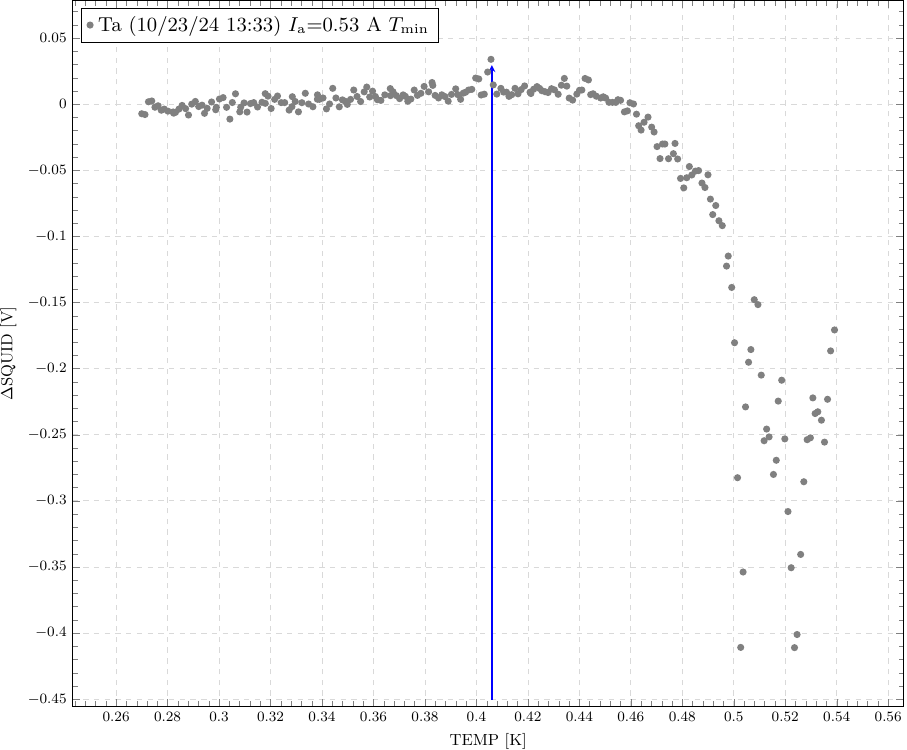}
\\
  \includegraphics*{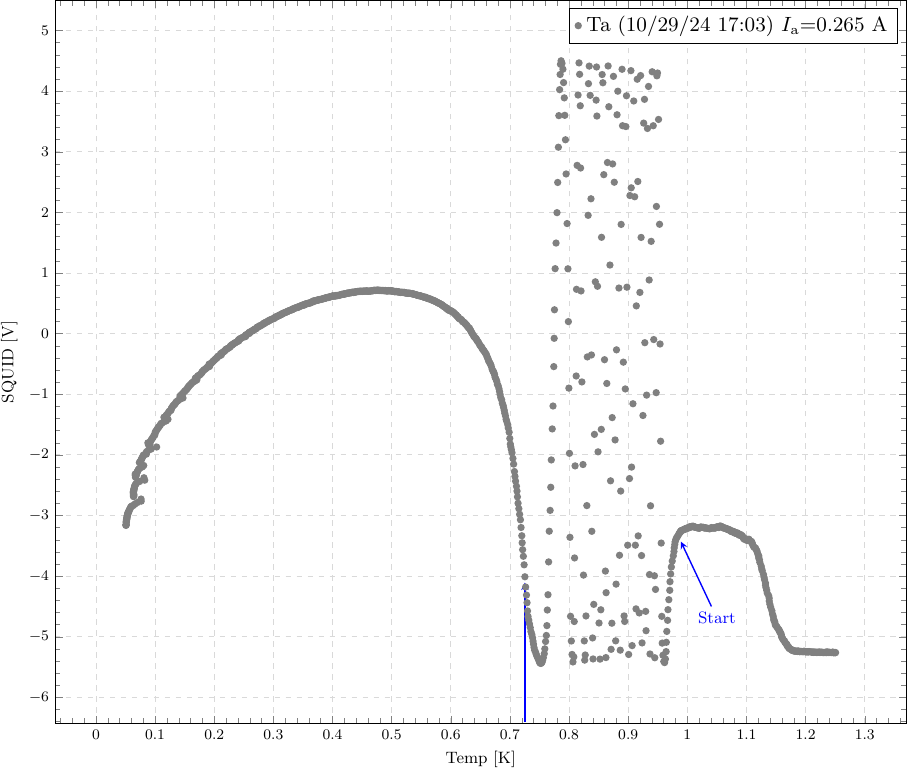}
&
  \includegraphics*{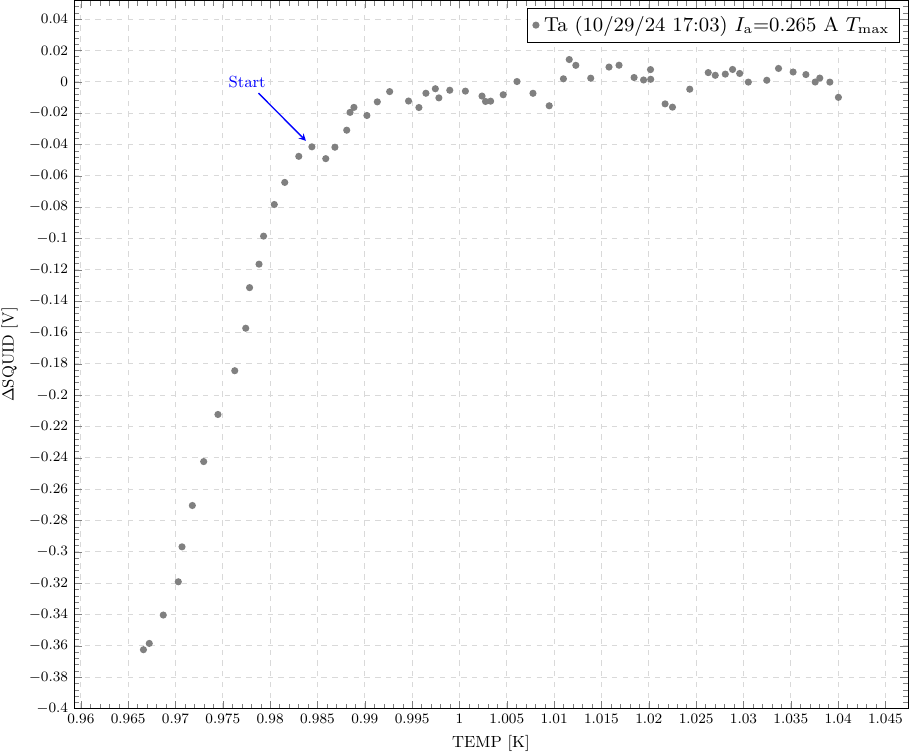}
&
  \includegraphics*{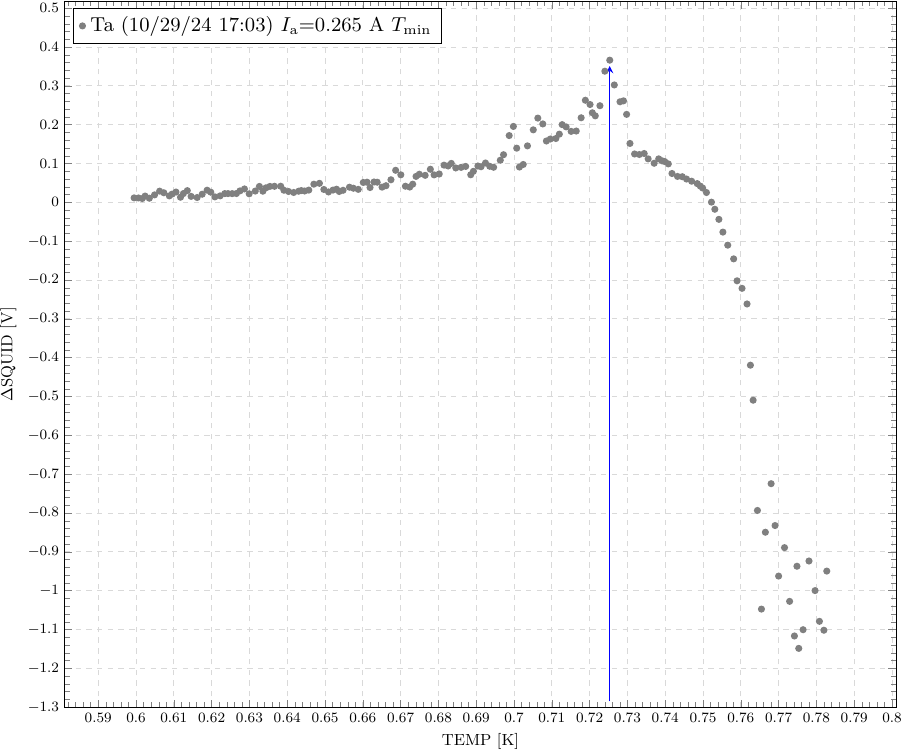}
\\
  \includegraphics*{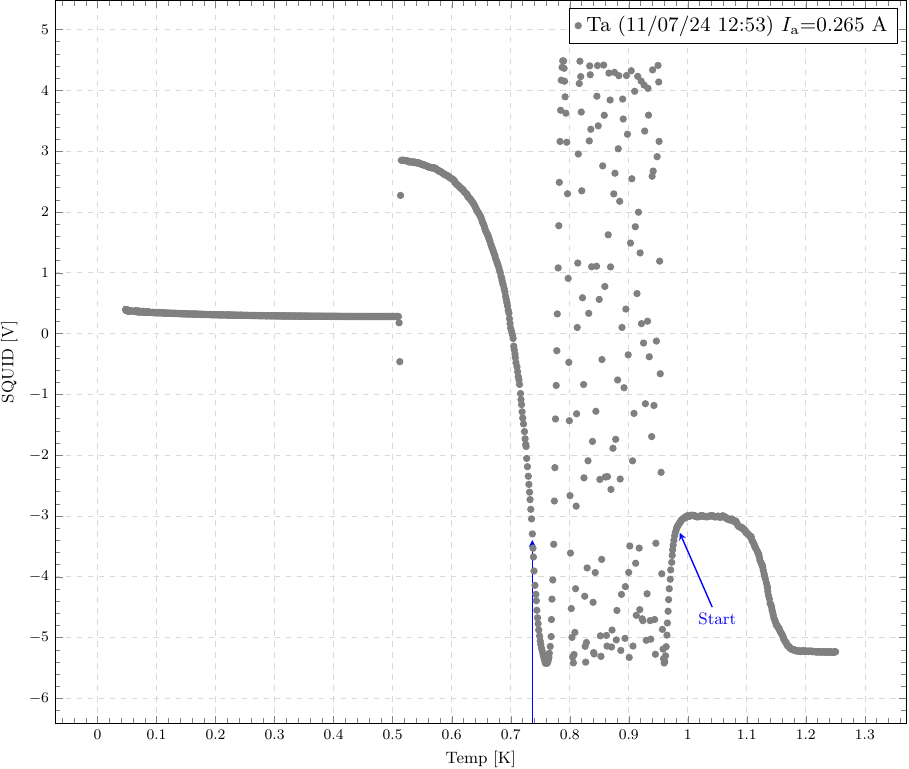}  
&
  \includegraphics*{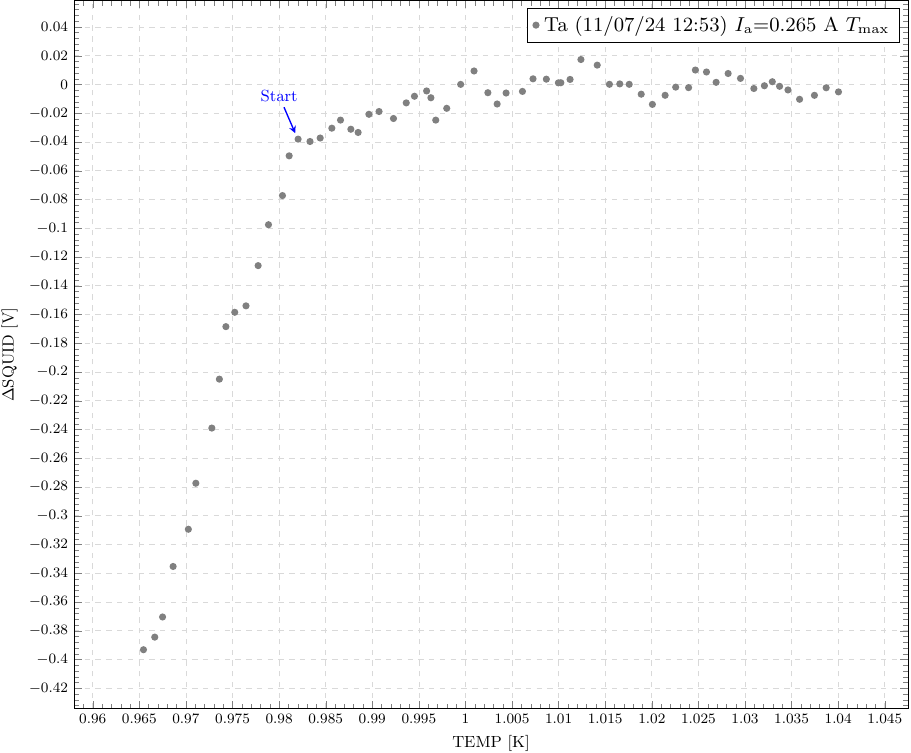}
&
  \includegraphics*{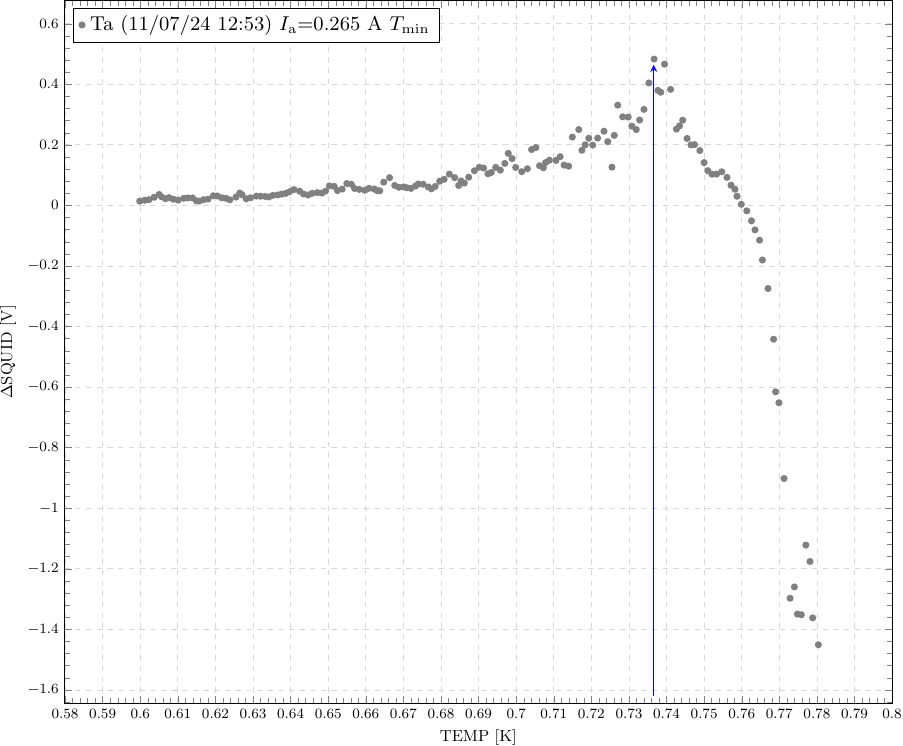}
\\
  \includegraphics*{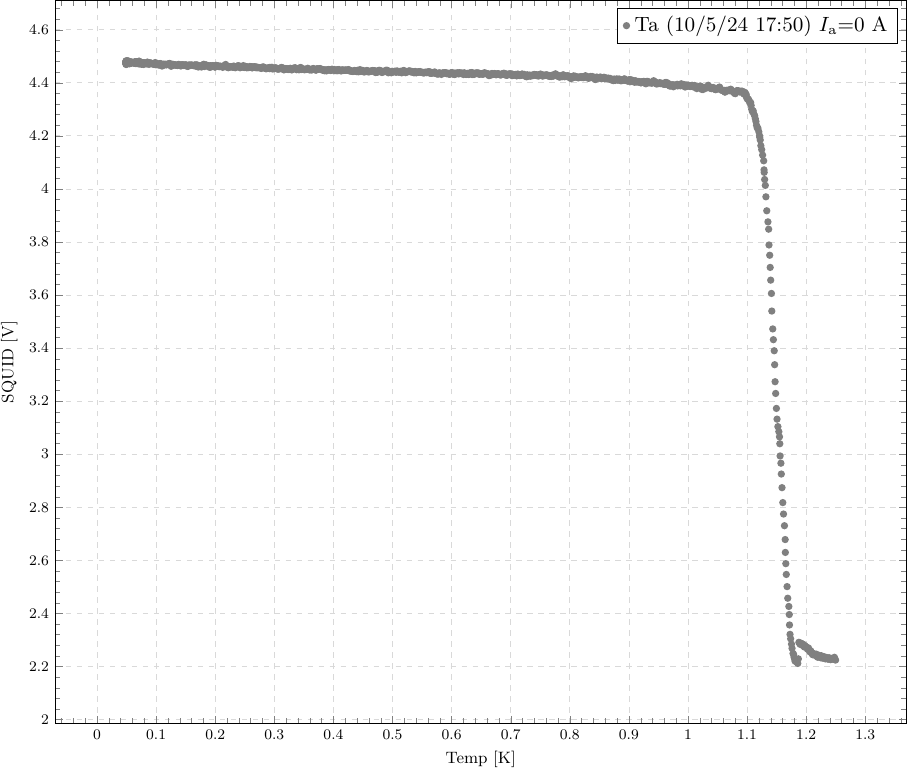}%node [right] {\color{cyan}
&
 \includegraphics*{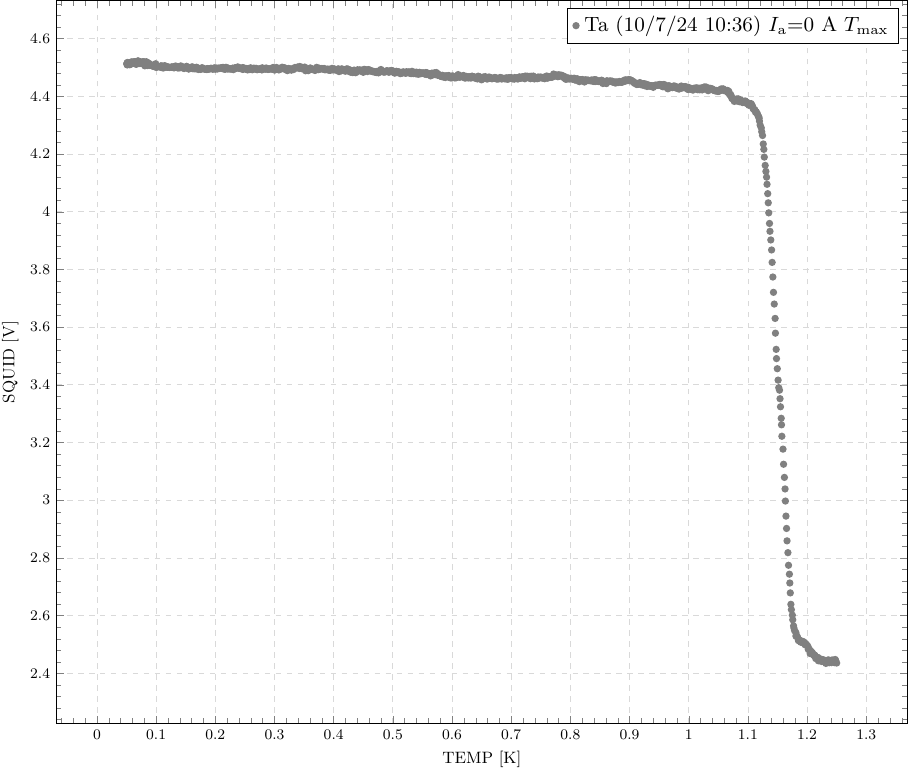}
&
\end{tabular}
}
\caption{SQUID data during the ramp down at $T<$1.25~K showing the beginning and completion of the Meissner effect at temperature of \Tmax~ and \Tmin~ respectively. Each data point on this plot represents one second and an averaging of 1000 SQUID raw data point. There is a lag in the temperature of the Aluminum booster core of about $\tau=0.194$~K (see \Eq{tau}). Using \Eq{criticalT_B} we can determine the magnetic field at the beginning of the Meissner effect and at the end. The magnetic field being proportional to the booster solenoid current, we find that the booster solenoid current was 10 larger below the critical temperature of aluminum than above.}\label{regenPlots}
\end{figure*}
\begin{figure*}
\centering
\pgfplotsset{
    small,
    legend style={
        at={(0.01,0.01)},
        anchor=south west,
    },
   }%
\resizebox{17 cm}{!}{
\begin{tabular}{ccc}

 \includegraphics*{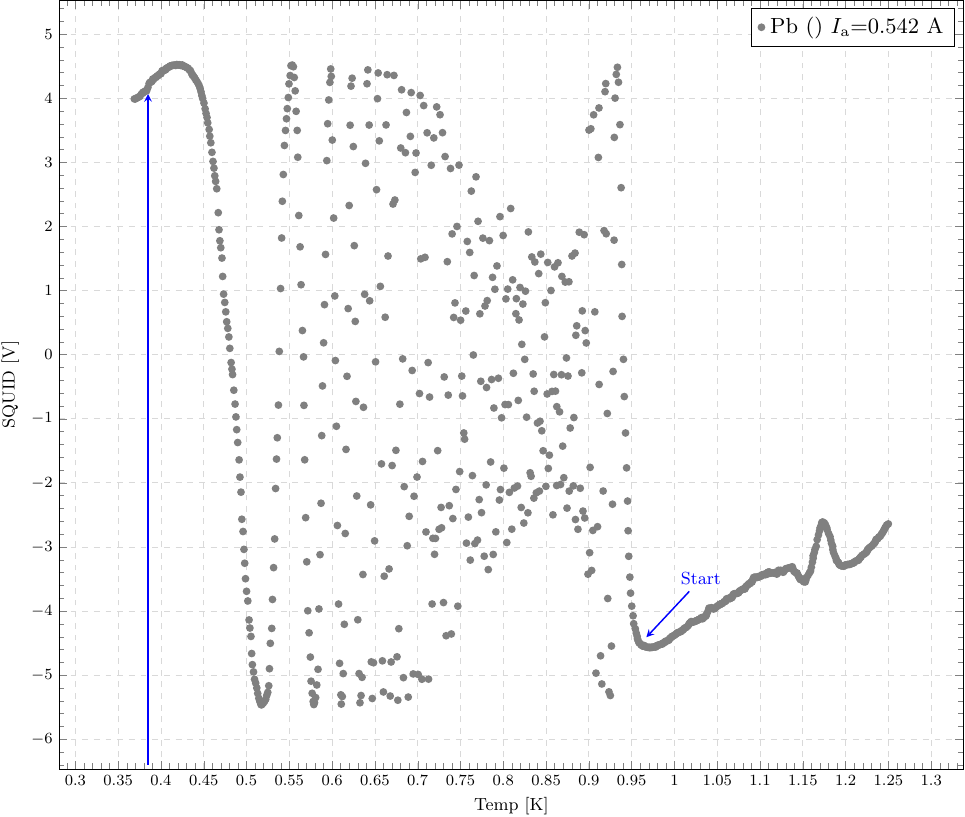}
&
 \includegraphics*{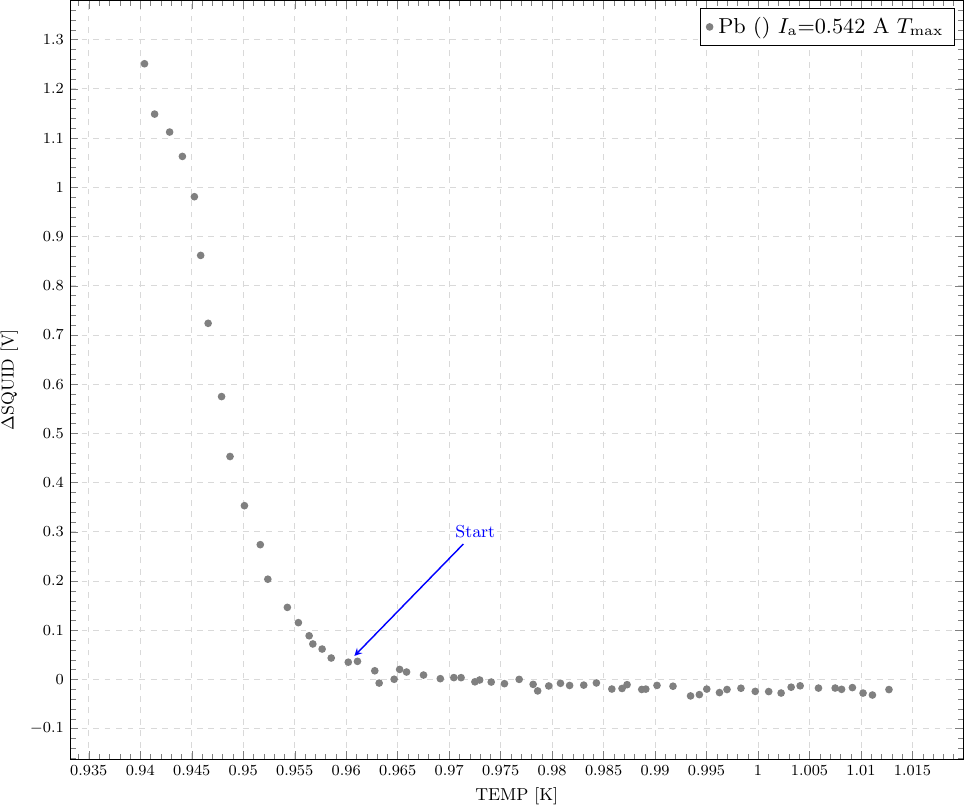}
&
 \includegraphics*{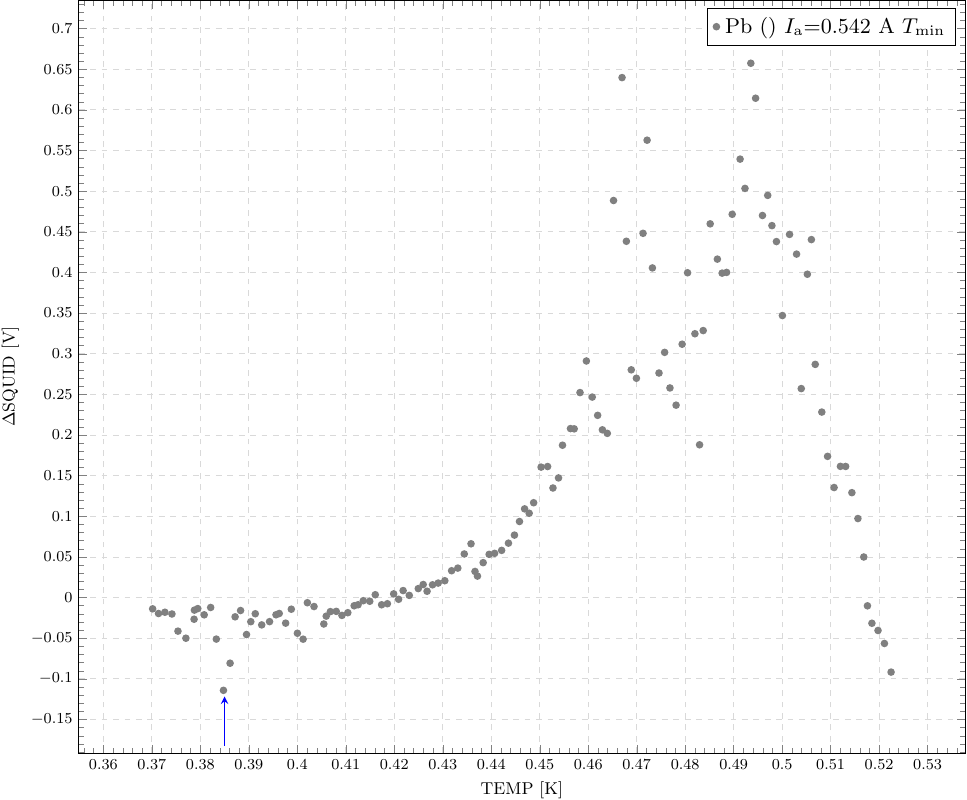}
\\
 \includegraphics*{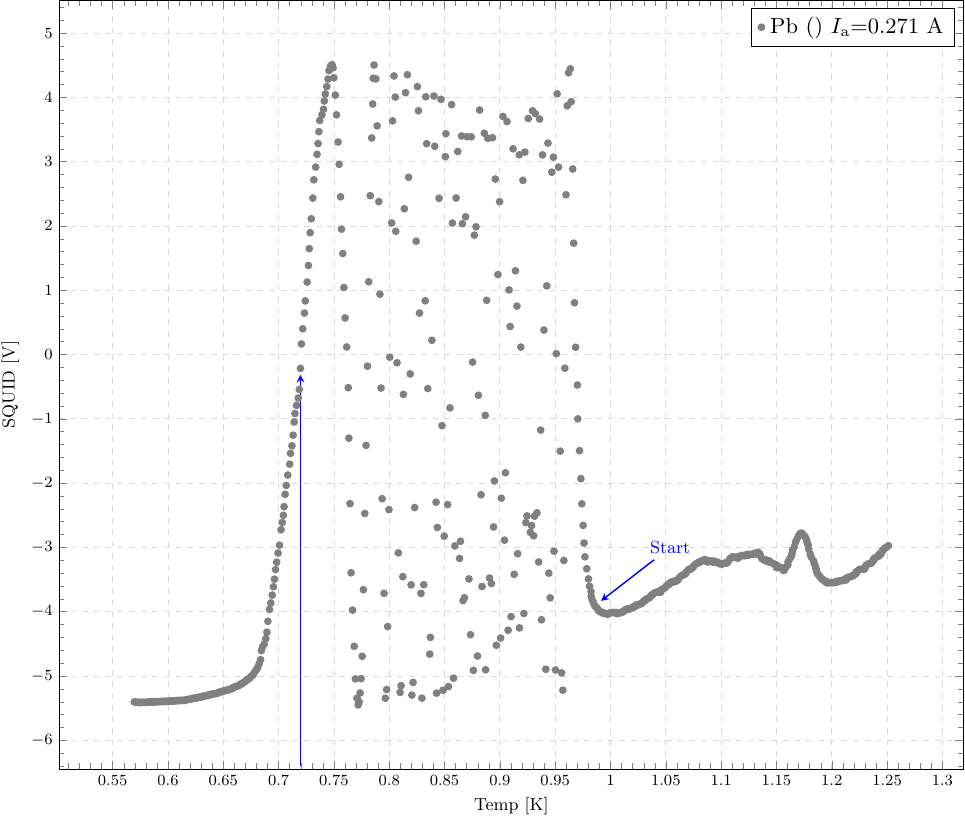}
&
 \includegraphics*{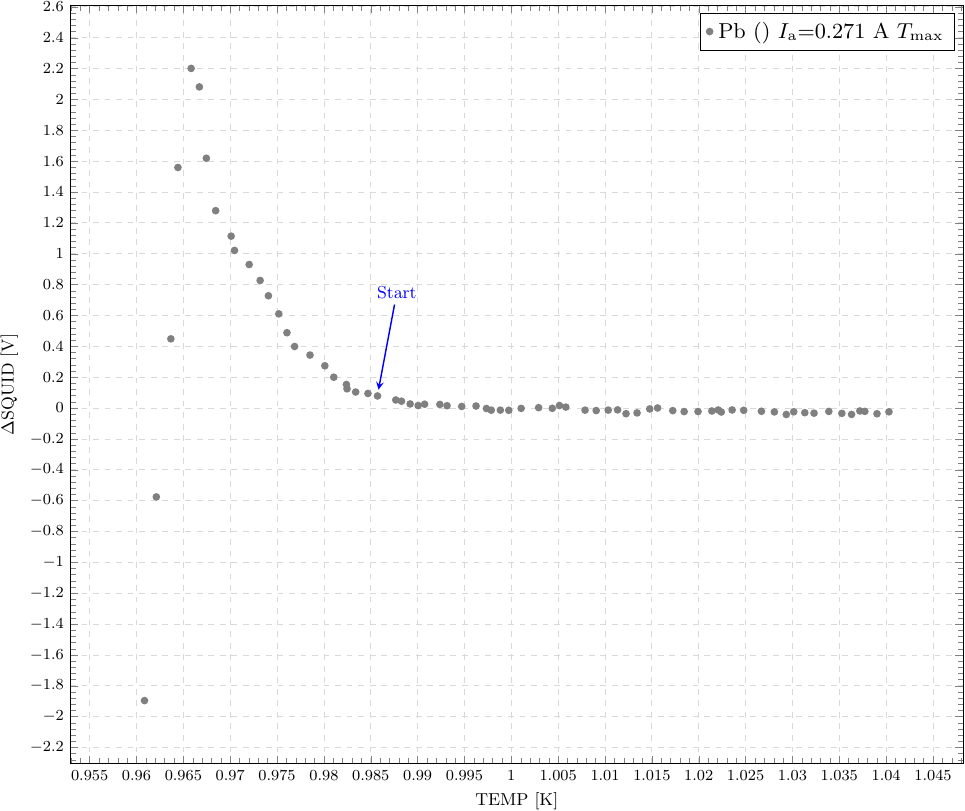}
&
 \includegraphics*{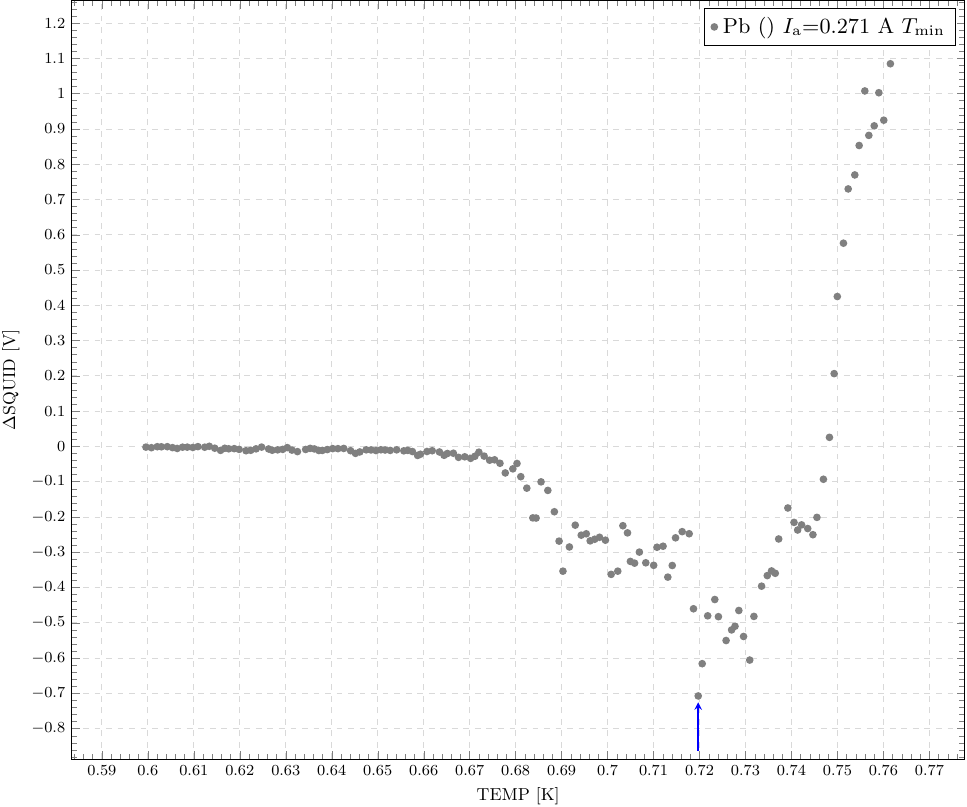}
\\
 \includegraphics*{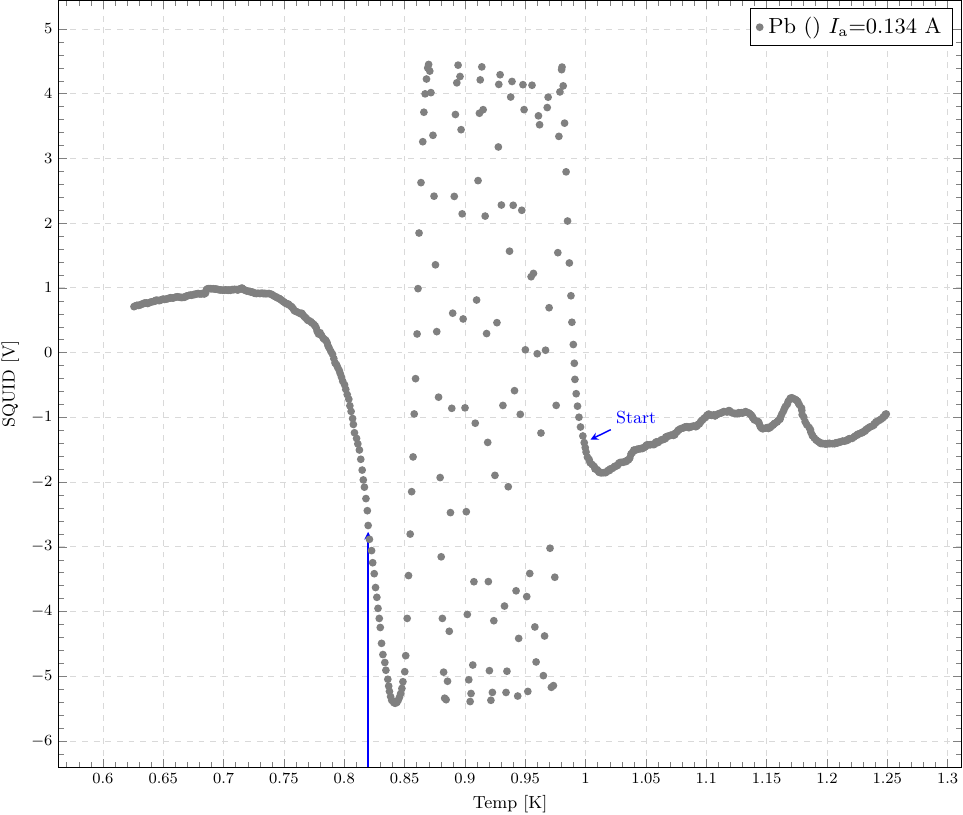}
&
 \includegraphics*{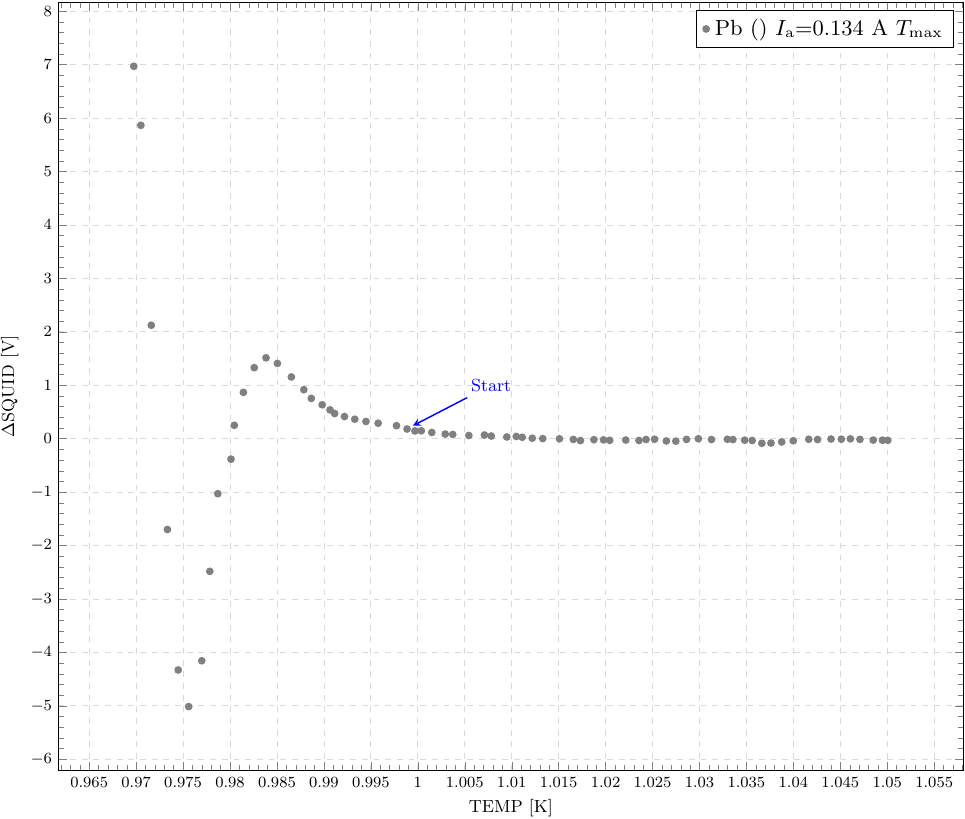}
&
 \includegraphics*{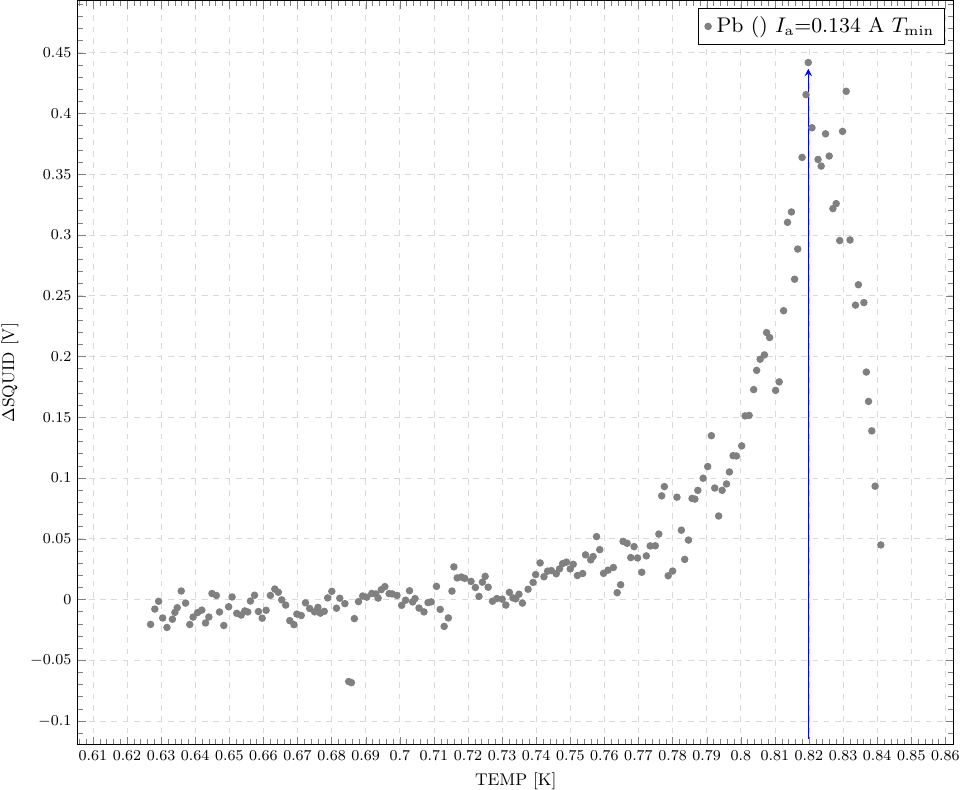}

\end{tabular}
}
%\vspace{-2.5cm}
\caption{REGEN plots at various time, $\text{\Ia}$ for Pb. Each data point is an average of the raw REGEN data points over one second.}\label{PbregenPlots}
\end{figure*}
\begin{center}
\begin{table*}[t]
\begin{tabular}{|l|l|l|l|l|l|l|l|l|l|}
\hline
              & \Ia (A) &   $T_\text{max}^\text{ADR}$~(K)  & \Tmax~(K)   &    $I_\text{sol,min}$(A)  &   $I_\text{min}^\text{num}$  & $T_\text{min}^\text{ADR}$~(K)  &   \Tmin~(K)  &  $I_\text{sol,max}$(A) & $I_\text{max}^\text{num}$ \\
\hline
Ta (10/14/24 10:29) & 530  &   0.960   & 1.154     & 0.37      &    0.33       &  0.417     &   0.611   &  3.66 & 3.54       \\
\hline
afterREG85 Ta  (10/23/24 13:33)  & 530 &  0.961    &  1.154    &     0.37       &   0.33    &     0.406      &   0.6    &  3.71  &  3.54       \\
\hline
REG65 Ta (10/29/24 17:03) & 265  &   0.984   &  1.177   &      0.19     &   0.17      &  0.725     &  0.919    &   2.04   &   1.77     \\
\hline
REG85 Ta (11/07/24 12:53) & 265  &  0.982    & 1.176     &     0.19      &    0.17     & 0.737      &   0.931   &  1.97    &    1.77    \\
\hline
Ta (10/5/24 17:50) & 0   &   1.18 &   &  --    & --    &  1.13      &     &  --   &   -- \\
\hline
Ta (10/7/24 10:36) & 0  &  1.18&     &  --     &   --         & 1.13     &        &  --   & --        \\
\hline
Pb  & 542  &   0.960   &  1.15   &     0.40      &   0.33      &  0.385     &  0.579    &   3.79   &   3.62     \\
\hline
Pb  & 271  &   0.986   &  1.1764   &     0.195      &    0.168     &  0.720     &   0.914   & 2.307    &     1.81   \\
\hline
Pb  & 134  &   0.999   &  1.1928   &    0.059       &  0.083       &  0.820     &  0.96  &  1.41    &   0.90     \\
\hline
\end{tabular}
\caption{Solenoid currents at the start and end of the Meissner effect for the element and date listed. Errors on the values of \Tmax ~and \Tmin ~can be taken as large enough to cover the range of values between the experimental and theoretical determinations.}\label{IsolVal}
\end{table*}
\end{center}

\subsection{Fixed temperature and fixed current data}\label{sec:fixedT}

Datasets have been created at various fixed temperatures and fixed currents. For the data analysis, the data points from different sets are compared and subtracted in order to assess and minimize systematics. The notation for the datasets and their properties are as follows
\begi
Each SQUID data point is an average of $10^5$ raw data points collected over 100~s and written into individual files at regular intervals of approximately 100.3~s. In this paper,  the words ``data point'' always refers to this average SQUID data point and not to a raw data point.
To each data point corresponds an average time taken to be the average of the raw time data points over which the raw SQUID data points were collected. In this paper, time always refers to this average time.
The data points in all datasets are always time-ordered with the earliest data point collected being the leftmost element and the last data point collected being the rightmost element.
To each data point corresponds an average {\it Pill Solenoid Current}  \ips~ taken to be the average of the raw \ips~ data points over which the raw SQUID data points were collected. In this paper, \ips~ always refers to this average current.
The absolute time of  collection of each data point is known for all datasets. However, for plotting and comparison  purposes, a relative time is used such that  the time corresponding to the first data point is set to 0 by subtracting the value of the first time data point from all the time data points.
$\text{\Ia}$ is the  current from the external  DC source applied to the main solenoid of the sample assembly.
$\text{\Isol}$ is the  current in the booster solenoid that creates the magnetic field in the penetration depth of the sample circuit.
\Is~ is the  current in the sample circuit.
\dIs is the change in the current in the sample circuit over a time $\Delta t$.
\Tset{Ta}=\{52.5, 65, 75, 85, 105\} is the set of fixed temperatures at which tantalum data was collected in units of mK.
\Iset{Ta}=\{0, 265, 530\} is the set of fixed $\text{\Ia}$ at which tantalum data was collected in units of mA. Data was also collected at \Ia=$-510$~mA in tantalum at a temperature of 65~mK.
\istc=\{$s_1, s_2,..., s_i,..., s_{N}$\} is the set of SQUID data points $s_i$ collected at an applied current $I_\text{a}$ (mA), a fixed temperature $T$ (mK), at time $t_i$ and FAA pill current $I_{\text{s},i}$ for a superconducting compound $C$ where $N$=$|$\istc$|$; the $s_i$ have units of V. For example \stc{530}{52.5}{Ta} represents the set of SQUID data points collected at an applied current of 530~mA at a temperature of 52.5~mK for tantalum. Therefore, each dataset \istc is a line in the 3-dimensional space of SQUID voltage, \ips~ and time.
\iitc=\{$i_{\text{ps}1}, i_{\text{ps}2},..., i_{\text{ps}N}$\} the set of average \ips~ data points that were collected at the same time as the corresponding data points in  \istc. \iitc is isomorphic with \istc ~and depends only on the value of $T$.
\ittc=\{$t_1, t_2,..., t_{N}$\} the set of average time data points\footnote{These are calculated from raw computer times collected by the ADR software. The offset between computer time and real time is of the order of 1~s over $10^5$~s.} at which  \istc ~and \iitc ~data points were collected. All three sets are isomorphic and in particular the cardinalities\footnote{The number of elements of a set.} are equal: $|$\ittc$|$=$|$\istc$|$=$|$\iitc$|$.
\dstc{I_{a1}}{I_{a2}}{T}{C}{m,n}$=\text{ \{}s_i~|~s_i=s_{2,i+n-1}-s_{1,i+m-1},~ s_{2,j}\!\in\!\mathbf{_{I_\text{a2}}\!S_T^C},~ s_{1,j}\!\in\!\mathbf{_{I_\text{a1}}\!S_T^C}$, $i=1,2,...,N_\text{min}$\} is the set of data points obtained by subtracting  a subset  of data points \dsstc{I_\text{a1}}{}{T}{C}{}$\subseteq\mathbf{_{I_\text{a1}}S_T^C}$ from a subset of data points \dsstc{I_\text{a2}}{}{T}{C}{}$\subseteq\mathbf{_{I_\text{a2}}S_T^C}$, where $m,n$ are selected to minimize the systematics as described below. $m\ge1$ is the  index of the element in $\mathbf{_{I_\text{a1}}S_T^C}$ that corresponds to the first element in the subset  \dsstc{I_\text{a1}}{}{T}{C}{} while $n\ge1$ is the  index of the element in $\mathbf{_{I_\text{a2}}S_T^C}$ that corresponds to the first element in the subset  \dsstc{I_\text{a2}}{}{T}{C}{}; $N_\text{min}$=$|$\dstc{I_\text{a2}}{}{T}{C}{}$|$-$n+1$ or $|$\dstc{I_\text{a1}}{}{T}{C}{}$|$-$m+1$, whichever is smaller.
\ditc{I_{a1}}{I_{a2}}{T}{C}{m,n}$=\text{ \{}I_{\text{ps},i}~|~I_{\text{ps},i}=I_{\text{ps},2,i+n-1}-I_{\text{ps},1,i+m-1},~ I_{\text{ps},2,j}\!\in\!\mathbf{_{I_\text{a2}}\!I_{\text{\bf{ps}},T}^C},~ I_{\text{ps},1,j}\!\in\!\mathbf{_{I_\text{a1}}\!I_{\text{\bf{ps}},T}^C}$ $i=1,2,...,N_\text{min}$\} is the set of data points obtained by subtracting  a subset  of data points in $\mathbf{_{I_\text{a1}}I_{\text{ps},T}^C}$ from a subset of data points in  $\mathbf{_{I_\text{a2}}I_{\text{ps},T}^C}$ where $m$, $n$ and $N_\text{min}$ are the same as in \dstc{I_{a1}}{I_{a2}}{T}{C}{m,n}.
$o\equiv n-m$ the offset between the selected data points in \dstc{I_\text{a2}}{}{T}{C}{} and \dstc{I_\text{a1}}{}{T}{C}{}.
\ei
\begin{figure}
\resizebox{8.75 cm}{!}{
\begin{tabular}{|c|c|c|c|c|c|}
\hline
          &   52.5~mK   &    65~mK   &   75~mK & 85~mK & 105~mK   \\
\hline
$-510$~mA &     $--$      &     29.8 (1071)   &    $--$   &      $--$         &        $--$         \\
\hline
0~mA & 14.7 (530)   &  14.9 (536)  & 18.4 (662) &  12.1 (437)  &  14.3 (515)   \\
\hline
265~mA &  16.7  (600) &  30.8 (1107)&   39.5 (1421)   &  48.0 (1728)  &   60.0 (2158) \\
\hline
530~mA & 15.5 (559)  & 30.5 (1096) &   40.0  (1438)  & 48.6  (1747)  &  61.0 (2194)  \\
\hline
\end{tabular}
}
\caption{The numbers at the intersection of the fixed currents (rows) and the fixed temperatures (columns) give the total number of hours of data collected at that fixed current and that fixed temperature with the tantalum sample, and, in parentheses, the total number of average SQUID data points. For example, over a period of 16.7 hours, 600 tantalum data points were collected at an applied current of 265~mA and at a temperature of 52.5~mK. A total of 17799 data points over 494.8 hours was collected at fixed temperature and current for tantalum.}\label{TaDatasets}\end{figure}
A fixed temperature in an ADR is achieved with a \ips$\ne0$. Indeed, adiabatically decreasing the spin entropy of the FAA salt pill with a magnetic field, implies an increase in the vibrational entropy of the pill and temperature. The fixed temperature can be maintained as long as the ambient heat absorbed by the pill does not exceed the amount needed to raise the pill temperature to the fixed temperature value. This means that as more ambient heat is absorbed by the FAA pill, the \ips~ decreases until it reaches 0. This fixed temperature mode of the ADR will be referred to as the REG mode.

For tantalum, the size of the experimental datasets at fixed temperature and fixed current $\text{\Ia}$  are given in \Tab{TaDatasets}. The numbers at the intersections of the currents (rows) and the temperatures (columns) give the run time and total number data points in parentheses at that fixed temperature and fixed current. For tantalum, sixteen \stc{I_a}{T}{Ta} datasets were experimentally created with a total number 17799 data points over 494.8 hours.

To determine the temperature dependence of the SQUID signal which is proportional to \dIs, a linear regression on time was performed for the 5 differenced datasets \dstc{265}{530}{T}{Ta}{~m,n} $\forall~T\in$~\Tset{Ta} using the REG mode of the ADR. As discussed at the end of \cref{expModel}, the systematic due to \ips~ should be minimized in differenced datasets while the systematic due to the time derivative of the temperature is null. The linear regressions on time will provide five rates of change of the differential SQUID signal at five fixed temperatures for an applied current difference of $I_\text{a2}-I_\text{a1}=265$~mA. According to \Eq{curieSol}, these rates of change should be inversely proportional to $T$ and proportional to the applied current \Isol=$M_\text{bm}$$\text{\Ia}$ where $M_\text{bm}$ is the mutual inductance between the booster and main solenoids.  \Ia ~will be used as a shorthand for \Isol ~in the description and analysis of the data. In \cref{sec:reverse}, the fixed temperature dataset \dstc{-510}{530}{65}{Ta}{~m,n} where the direction of $\text{\Isol}$ is reversed is described; using that reverse field dataset, a sixth linear regression slope with $I_\text{a2}-I_\text{a1}=1040$~mA is included in the final analysis of the $1/T$ dependence.

Complete datasets were collected for \stc{530}{T}{Ta} and  \stc{265}{T}{Ta}; the word `complete' means that the collection of raw data in the REG mode was started after a REGEN was performed and continued to be collected until \ips~ was too small to maintain a fixed temperature. For a fixed temperature $T$, the difference in the number of data points between \stc{530}{T}{Ta} and \stc{265}{T}{Ta} was always $<$40 out of 600 to 2194 data points. Hence  $o=m-n$ is typically at most a few dozens and \itc{530}{T}{Ta}$\cong$\itc{265}{T}{Ta} $\forall~T\in$~\Tset{Ta}. \stc{-510}{65}{Ta} is also a complete dataset. At \Ia=0~mA, only \stc{0}{52.5}{Ta} is a complete dataset while for $T\ne52.5$, between 437 and 662 data points were collected, just enough to establish the dependence of the rate of change on $\text{\Ia}$ using \dstc{0}{530}{T}{Ta}{o} and \dstc{0}{265}{T}{Ta}{o}. Using the \stc{0}{52.5}{Ta} dataset, a seventh linear regression slope with $I_\text{a2}-I_\text{a1}=530$~mA is included in the final analysis of the $1/T$ dependence.

The standard deviation on a fixed temperature is typically about 10~$\mu$K and nearly time-independent as seen in the plots of \cref{app:Tstddev}. These small temperature fluctuations are not experimentally significant as confirmed by the data at \Ia=0~mA. Hence, they also do not create any significant power emission, and cannot create the temperature dependence observed.

\subsubsection{Ta: dependence of the sample current on the temperature}\label{sec:1overT}

As mentioned previously, \dstc{265}{530}{T}{Ta}{~m,n} is used to extract the dependence of the rate of change on temperature. Since the \ips~ does not depend on the main solenoid current \Ia, one can control for any \ips~ systematic by subtracting the dataset collected at fixed $T$ and \Ia=530~mA from the dataset collected at the same fixed $T$ but at \Ia=265~mA. The cardinalities of \stc{530}{T}{Ta} and {265}{T}{Ta} were different, but it was found that these differences were quite small for all datasets taken at the same fixed temperature $T$. To assess the impact that \stc{265}{T}{Ta} is not one-to-one with \stc{530}{T}{Ta}, $m,n$ were varied using the following procedure
\begi
Find the $m=\bar{m},n=\bar{n}$   that minimize    systematics shared by \stc{265}{T}{Ta} and  \stc{530}{T}{Ta} in \dstc{265}{530}{T}{Ta}{\bar{m},\bar{n}}
Plot the corresponding $\Delta$\ips=\ditc{265}{530}{T}{Ta}{\bar{m},\bar{n}}
Recompute \dstc{265}{530}{T}{Ta}{~m,n} and  \ditc{265}{530}{T}{Ta}{~m,n} for different values of $m,n$  to verify robustness against these offset variations between the two datasets being subtracted.
The linear regression rate of change corresponding to \dstc{I_{a1}}{I_{a2}}{T}{ele}{\bar{m},\bar{n}} is denoted \gtc{I_{a2}-I_{a1}}{T}{ele}. The linear regression slope of \dstc{265}{530}{T}{Ta}{\bar{m},\bar{n}} is used to  produce a new dataset of linear model data points \mtc{265}{T}{Ta}=\{$s~|~s_n$=\gtc{265}{T}{Ta}$n\Delta~t_n+b, n=1,2,...,N$\} where $b$ is the linear regression intercept, $N$=$|$\dstc{265}{530}{T}{Ta}{~m,n}$|$ ~and $\Delta t_n\in\text{\{}\Delta t~|~\Delta t_n=t_{n+1}-t_n,~t_n\in\text{\ittc},~n=1,2,...,N-1,~\Delta t_N=\Delta t_{N-1}\text{\}}$. Note that in \mtc{265}{T}{Ta}, the bottom left 265 label represents \Ia=530-265=265. The explanatory power of the model is given by the corresponding $R^2$.
\ei

\begin{figure*}
\centering
\resizebox{16 cm}{!}{
\begin{tabular}{ccc}

 \includegraphics[width=16cm]{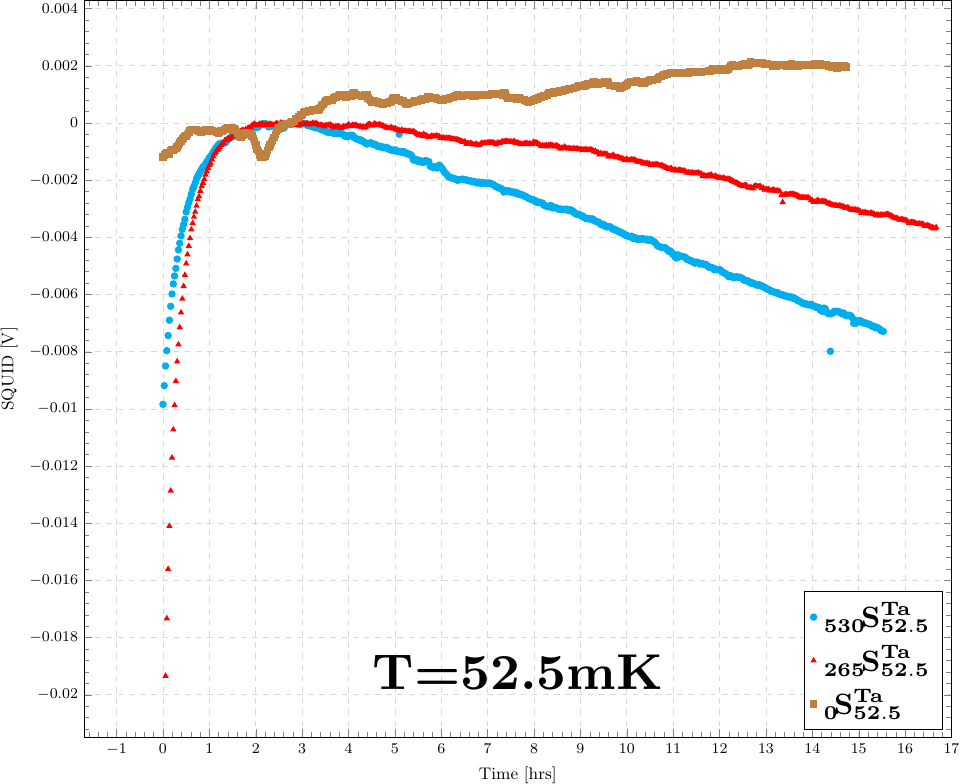}
&
 \includegraphics[width=16cm]{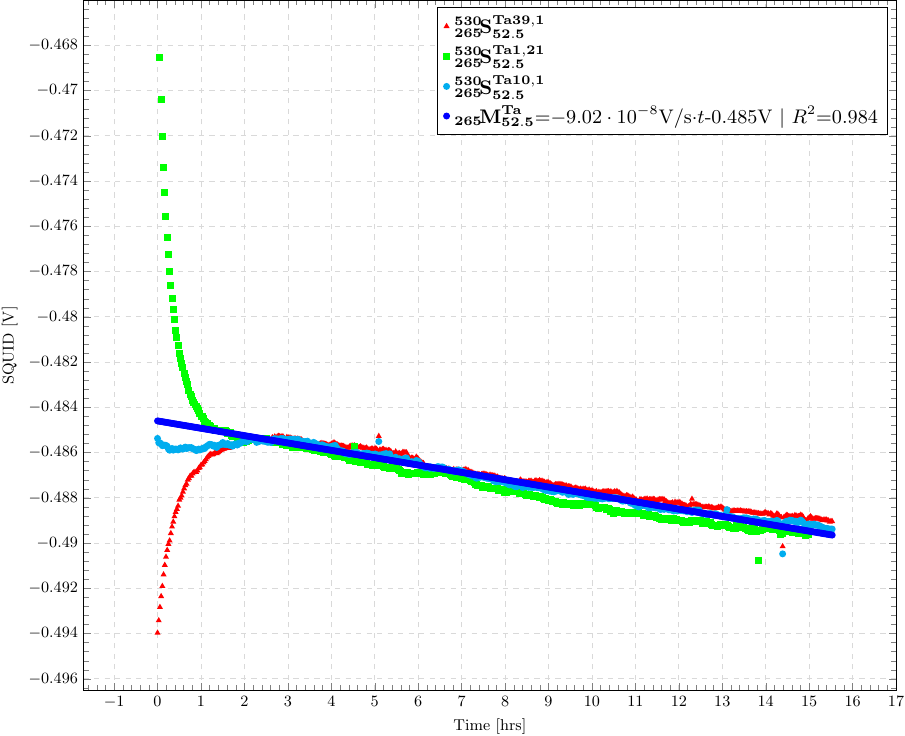}
&
 \includegraphics[width=17.1cm]{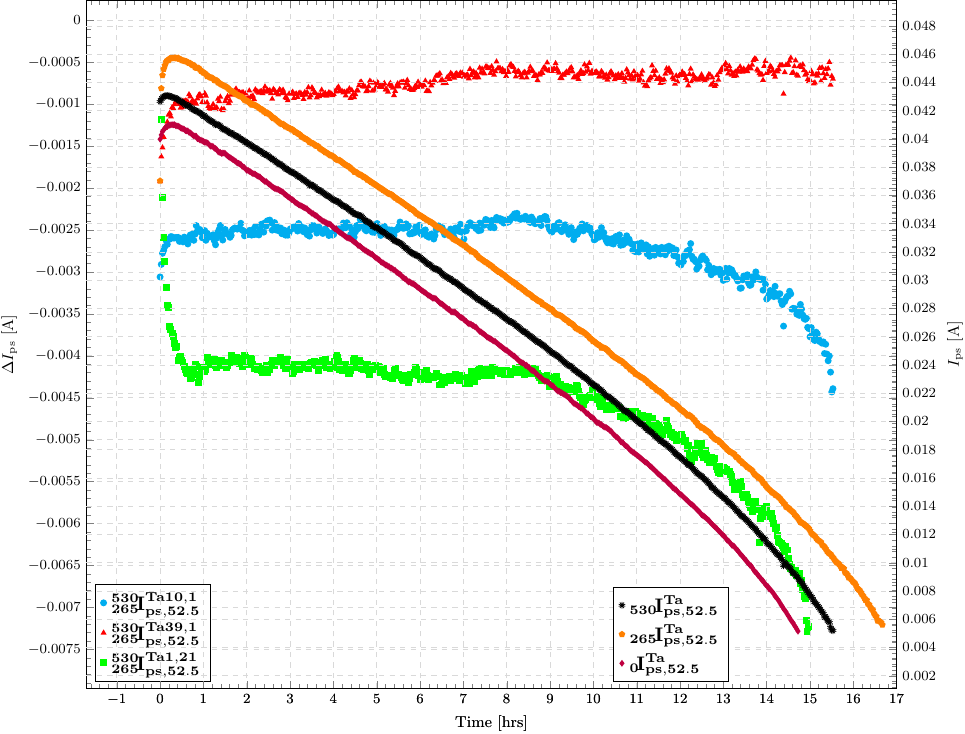}
\\
 \includegraphics[width=16cm]{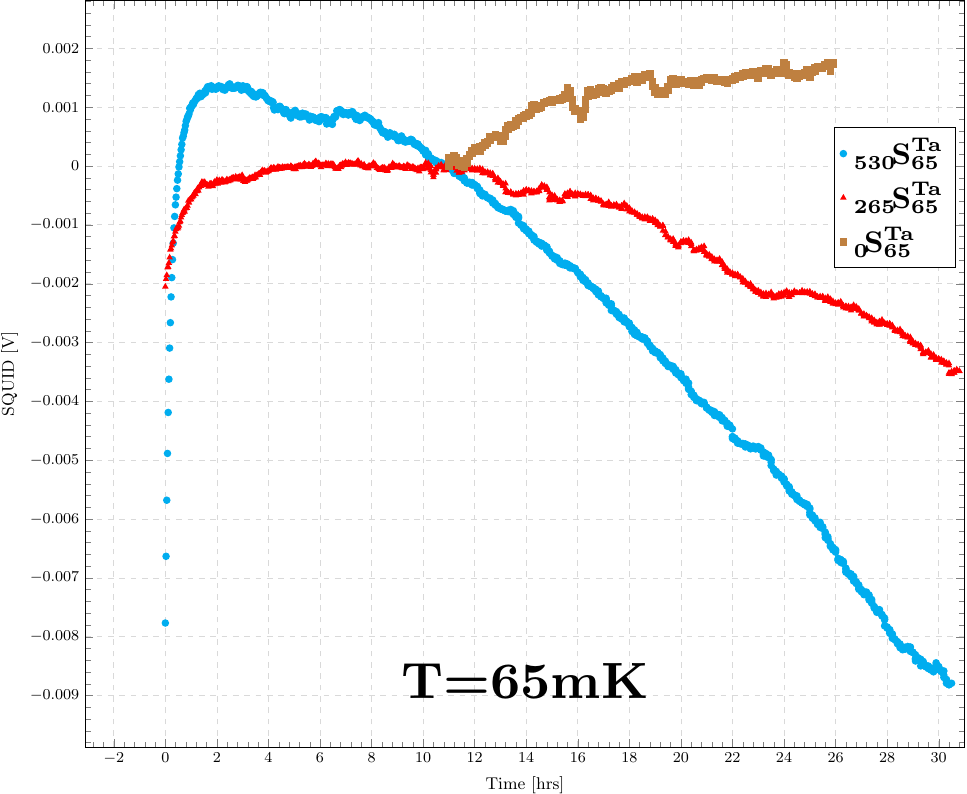}
&
 \includegraphics[width=16cm]{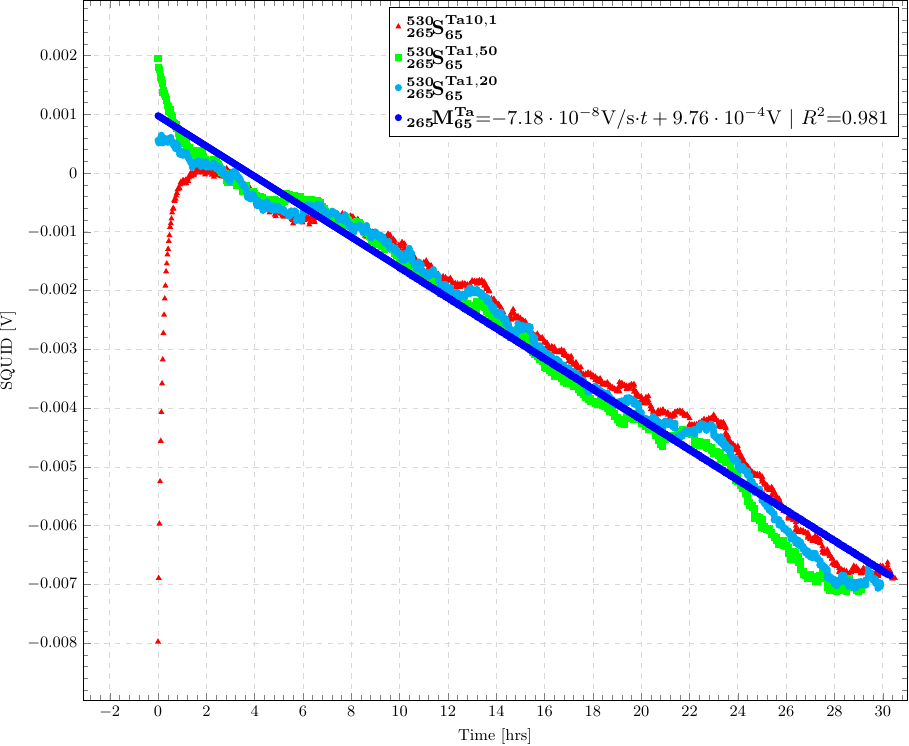}
&
 \includegraphics[width=17.1cm]{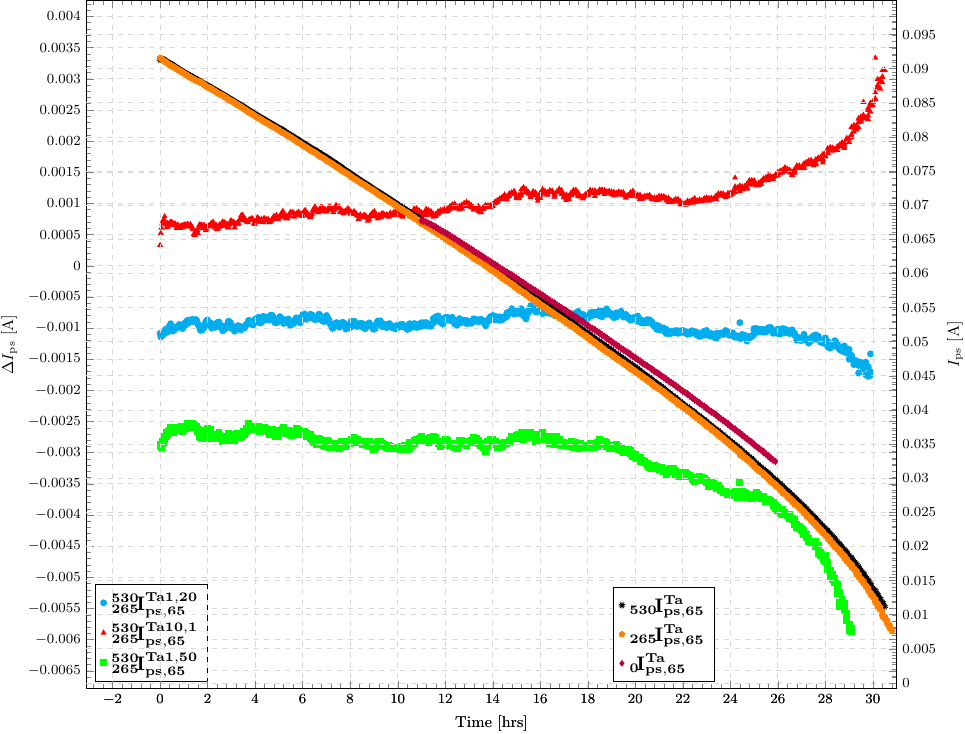}
\\
 \includegraphics[width=16cm]{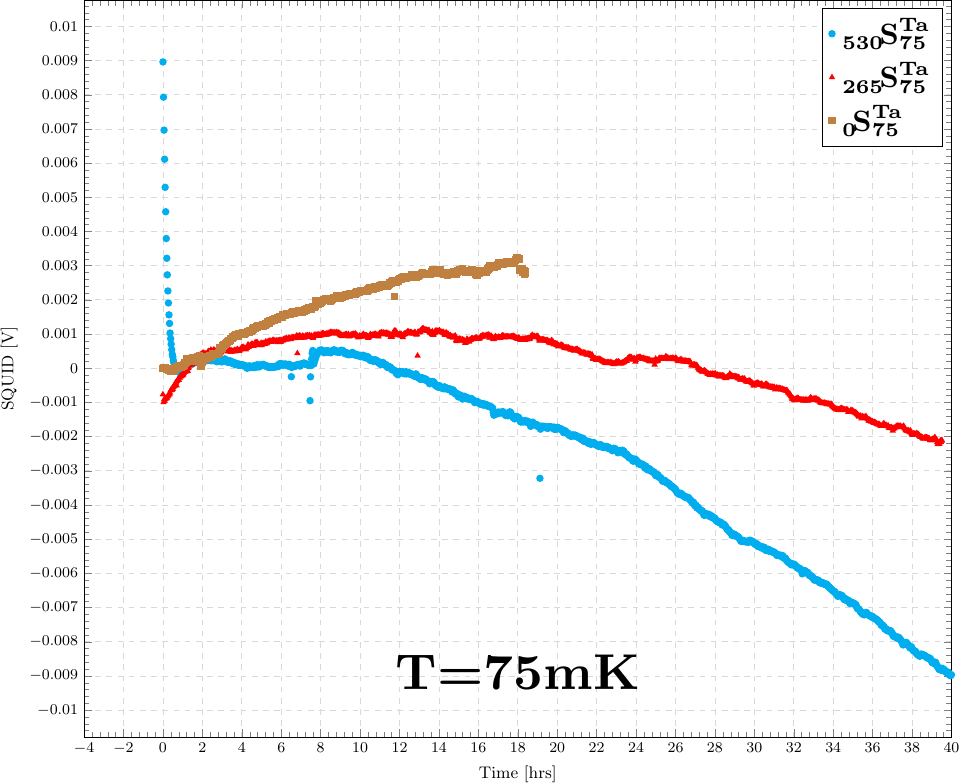}
&
 \includegraphics[width=16cm]{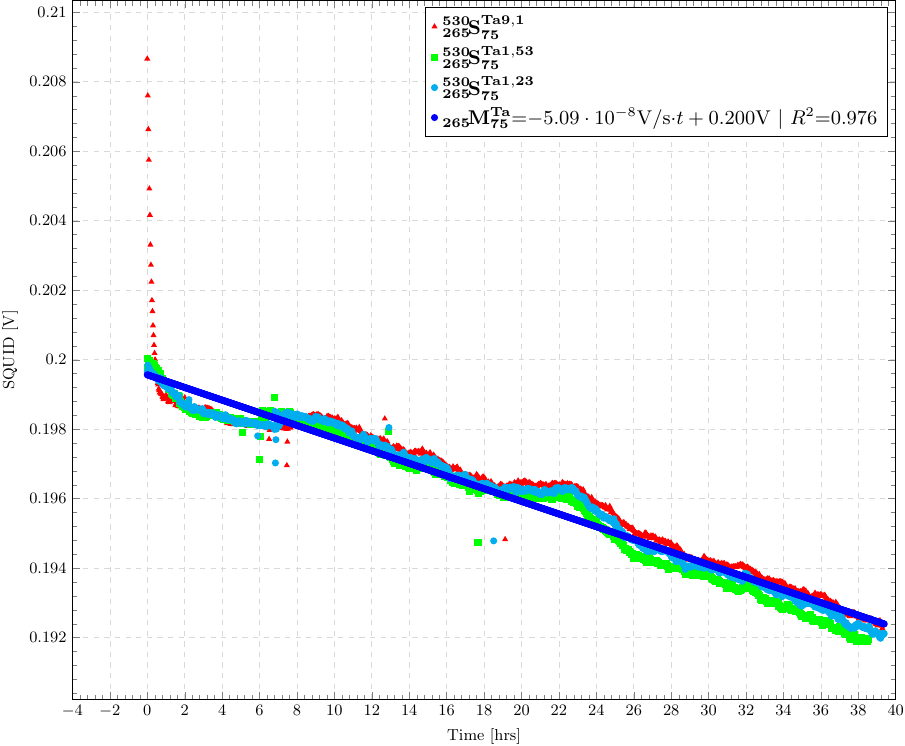}
&
 \includegraphics[width=17.1cm]{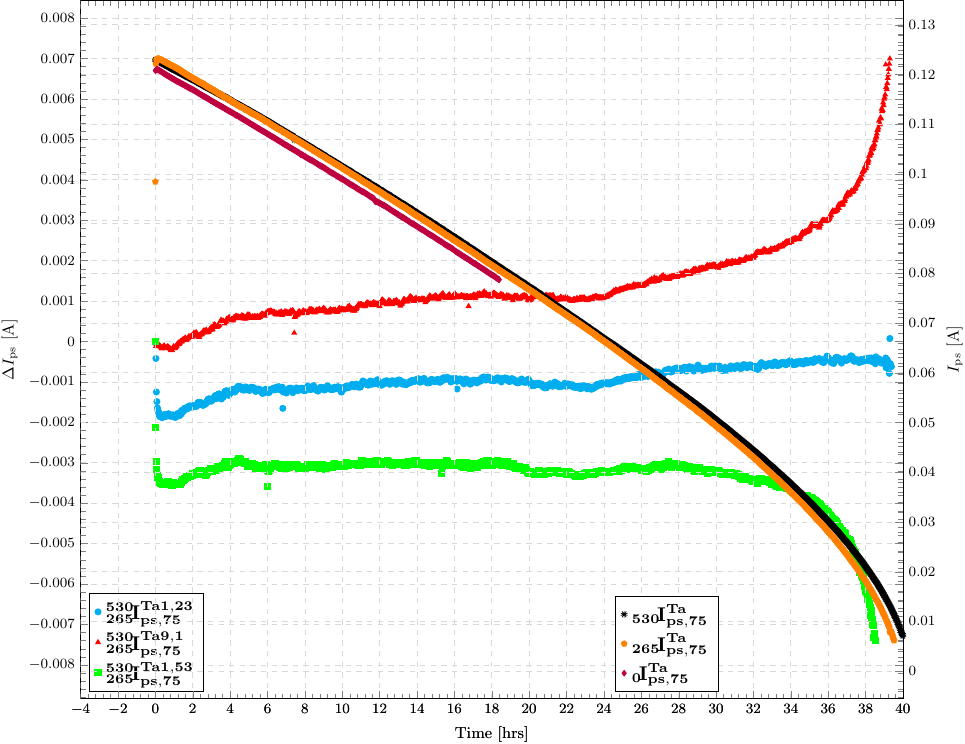}
\\
 \includegraphics[width=16cm]{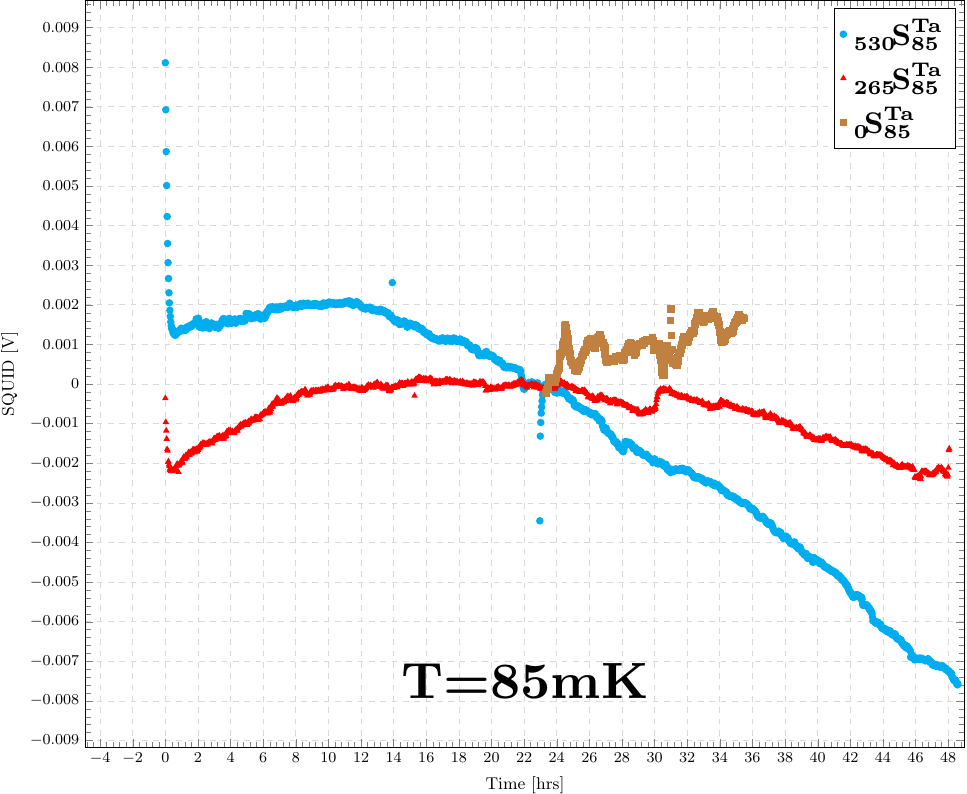}
&
 \includegraphics[width=16cm]{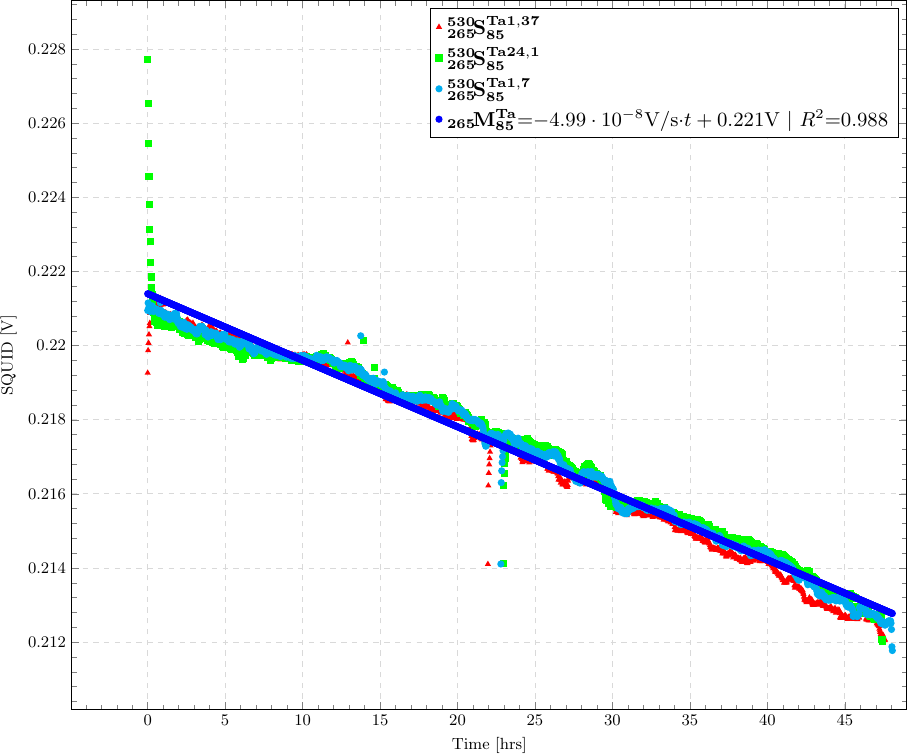}
&
 \includegraphics[width=17.1cm]{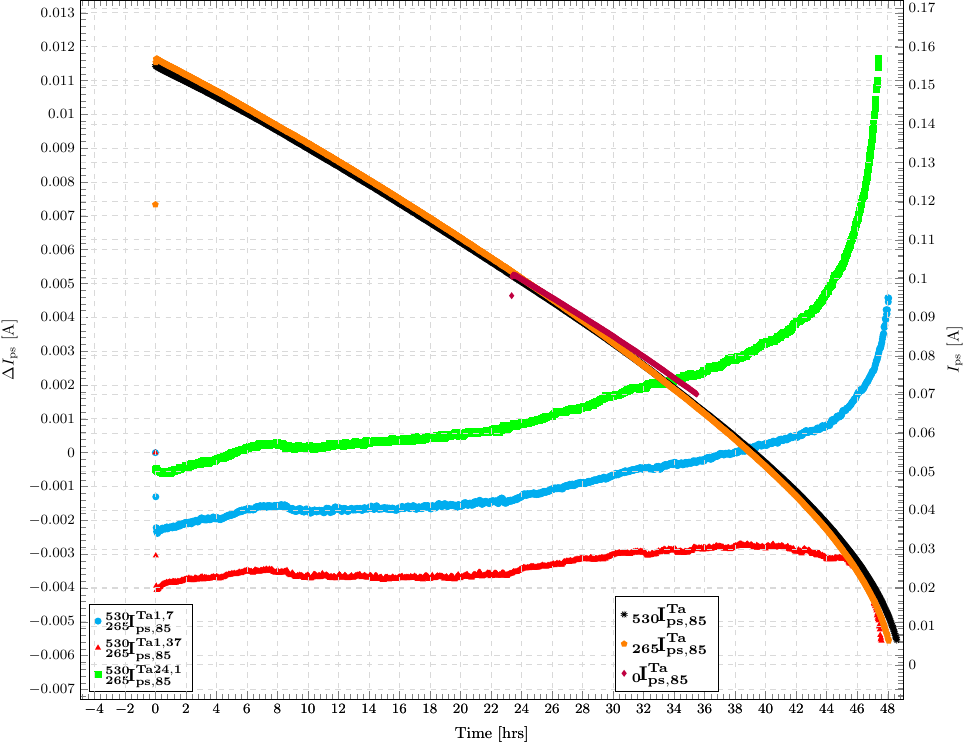}
\\
 \includegraphics[width=16cm]{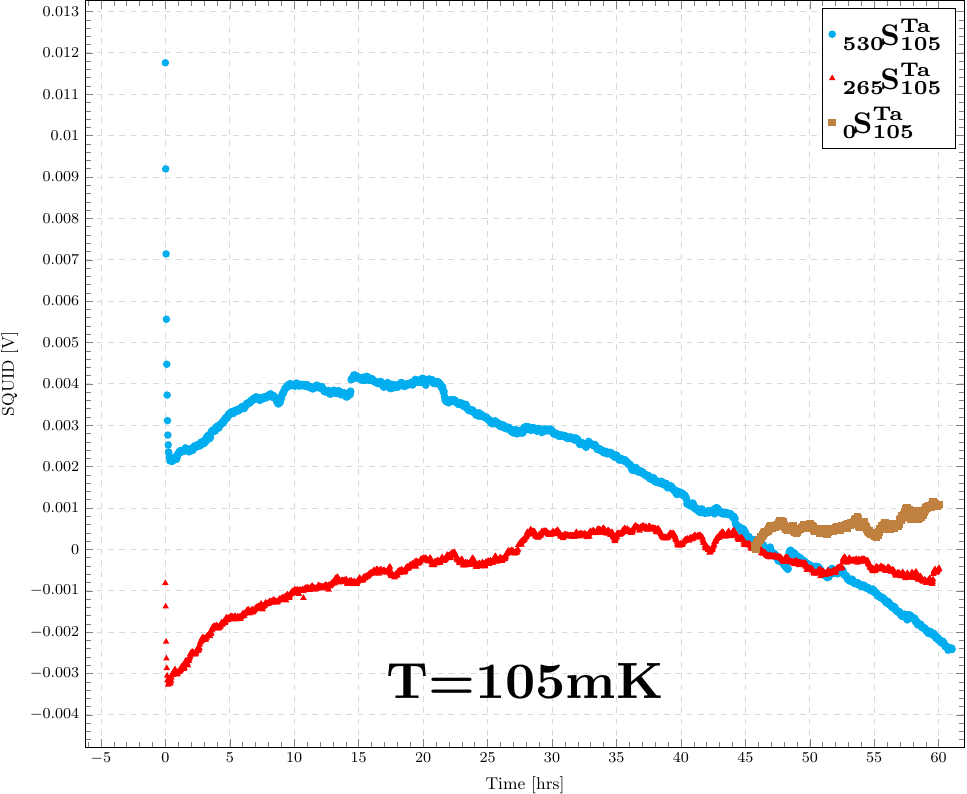}
&
 \includegraphics[width=16cm]{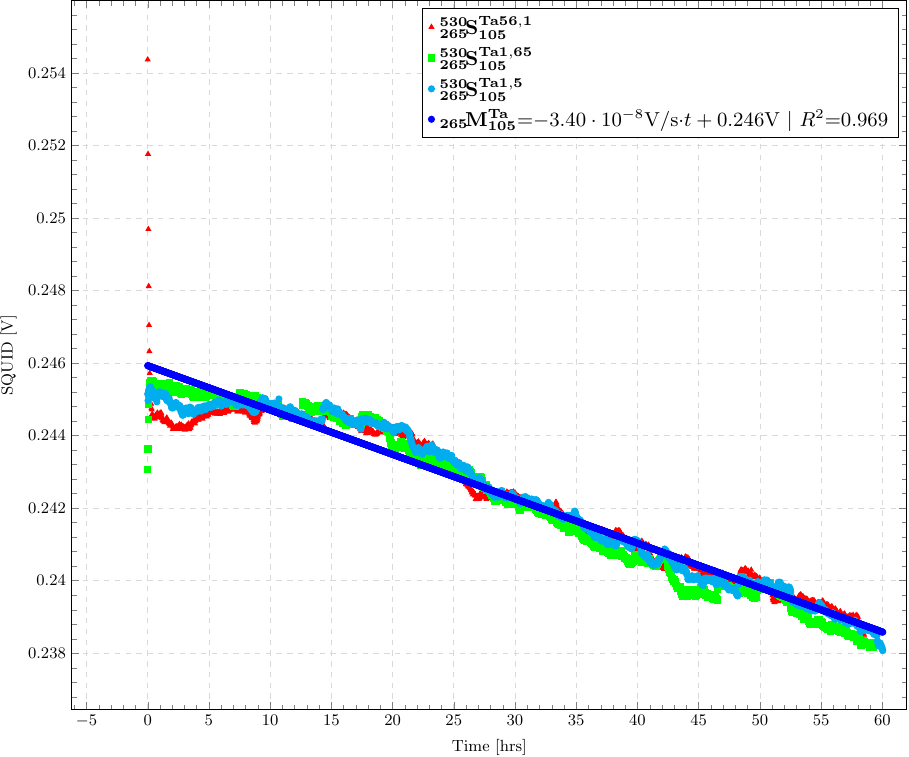}
&
 \includegraphics[width=17.1cm]{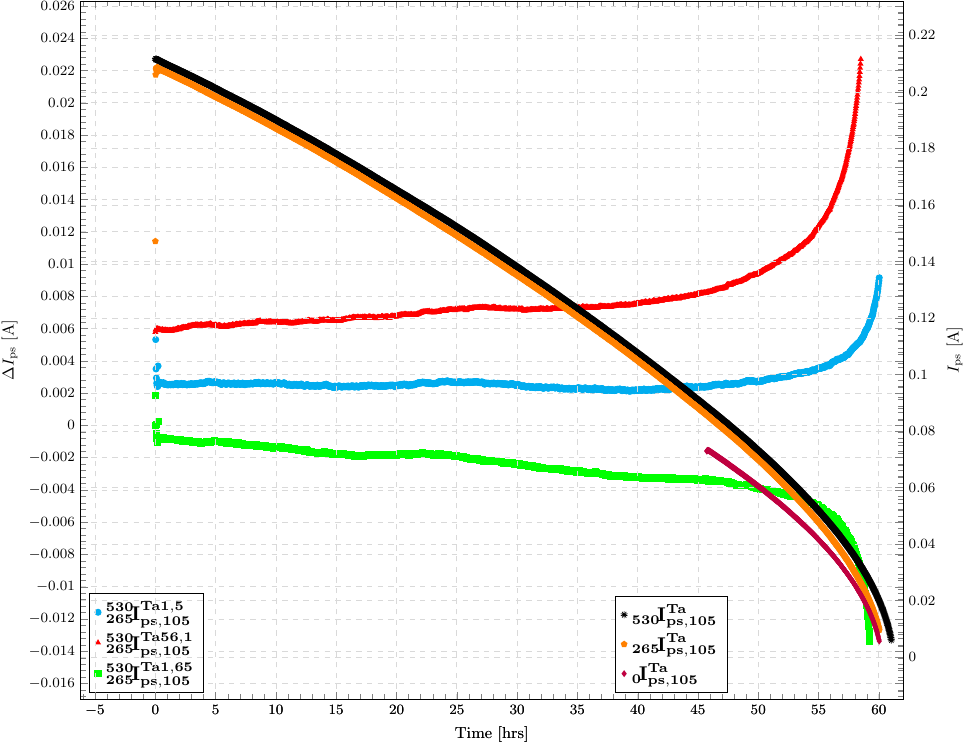}

\end{tabular}
}
%\vspace{-2.5cm}
\caption{Tantalum SQUID and PS I data at fixed temperature and current data. The rows are ordered from top to bottom in increasing temperature of the elements in \Tset{Ta}. The $1^\text{st}$ column shows plots of \stc{530}{T}{Ta}, \stc{265}{T}{Ta}, and \stc{0}{T}{Ta}, the $2^\text{nd}$ column shows plots of \dstc{265}{530}{T}{Ta}{~m,n} at different values of $m,n$, and the $3^\text{rd}$ column shows plots of \ditc{265}{530}{T}{Ta}{~m,n} and \itc{I_a}{52.5}{Ta} \Ia$\in$\Iset{Ta}. The time $t$=0 represents the moment when the data collection began for each dataset, except for the \stc{0}{T}{Ta} with $T$=65,85,105~mK: these datasets are plotted with a starting time found by converting the corresponding starting \ips~ into a time value by using the isomorphism between \itc{I_a}{T}{Ta} and \ttc{I_a}{T}{Ta}.}\label{fixedTempCurr-T}
\end{figure*}
\hspace*{-0.375cm}To implement the above procedure, 16728 data points collected over 465 hours and plotted in \Fig{fixedTempCurr-T}. The datasets in each row were collected at the 5 fixed temperatures $T\in$\Tset{Ta} and 3 fixed current \Ia$\in$\Iset{Ta} starting with the first row at $T$=52.5~mK. Although the first column shows the \Ia=0~mA plot, only \stc{0}{52.5}{Ta} is used in this section while the \stc{0}{T}{Ta} with $T\ne52.5$~mK are not used until \cref{sec:Ia}.  \Fig{fixedTempCurr-T} has three columns: 
\begi
the 1$^\text{st}$column plots \stc{530}{T}{Ta}, \stc{265}{T}{Ta}, and \stc{0}{T}{Ta} as a function of time. The thermal contact between the salt pills and their heat sink during REGEN\footnote{The heat sink during REGEN is  the 3~K stage.}, as well as the thermal contact between the sample assembly and the 50~mK rod are the most important factors in achieving long hold times at a fixed temperature. The quality of the thermal contacts reduces temperature noise and the likelihood of temperature jumps during the data collection (see \cref{app:Tstddev}). These jumps occur during both fixed temperature and zero-field runs. The thermal noise and jumps are likely due to Kapitza thermal resistances~\cite{Kapitza1941} at the boundary between the sample assembly and the 50~mK rod, as well as at the boundaries within the sample assembly itself, and possibly between the pill and the pill suspension.  Two features stand out in the 1$^\text{st}$column:
\up
Each complete dataset has a `tail' at early times:
\up
The tail does not occur in the  0~A dataset at 52.5~mK and therefore appears to require \Ia$\ne$0.
It is hypothesized that the tail is due to a lag between the displayed temperature of the 50~mK stage and the actual temperature of the booster solenoid as noted earlier in our discussion on the critical temperature of the Al-booster core during REGEN.
\down
Each dataset has a maximum that moves to later times as the fixed temperature increases or $\text{\Ia}$ decreases: This is likely due to the interplay between the decreasing temperature-dependent \Is~ and the lag between the displayed temperature of the 50~mK stage and the actual temperature of the booster solenoid.
\down
The 2$^\text{nd}$column plots \dstc{265}{530}{T}{Ta}{~m,n} for various $m,n$ as a function of time. The cyan plot always corresponds to \dstc{265}{530}{T}{Ta}{\bar{m},\bar{n}}; its associated linear regression slope, \gtc{265}{T}{Ta} and $R^2$  reveals that the linear model explains between 96.9\% and 98.8\% of the \dstc{265}{530}{T}{Ta}{\bar{m},\bar{n}} data points at any fixed $T$.  The linear regression slopes for all the \dstc{265}{530}{T}{Ta}{~m,n} datasets are given in \Tab{gammaR2}. The stability of the last $\sim$4~hours of data points of the \dstc{265}{530}{T}{Ta}{~m,n} datasets is striking in light of the large, rapid swings of the corresponding last $\sim$4~hours of data points of \ditc{265}{530}{T}{Ta}{~m,n} in the third column when \ips$\rightarrow0$: it is an indication of the robustness of \dstc{265}{530}{T}{Ta}{~m,n} against variations of the salt pill current \ips.
The 3$^\text{rd}$column plots the \ditc{265}{530}{T}{Ta}{~m,n} corresponding to the \dstc{265}{530}{T}{Ta}{~m,n} of the 2nd column. Also plotted are \itc{530}{T}{Ta}, \itc{265}{T}{Ta}, and \itc{0}{T}{Ta}. It is seen that \itc{530}{T}{Ta} and \itc{265}{T}{Ta} nearly coincide at all times; this observation is  least true at $T$=52.5~mK but even there $\Delta$\ips~ is of the order of 5-8~mA at any given point in time.
\ei
\hspace*{-0.375cm}Each row of \Fig{fixedTempCurr-T} is now discussed in turn:
\begin{itemize}[align=left, leftmargin=*, labelwidth=*, labelsep=0pt]
\item[{\bf52.5~mK:}]\stc{0}{52.5}{Ta} is the only complete dataset collected at \Ia=0~mA. The \dstc{265}{530}{T}{Ta}{\bar{m},\bar{n}} plots all seem to converge at early time to the 100$^\text{th}$ index, and as such the linear regression was performed for \dstc{265}{530}{T}{Ta}{\bar{m},\bar{n}} elements with index 100$\rightarrow N_\text{min}$. The plots of \itc{0}{52.5}{Ta},  \itc{265}{52.5}{Ta}, and \itc{530}{52.5}{Ta} have nearly identical shapes the main difference being that the \ips~ could maintain the fixed temperature of 52.5~mK for different time intervals: 14.7~h, 16.7~h and 15.5~h respectively.
\item[{\bf65~mK:}]\stc{0}{65}{Ta} is not a complete dataset and the starting time of the plot of \stc{0}{65}{Ta}  was obtained by converting  the starting  \ips~ (the first element of \itc{0}{65}{Ta}) to a starting time using  the isomorphism between \itc{530}{65}{Ta} and \ttc{530}{65}{Ta}. \stc{0}{65}{Ta} was obtained after a run at 50~mK and before a run at 75~mK that were never used because longer runs were obtained at 52.5~mK and 75~mK.
\item[{\bf75~mK:}]\stc{0}{75}{Ta} was collected  after a REGEN that followed the end of the \stc{0}{105}{Ta} run; it was stopped after 18.4 hours
\item[{\bf85~mK:}]\stc{0}{85}{Ta} was collected immediately after the 75~mK run, and ran for 12.1~hours after which a REGEN was begun to collect a complete \stc{0}{52.5}{Ta} dataset.
\item[{\bf105~mK:}]\stc{0}{105}{Ta} data was collected after a run at 85~mK that was never used because it only had 6 hours of data, and ran until the \ips$\cong0$~A. It is noted that the \itc{0}{105}{Ta} has a relatively large mismatch with \itc{265}{105}{Ta} and \itc{530}{105}{Ta} for data points subtracted at the same time compared to other temperatures. At 105~mK, the change in $o$ spans 120 meaning that we could compare elements that would have been collected over 3.5 hours apart had they been taken during the same run, and could have differed by up to 37~mA (as seen in the third column) without significantly changing the slope of the \gtc{265}{T}{Ta}{~m,n} as seen in the last column of \Tab{gammaR2}
\end{itemize}
\begin{center}
\begin{table*}[t]
\begin{tabular}{|l|l|l|l|l|l|}
\hline
  $T$(mK) [$\frac{1}{T}$(K$^{-1}$)]     &   52.5  [19.0]   &   65  [15.4]  &   75 [13.3]~K & 85 [11.8] & 105 [9.52]  \\
\hline
$\gamma_\text{max}$                            &  -8.46 (39,1) $100:559$ &  -6.95 (10,1) $50:1096$  & -4.78 (9,1) $15:1413$ &  -4.79 (24,1) 17:1705 &   -3.16 (56,1) 9:2103  \\
\hline
\gbar{265}{T}{Ta}($10^{-8}\text{V/s}$) &  -9.02 (10,1) $100:559$ &  -7.18 (1,20) $1:1077$ & -5.09  (1,23) $1:1416$ &  -4.99 (1,7) 1:1728  &   -3.40 (1,5) 1:2158  \\
\hline
$\gamma_\text{min}$              &  -9.54  (1,21)  $100:539$ &  -7.64  (1,50) $50:1047$ & -5.21 (1,53)  $1:1386$  &  -5.21 (1,37) 17:1711   &   -3.70 (1,65) 9:2130 \\
\hline
$R^2$ &  0.984   & 0.981   & 0.976  &  0.988   &   0.969  \\
\hline
\end{tabular}
\caption{The intersection between a particular slope and a particular temperature/[inverse temperatures] provides the value of the slope, \gbar{265}{T}{Ta}, the (m,n) used for that slope, and the range of indexes, $i:j$ used for the linear regression. The first index $i$ was chosen to exclude data points where the influence of the early time tail is visible. The middle \gbar{265}{T}{Ta}  is obtained from \dstc{265}{530}{T}{Ta}{\bar{m},\bar{n}} and the last row is its corresponding $R^2$. The \gtc{265}{T}{Ta} obtained from \dstc{265}{530}{T}{Ta}{~m,n}  $m\ne\bar{m}$, $n\ne\bar{n}$, namely $\gamma_\text{max}$ and $\gamma_\text{min}$, are used to assess the robustness of the slopes against offsets $o=\pm30$ that correspond to time offsets of 50 minutes for $T\ne105$~mK and $o=\pm60$  for $T=105$~mK. The $p$-value and the standard deviation were  0 and $<0.01$ respectively for all the slopes in this table. }\label{gammaR2}
\end{table*}
\end{center}

At $T=52.5$~mK there are two independent, complete, differenced datasets that can be used in the $1/T$ analysis, \dstc{265}{530}{T}{Ta}{\bar{m},\bar{n}} being one of them; the other two possibilities are \dstc{0}{530}{52.5}{Ta}{\bar{m},\bar{n}} and \dstc{0}{265}{52.5}{Ta}{\bar{m},\bar{n}}. \dstc{0}{530}{52.5}{Ta}{\bar{m},\bar{n}} was chosen because it has a larger signal to noise ratio than \dstc{0}{265}{52.5}{Ta}{\bar{m},\bar{n}} and the mean amplitude of the remaining noise in \dstc{0}{530}{52.5}{Ta}{\bar{m},\bar{n}} will be further halved as seen shortly.
The \dstc{0}{530}{52.5}{Ta}{\bar{m},\bar{n}} does not have a tail so $\bar{m},$ and $\bar{n}$ cannot be determined by minimizing the tail. Instead, three cases were considered: $m=1,n=1$, $m=30,n=1$, and $m=1,n=30$. The corresponding slopes and $R^2$ were
\begi
$m=1,n=1:~ \text{\gtc{530}{52.5}{Ta}}=-2.037\cdot10^{-7}\text{ V/s with } R^2=0.992$
$m=30,n=1:~ \text{\gtc{530}{52.5}{Ta}}=-2.015\cdot10^{-7}\text{ V/s with } R^2=0.991$
$m=1,n=30:~ \text{\gtc{530}{52.5}{Ta}}=-2.058\cdot10^{-7}\text{ V/s with } R^2=0.993$
\ei
The slopes and $R^2$ are all extremely similar with a mean \gbar{530}{52.5}{Ta}=$-2.03\cdot10^{-7}$~V/s. In order to include that slope in the \gbar{265}{T}{Ta} analysis consistent with the underlying hypothesis, it is divided by the ratio of currents, namely 530/265=2 with the result $-1.01\cdot10^{-7}$~V/s; this division by two halves the magnitude of the remaining noise in the differenced data. This division by the ratio of currents is justified in \cref{sec:Ia} where it is shown that \gbar{265}{T}{Ta} is proportional to \Isol.

\begin{figure}
\centering
\resizebox{8 cm}{!}{
  \includegraphics*{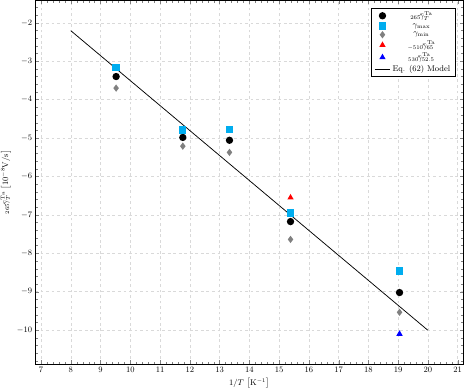}
}
\caption{\gbar{265}{T}{Ta} data slopes plotted against the inverse temperature and the corresponding linear regression model \eq{gammabar}. Also plotted are the $\gamma_\text{max}$ and $\gamma_\text{min}$ slopes systematically above and below the \gbar{265}{T}{Ta} data points respectively. The blue triangle corresponds to  \gbar{530}{52.5}{Ta}/2, while the red triangle corresponds to the data point from the reverse field with \Ia=$-510$~mA at $T$=65~mK, see \cref{sec:reverse}.}\label{gammaTTaMod}
\end{figure}
In \Fig{gammaTTaMod}, the five \gbar{265}{T}{Ta} are scatter-plotted against $1/T$. The blue squares correspond to the $\gamma_\text{max}$ slopes of the \dstc{265}{530}{T}{Ta}{~m,n} with $m=\bar{m}+30$, in other words, the subtracted data was shifted forward in time 3000~s compared to $_{265}\!\bar{\gamma}_{T}^{\text{Ta}}$. All of these points are shifted up. In contrast, the gray diamonds correspond to the $\gamma_\text{min}$ slopes where the subtracted data was shifted 3000~s backward in time compared to $_{265}\!\bar{\gamma}_{T}^{\text{Ta}}$. All of these points are shifted down. It is seen that shifting the subtracted data in time  results in a systematic upward or downward motion of the data points in a way that is not random. The 75~mK data point stands out somewhat, especially when compared to the model \Eq{gammabar} discussed below. This may be due to a slightly greater residue of systematics at early times in \dstc{265}{530}{75}{Ta}{1,23}. The linear regression $p$-value\footnote{This is the $p$-value for the null hypothesis that there is no $1/T$ dependence.} and standard deviation are the smallest for the $_{265}\!\bar{\gamma}^{\text{Ta}}$ slope, consistent with our criterion of choosing the $(\bar{m},\bar{n})$ that minimize early time systematics.

The linear regression model of these \gbar{265}{T}{Ta} including the \gbar{530}{52.5}{Ta} and reverse field data points in \Fig{gammaTTaMod} is
\begin{align}\label{gammabar}
&\text{\gtc{265}{T}{Ta}}\!=\!\frac{_{265}\!\bar{\gamma}^{\text{Ta}}}{T}+2.99\cdot10^{-8}~\frac{\text{V}}{\text{s}} ~|~R^2=0.960\\ \label{gammabarM0}
& _{265}\!\bar{\gamma}^{\text{Ta}}\!=\!(-5.95\pm0.55)\cdot10^{-9}~\frac{\text{V}\cdot\text{K}}{\text{s}},~p=0.0017\\ \label{gammabarM}
&\text{\gbar{265}{}{Ta}}\!=\!(-6.50\pm0.59)\cdot10^{-9}\frac{\text{V}\cdot\text{K}}{\text{s}}, ~p=0.00011 \\
&\text{\gbar{265}{}{Ta}}\!=\!(-5.95\pm5.63)\cdot10^{-9}~\frac{\text{V}\cdot\text{K}}{\text{s}}\ne0,~99.8\text{\%}~\text{CL}\\
&\text{\gbar{265}{}{Ta}}\!=\!(-6.50\pm6.09)\cdot10^{-9}\frac{\text{V}\cdot\text{K}}{\text{s}} \ne0,~99.985\text{\%}~\text{CL},
\end{align}
where \Eq{gammabarM0} is the \gbar{265}{}{Ta} obtained when the \dstc{0}{530}{52.5}{Ta}{\bar{m},\bar{n}} and \dstc{-510}{530}{65}{Ta}{\bar{m},\bar{n}} slopes are excluded while \Eq{gammabarM} is the \gbar{265}{}{Ta} when they are included; the $p$-value decreases by a factor of 15 when these two points are included. \Eq{gammabarM} is the value used to calculate the nuclear EDM of tantalum.  The $R^2$ is consistent with \gbar{265}{T}{Ta} being inversely proportional to the temperature and it is consistent with ${\cal{E}}_L\ne0$ in \Eq{betaSlope} at the 99.985\% confidence level. Before concluding that \gbar{265}{}{Ta} is the parameter ${\cal{E}}_L$ in \Eq{betaSlope} at a fixed \Ia, two other conditions must be satisfied:
\begin{enumerate}
\item \gbar{\text{\Ia}}{T}{Ta} at fixed $T$ must be shown to be linearly proportional to \Isol. This proportionality is visually suggested in the plots of the first column of \Fig{fixedTempCurr-T} where the \stc{0}{T}{Ta} plot is above the  \stc{265}{T}{Ta} plot which is itself above the \stc{530}{T}{Ta} plot at all temperatures $T$.
\item The Pb and tantalum zero-field data points should be correlated since they were collected with the same equipment and solenoids. When the contribution from the electrization field is subtracted from the tantalum data using \gbar{265}{}{Ta} inserted in \Eq{intdidt}, that correlation should improve, since $\Fedm{ele}$ is much weaker in Pb.
\end{enumerate} 
\subsubsection{Ta: dependence of the sample current on the applied current}\label{sec:Ia}
\begin{center}
\begin{table*}[t]
\begin{tabular}{|l|l|l|l|l|}
\hline
  $T$(mK)     &   65    &   75  & 85  & 105  \\
\hline
\itc{0}{T}{Ta} (A) [$m$]  & 0.0677 [1] & 0.121 [4]  & 0.101 [6] &      0.0731 [4] \\
\hline
\itc{530}{T}{Ta} (A) [$i$]   & 0.0677 [396,426,366] & 0.119 [69,99,129] & 0.101 [840,870,810] & 0.0881 [1648,1618,1588] \\
\hline
\itc{265}{T}{Ta} (A) [$j$] & 0.0687 [377,407,347] & 0.121 [47,77,107]   & 0.102 [834,864,804] & 0.0857 [1644,1614,1584]\\
\hline
\end{tabular}
\caption{At the intersection of the temperatures columns and the \itc{0}{T}{Ta} row is the value of the largest \ips~ at that temperature and the corresponding \itc{0}{T}{Ta} element index. For example, at 65~mK, the largest \ips~ is 0.0677~A which is the 1st element of \itc{0}{T}{Ta}. The second row contains the closest \ips~ in\itc{530}{T}{Ta} and the corresponding element index $i$ at each $T$ with $[i-30,i,i+30]$. The $i\pm30$ are to test robustness of the results against offsets, exactly as was done in \cref{sec:1overT}. The third row uses $j=\bar{m}-\bar{n}+i$ to find the corresponding elements in \itc{265}{T}{Ta} and the value of the $j$th element of \itc{265}{T}{Ta} is shown.}\label{ipsVal}
\end{table*}
\end{center}
Since \Isol=$M_\text{bm}$\Ia, the dependence of \Is~ on $\text{\Ia}$ will be  used as a proxy for the dependence of \Is~ on \Isol. In this section, the dependence of \Is~ on $\text{\Ia}$ is determined by subtracting \stc{0}{T}{Ta} from both \stc{265}{T}{Ta} and  \stc{530}{T}{Ta}. Since $|$\stc{0}{T}{Ta}$|$$<$\stc{265}{T}{Ta}, $|$\stc{0}{T}{Ta}$|$$<$\stc{530}{T}{Ta} $\forall T\in$\Tset, the following procedure will be used to select the subsets \dsstc{I_a}{}{T}{Ta}{}$\in$\stc{I_a}{T}{Ta} where $|$\dsstc{I_a}{}{T}{Ta}{}$|$=$|$\stc{0}{T}{Ta}$|$ and \Ia=265~mA, 530~mA
\begin{itemize}
\item To calculate \dstc{0}{265}{T}{Ta}{~m,j} and \dstc{0}{530}{T}{Ta}{~m,i}, consistency with the results from \cref{sec:1overT} imposes a constraint on $i$ and $j$:  the linear regression slopes $\gamma^{5300}_T$ and $\gamma^{2650}_T$ calculated from \dstc{0}{530}{T}{Ta}{~m,i} and  \dstc{0}{265}{T}{Ta}{~m,j} respectively should approximately satisfy
\beq\label{consist}
\gamma^{5300}_T-\gamma^{2650}_T\cong\text{\gbar{265}{T}{Ta}}
\eeq
Since \gbar{265}{T}{Ta} was calculated from \dstc{265}{530}{T}{Ta}{~\bar{m},\bar{n}}, this translates into the relation 
\beq\label{jimn}
j-i=\bar{m}-\bar{n}
\eeq
\item \stc{0}{52.5}{Ta} is the only complete dataset collected with \Ia=0~mA and that case is treated differently from the other incomplete cases \stc{0}{T\ne52.5}{Ta} where 
\item At $T$=52.5~mK 
\begin{itemize}
\item Three $m,i$ are chosen
\begin{itemize}
\item \dstc{0}{530}{T}{Ta}{~1,1}, where \stc{0}{52.5}{Ta} is subtracted from the first 530 data points of \stc{530}{52.5}{Ta}
\item \dstc{0}{530}{T}{Ta}{~1,30}, where \stc{0}{52.5}{Ta} is subtracted from the last 530 data points of \stc{530}{52.5}{Ta} 
\item \dstc{0}{530}{T}{Ta}{~30,1}, where the last 500 points of \stc{0}{52.5}{Ta} are subtracted from the first 500 data points of \stc{530}{52.5}{Ta}
\item to determine the three $j$'s for \Ia=265, use \Eq{jimn} leading to the differenced datasets: \dstc{0}{265}{T}{Ta}{~1,10}, \dstc{0}{265}{T}{Ta}{~1,39},  \dstc{0}{265}{T}{Ta}{~30,10}
\end{itemize}
\item Three slopes at \Ia=530 are obtained from  \dstc{0}{530}{52.5}{Ta}{m,i} and their mean is denoted $\gamma^{5300}_{52.5}$.
\item Three slopes at \Ia=265 are obtained from   \dstc{0}{265}{52.5}{Ta}{m,j} and their mean is denoted $\gamma^{2650}_{52.5}$.
\item Verify the consistency requirement \Eq{consist}.
\item Calculate the ratio of the average values, $\gamma^{5300}_{52.5}/\gamma^{2650}_{52.5}$.
\end{itemize}
\item At $T\neq$52.5~mK the $i,j$ of \Tab{ipsVal} are used. $i$ corresponds to the index of the $I_{\text{ps},i}\in\text{\itc{530}{T}{Ta}}$ that is closest to the first element in \itc{0}{T}{Ta} (which is also its largest element); $j$ can then be found with \Eq{jimn}. All the elements $\in$\stc{0}{T}{Ta} are used in the differenced dataset \dstc{0}{I_a}{T}{Ta}{m,n}. The largest $\Delta$\ips~ is 13~mA at $T=105$ which is less than the $\Delta$\ips=23~mA observed at late times in  \ditc{265}{530}{105}{Ta}{56,1} in \Fig{fixedTempCurr-T} that had no visible impact on \dstc{265}{530}{105}{Ta}{56,1}.
\end{itemize}

\begin{figure*}
\centering
\resizebox{11 cm}{!}{
\begin{tabular}{cc}
  \includegraphics*{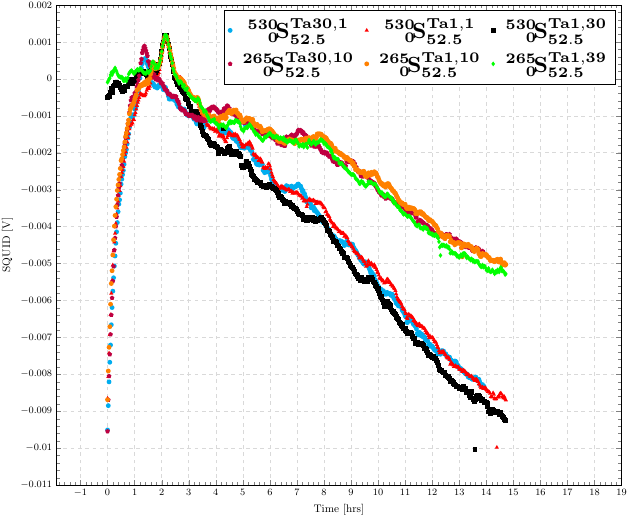}
&
  \includegraphics*{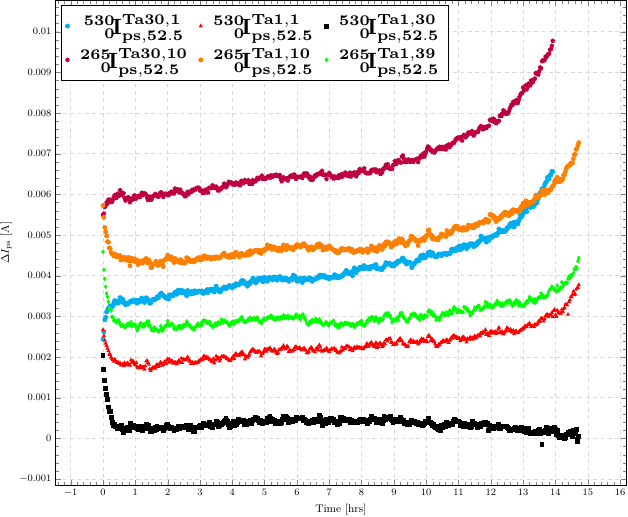}
 \\
  \includegraphics*{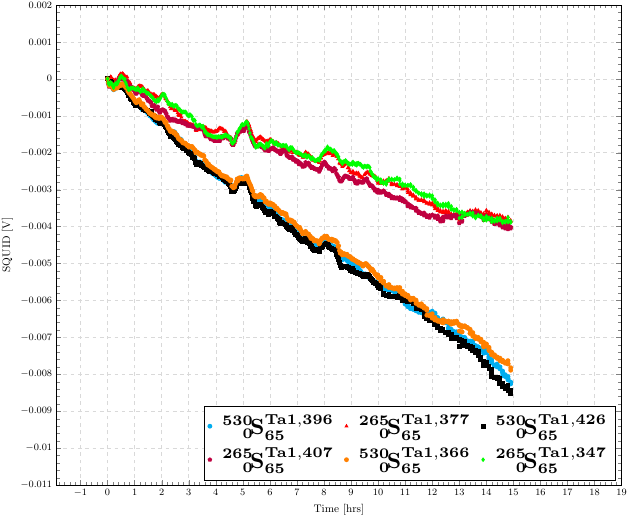}
&
  \includegraphics*{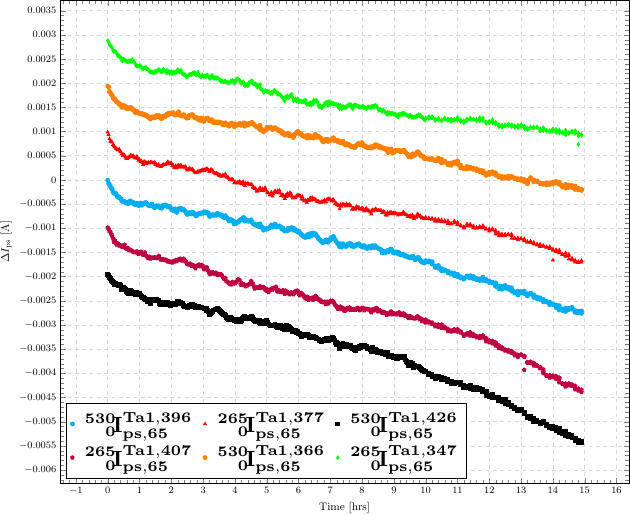}
\\
  \includegraphics*{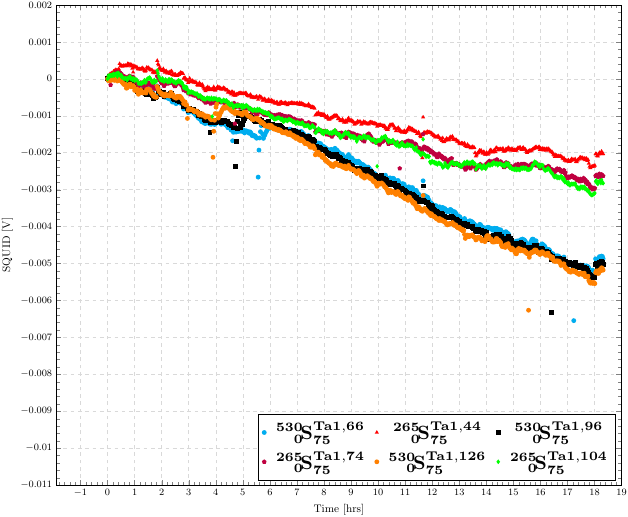}
&
  \includegraphics*{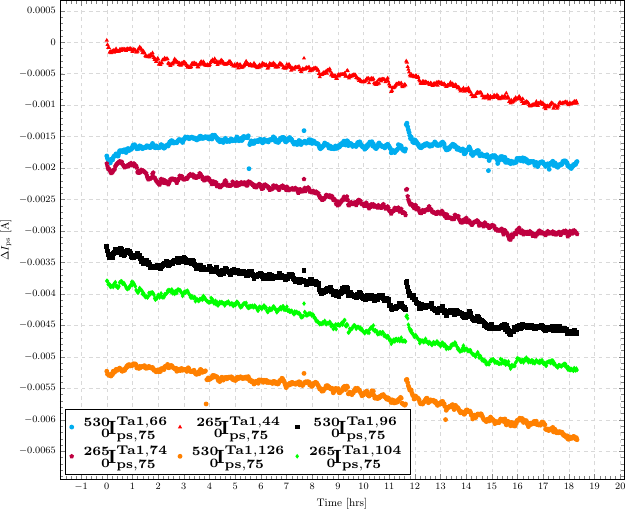}
\\
  \includegraphics*{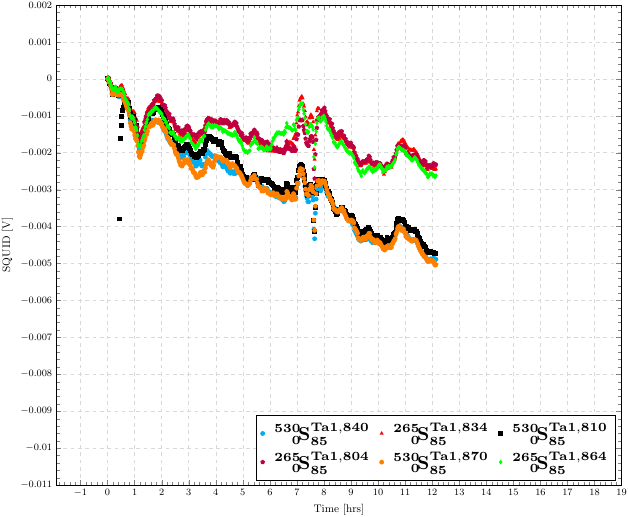}
&
  \includegraphics*{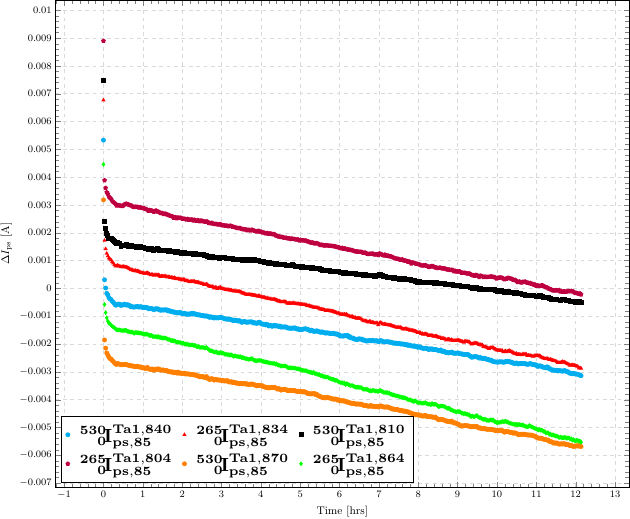}
\\
  \includegraphics*{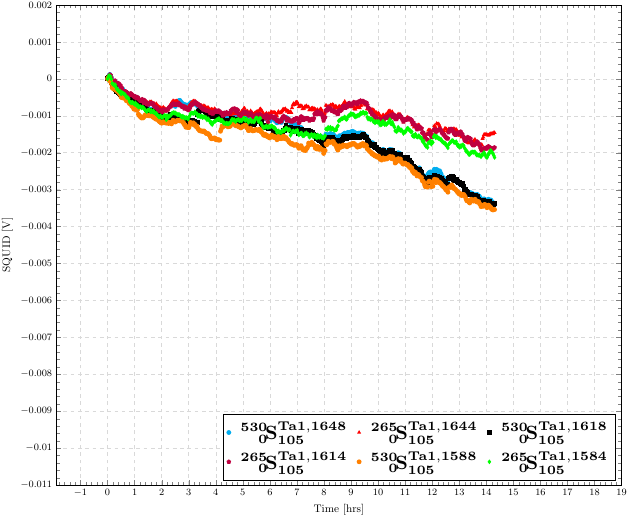}
&
  \includegraphics*{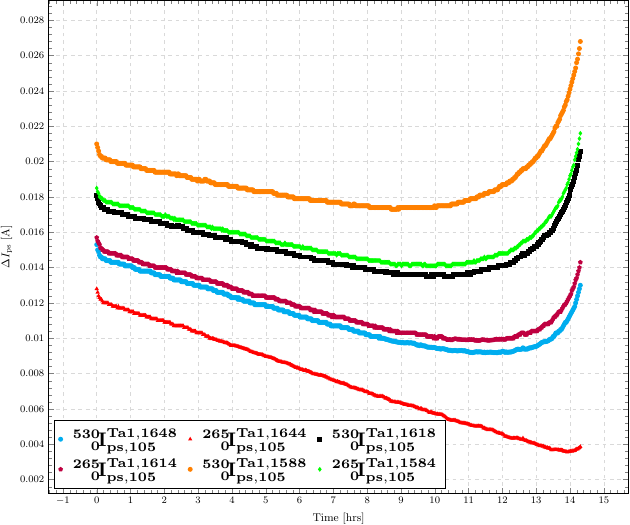}
\end{tabular}
}
%\vspace{-2.5cm}
\caption{Tantalum SQUID and PS I data showing the proportionality of the data with the applied current. SQUID data points (\dstc{0}{530}{75}{Ta}{1,66})$_{200}$,  (\dstc{0}{530}{75}{Ta}{1,96})$_{170}$,  (\dstc{0}{530}{75}{Ta}{1,126})$_{140}$ were removed to give a more  limited range to the $T$=75~mK plots. The SQUID axes all have the same minimum and maximum values and time range for ease of visual comparison between different temperature plots.}\label{SpropIa}
\end{figure*}
In the first column of \Fig{SpropIa}, the range of the axes is the same at all temperatures in order to easily see the increase of the slopes of \dstc{0}{530}{T}{Ta}{m,i} and \dstc{0}{265}{T}{Ta}{m,j} as $T$ increases. This is consistent with the inverse temperature dependence concluded at the end of the last subsection, although the plots in \Fig{SpropIa} are noisier because \dstc{0}{530}{T}{Ta}{m,i} and \dstc{0}{265}{T}{Ta}{m,j} have significantly fewer elements than \dstc{265}{530}{T}{Ta}{m,n} especially at $T>65$~mK.
\begin{center}
\begin{table*}[t]
\begin{tabular}{|l|l|l|l|l|l|l|l|l|l|l|}
\hline
  & $\gamma_{52.5}$  & $\sigma_{52.5}$  & $\gamma_{65}$ & $\sigma_{65}$  & $\gamma_{75}$  &  $\sigma_{75}$  & $\gamma_{85}$  &  $\sigma_{85}$  & $\gamma_{105}$   & $\sigma_{105}$  \\
\hline
530~mA&-2.04E-07&8.58E-10&-1.42E-07&3.92E-10&-8.92E-08&4.36E-10&-9.10E-08&1.32E-09&-5.05E-08&6.90E-10\\
\hline
265~mA&-1.11E-07&1.06E-09&-7.11E-08&4.71E-10&-4.85E-08&2.89E-10&-3.55E-08&1.46E-09&-2.35E-08&6.73E-10\\
\hline
530~mA&-2.06E-07&8.43E-10&-1.49E-07&4.63E-10&-8.54E-08&4,27E-10&-9.22E-08&1.26E-09&-5.11E-08&6.98E-10\\
\hline
265~mA&-1.17E-07&1.06E-09&-7.43E-08&4.93E-10&-4.55E-08&2.81E-10&-3.65E-08&1.24E-09&-2.17E-08&6.41E-10\\
\hline
530~mA&-2.02E-07&9.7E-10&-1.45E-07&4.39E-10&-8.17E-08&4.39E-10&-9.32E-08&1.13E-09&-5.30E-08&7.39E-10\\
\hline
265~mA&-1.07E-07&1.16E-09&-7.32E-08&4.84E-10&-4.23E-08&2.66E-10&-3.58E-08&1.45E-09&-1.99E-08&5.88E-10\\
\hline
\end{tabular}
\caption{Table of slopes of all the plots in \Fig{SpropIa} and their standard deviations.   }\label{ratioTab}
\end{table*}
\end{center}
\begin{center}
\begin{table*}[t]
\begin{tabular}{|l|l|l|l|l|l|l|l|l|l|l|}
\hline
  & $r_{\gamma_{52.5}}$  & $\sigma_{52.5}$  & $r_{\gamma_{65}}$ & $\sigma_{65}$  & $r_{\gamma_{75}}$  &  $\sigma_{75}$  & $r_{\gamma_{85}}$  &  $\sigma_{85}$  & $r_{\gamma_{105}}$   & $\sigma_{105}$  \\
\hline
ratio 1& 1.84 & 1.92E-02 & 1.99 & 1.43E-02 & 1.84 & 1.42E-02 & 2.56 & 1.11E-01 & 2.15 & 6.81E-02 \\
\hline
ratio 2& 1.76 & 1.76E-02 & 2.00 & 1.47E-02 & 1.88 & 1.49E-02 & 2.53 & 9.27E-02 & 2.35 & 7.64E-02 \\ 
\hline
ratio 3& 1.89 & 2.26E-02 & 1.98 & 1.44E-02 & 1.93 & 1.60E-02 & 2.60 & 1.10E-01 & 2.66 & 8.67E-02 \\
\hline
mean &1.83&  &1.99&  &1.88&  &2.56&  &2.38 & \\
\hline
range & 0.05 &  & 0.02 &  & 0.09 &  &0.07 &  &0.51 & \\
\hline
\end{tabular}
\caption{The first three rows provide the three possible slope ratios at each $T\in$\Tset{Ta} in \Fig{SpropIa}.  The fourth row is the mean ratio at each $T$ while the last row displays the range of the three slopes. It is noted that the ratios at $T\ne105$~mK are reasonably robust against the different offsets with a range$<$2\% of the mean. }\label{slopeRTab}
\end{table*}
\end{center}
\begin{center}
\begin{table*}[t]
\begin{tabular}{|l|l|l|l|l|l|}
\hline
  & 52.5~mK   & 65  & 75& 85  & 105  \\
\hline
$1^\text{st}~\gamma^{5300,~i}_T-\gamma^{2650,~j}_T$ & -9.28E-08 & -7.05E-08 & -4.07E-08 & -5.55E-08 & -2.70E-08 \\
\hline
$2^\text{nd}~\gamma^{5300,~i}_T-\gamma^{2650,~j}_T$ & -8.89E-08 & -7.44E-08 & -3.98E-08 & -5.57E-08 & -2.93E-08 \\
\hline
$3^\text{rd}~\gamma^{5300,~i}_T-\gamma^{2650,~j}_T$ & -9.50E-08 & -7.16E-08 & -3.94E-08 & -5.74E-08 & -3.30E-08 \\
\hline
mean[$\gamma^{5300}_T-\gamma^{2650}_T$] & -9.22E-08 & -7.21E-08 & -4.00E-08 & -5.62E-08 & -2.98E-08 \\
\hline
\end{tabular}
\caption{Table of the three consistency check of \Eq{consist}  at all $T\in$\Tset{Ta} and their average in the last row.   }\label{consistTab}
\end{table*}
\end{center}
In order to establish the dependence of \gtc{\text{\Ia}}{T}{Ta} on \Ia, $\text{\gtc{\text{530}}{T}{Ta}}^{m,i}$ and $\text{\gtc{\text{265}}{T}{Ta}}^{m,j}$ for \dstc{0}{265}{T}{Ta}{m,j} and \dstc{0}{530}{T}{Ta}{m,j} are calculated where $i$ and $j$ are related by \Eq{jimn}. The ratios
\beq
\rho^{\text{Ta},m,i,j}_T=\frac{\text{\gtc{\text{530}}{T}{Ta}}^{m,i}}{\text{\gtc{\text{265}}{T}{Ta}}^{m,j}}~\forall T\in\text{\Tset{Ta}}~\text{at the various $i,j$}.
\eeq
are next calculated. A weighted mean of the ratios will then provide an overall ratio $\rho^\text{Ta}$ that, if \gtc{\text{\Ia}}{T}{Ta} is proportional to \Ia, should yield $\rho^{\text{Ta}}\cong2$.

The  slopes were calculated via linear regression for all \dstc{0}{530}{T}{Ta}{m,i} and \dstc{0}{265}{T}{Ta}{m,j} datasets (a total of 30 datasets) and are listed in \Tab{ratioTab}. The corresponding 15 ratios, with standard deviations, are given in \Tab{slopeRTab}. The ratios are all between 1.76 and 2.66; Note that the largest standard deviations are associated with the noisiest data at $T$=85~mK,105~mK while the smallest standard deviation is at $T$=65~mK. The ranges on the last row show a similar pattern: the largest range is at $T$=105~mK while the smallest is at  $T$=65~mK.

The possibility of a linear relationship between the slope ratios and the temperature is considered. The linear regression of the ratios on the temperatures yields a $R^2$=0.60 and a $P$-value of $0.124>0.05$ leading to the acceptance of the null hypothesis of no linear relationship of the ratios with temperatures. Hence, an average of the slope ratios is considered. 

The overall mean ratio $\rho^\text{Ta}$ can be calculated with inverse-variance weighting. Since the linear regression does not account for the dependence of \dstc{0}{530}{T}{Ta}{~m,i} and \dstc{0}{265}{T}{Ta}{~m,j} on $i$ and $j$, three results for the overall mean ratio can be compared
\begin{enumerate}
\item the inverse-variance weights equal the square of the inverse standard deviations in \Tab{slopeRTab}
\begin{equation}\label{eqratioa}
\rho^\text{Ta}=1.92
\end{equation}
\item the inverse-variance weights equal the square of the inverse ranges given in the last row of \Tab{slopeRTab}
\begin{equation}\label{eqratiob}
\rho^\text{Ta}=2.00
\end{equation}
\item The straight average without weights gives the result
\begin{equation}\label{eqratioc}
\rho^\text{Ta}=2.13
\end{equation}
\end{enumerate}
These three results are consistent with the hypothesis that \gtc{\text{\Ia}}{T}{Ta}$\propto$$\text{\Ia}$  and their difference can be used as an uncertainty estimate for the overall ratio. To test the significance of this result, we  use as the null hypothesis that there is no dependence on $\text{\Ia}$ and that $\rho^\text{Ta}=1$ as opposed to the alternative hypothesis $\rho^\text{Ta}=2$. The F-statistic=13.4 yielding a $P$-value=0.014. If the reverse field slope ratio is included, the $P$-value falls to 0.006 which allows us to reject the null hypothesis at the 99\% confidence level.

It remains to verify the consistency of our slopes with \Eq{consist}. The $\gamma^{5300,~i}_T-\gamma^{2650,~j}_T$ are  provided in \Tab{consistTab} for all $T$ where $i,j$ are related by \Eq{jimn} and the last row gives the mean at each temperature. Comparing the results in the last row of \Tab{consistTab} with \gbar{265}{T}{Ta} in \Tab{gammaR2}, it is seen that they are in excellent agreement with \Eq{consist} at $T=52.5$~mK and 65~mK. At $T=85$~mK and 105~mK there is reasonable agreement with \Eq{consist} considering the noisiness  and short collection times of the \stc{0}{T}{Ta} data points at these temperatures. The worst agreement is for $T$=75~mK. This is related to our earlier observation in \Fig{gammaTTaMod} that there may be a greater residue of systematics at early times in \dstc{265}{530}{75}{Ta}{1,23}. This residue is amplified in \dstc{0}{530}{75}{Ta}{m,i} and \dstc{0}{530}{75}{Ta}{m,j} because a large fraction of \stc{0}{75}{Ta} is taken at the earliest times as seen in the middle row/first column of \Fig{fixedTempCurr-T}.

\begin{figure*}
\centering
\resizebox{17 cm}{!}{
\begin{tabular}{ccc}
 \includegraphics*{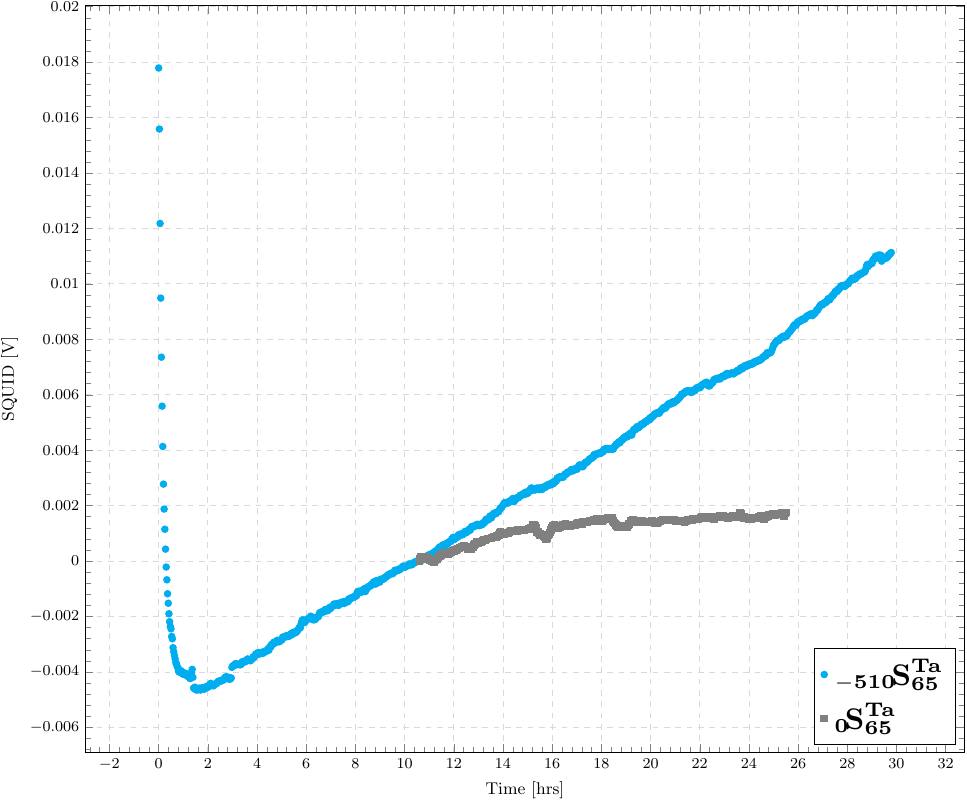}
&
  \includegraphics*{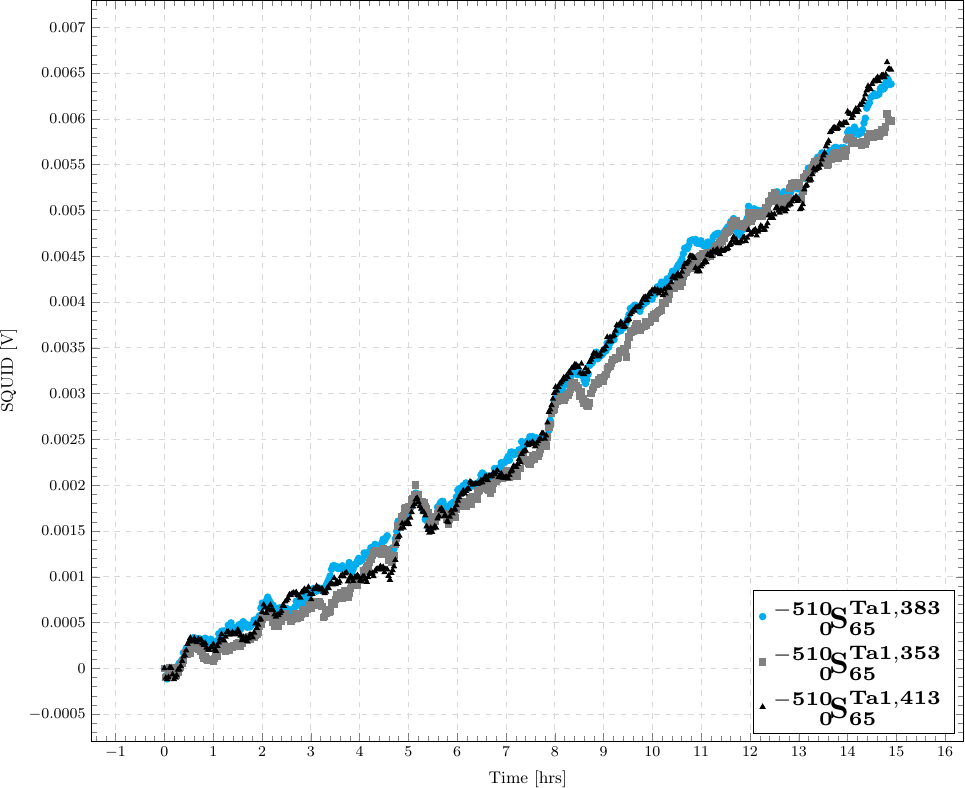}
&
  \includegraphics*{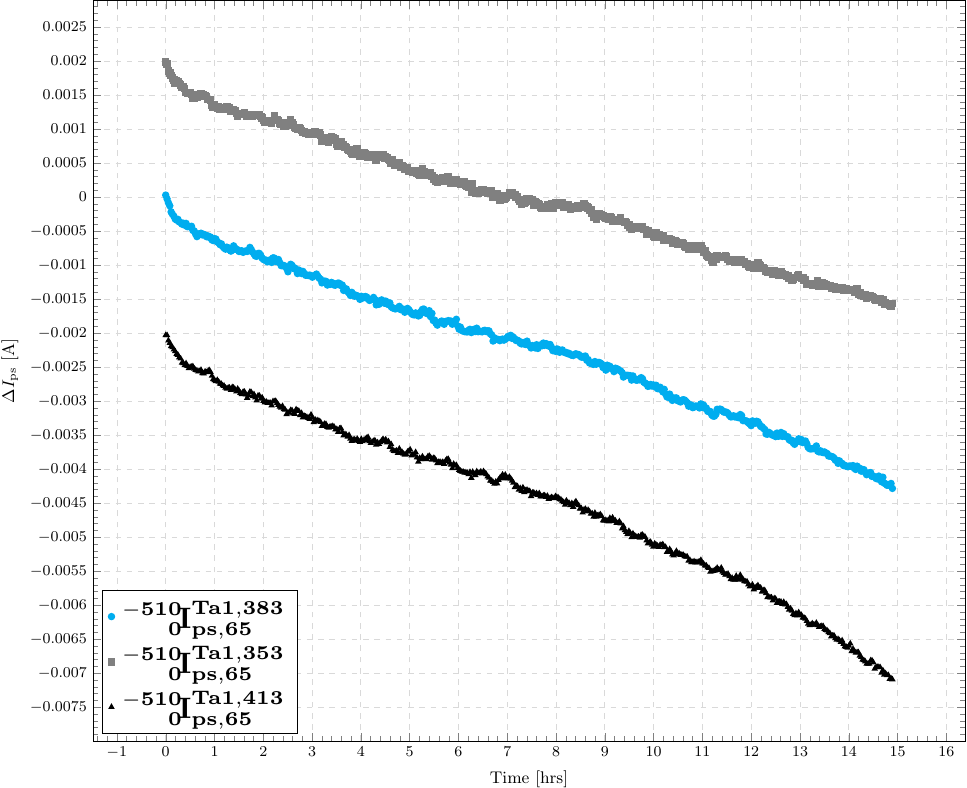}
\\
  \includegraphics*{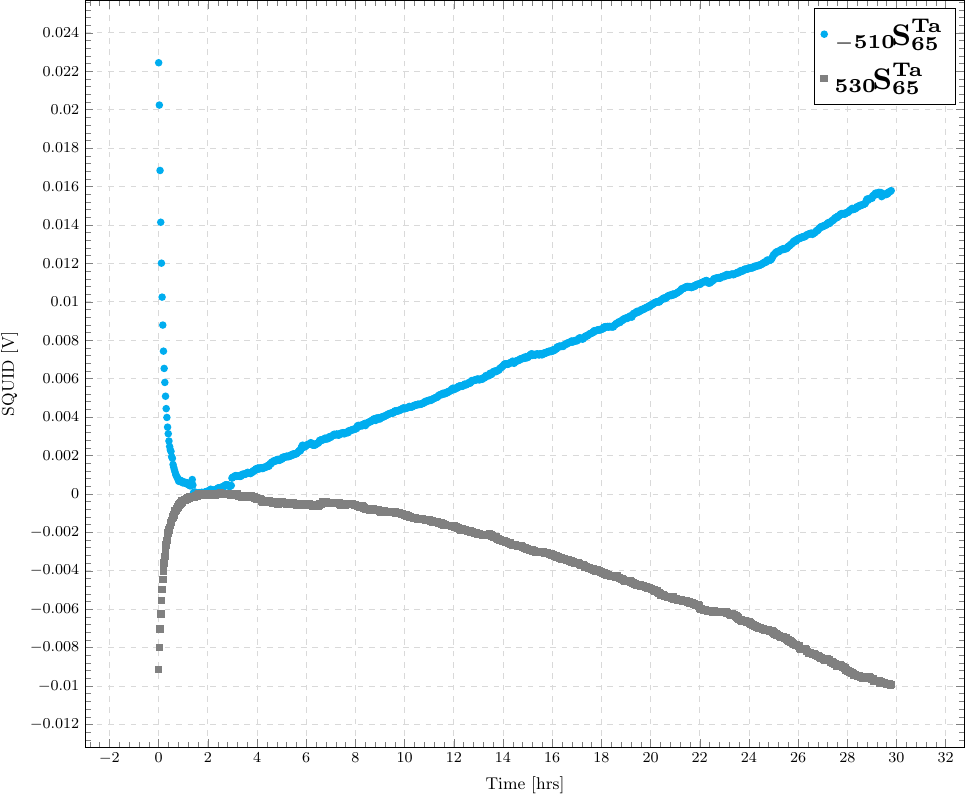}
&
  \includegraphics*{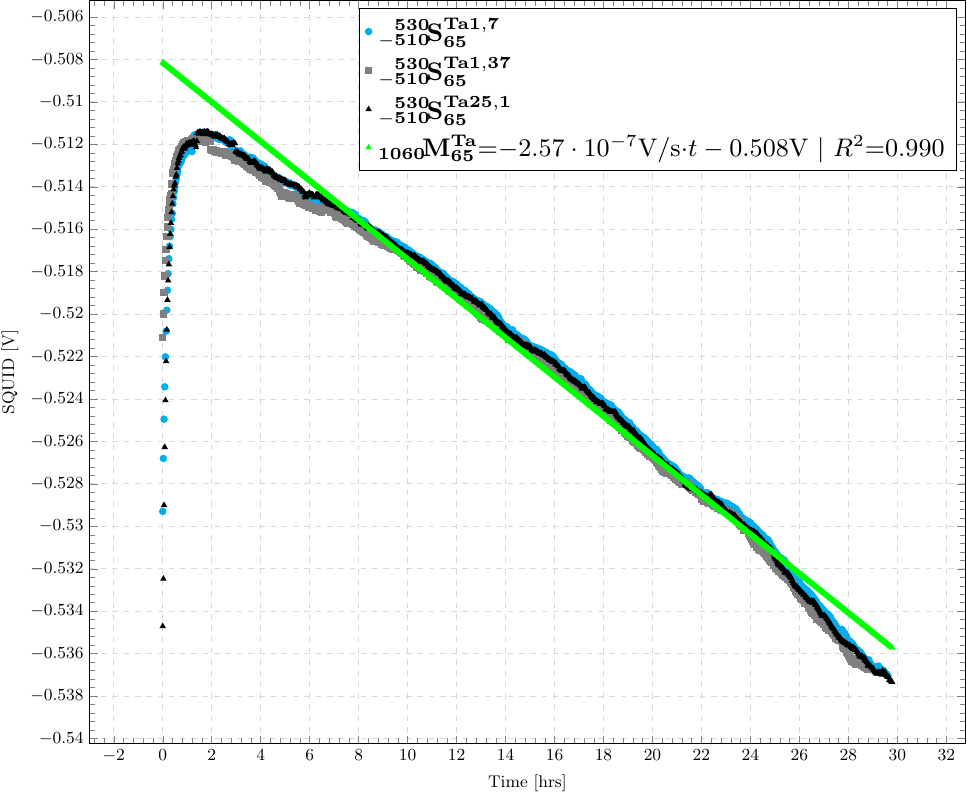}
&
  \includegraphics*{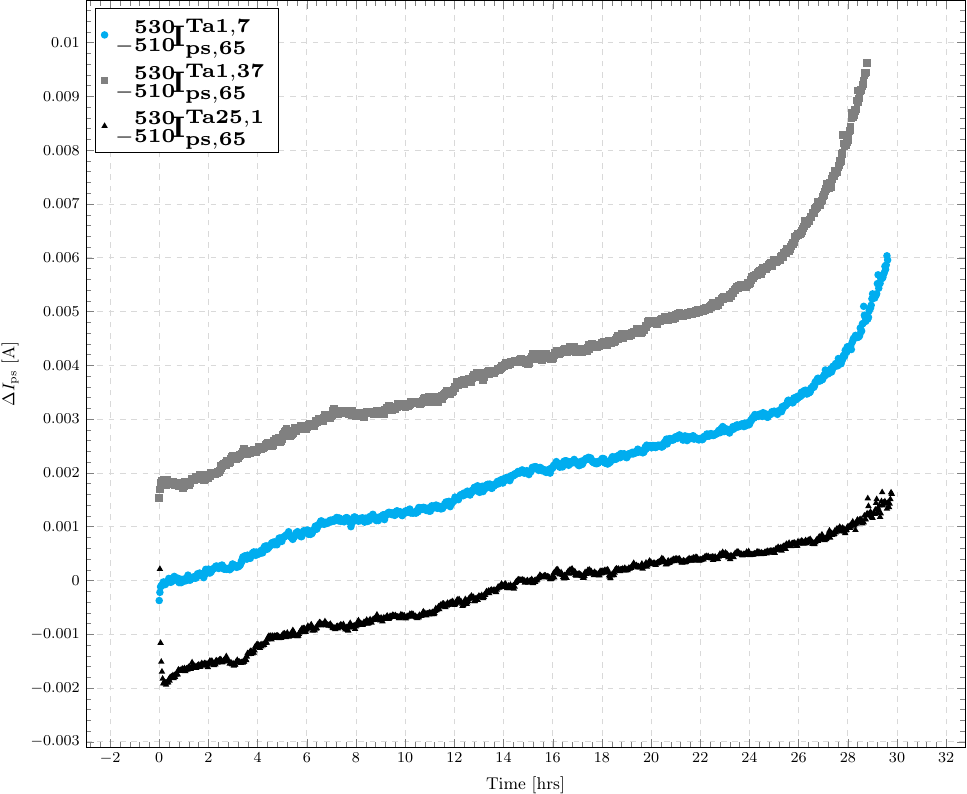}
\end{tabular}
}
\caption{Fixed temperature plots of Ta with a reverse current \Ia=$-510$~mA, $T$=0.065~K. The first row makes a comparison to the 0~A data while the second row compares to the \Ia=+530~mA data. The model linear regression was performed for data points $99\rightarrow1065\in$ \stc{-510}{\text{\bf{65}}}{Ta}.}\label{RevB}
\end{figure*}

\subsubsection{Ta: reverse field data}\label{sec:reverse}

The last fixed temperature experimental run with the tantalum sample wire was at $T$=0.065~K with a reverse current $\text{\Ia}$=~$-510$~mA. This complete dataset was made in order to verify that changing the direction of the sample solenoid magnetic field would change the sign of the slope compared to the fixed current plots with $\text{\Ia}>0$; this is in fact what was observed as seen in the first column \stc{-510}{65}{Ta} of \Fig{RevB} which also includes the \stc{0}{65}{Ta}, \stc{530}{65}{Ta} plots from \Fig{fixedTempCurr-T} in the first and second row respectively.
\begin{itemize}
\item In the first row, the center plot shows the subtraction of the 0~A data from the 0.53~A data: \dstc{0}{-510}{\text{\bf{65}}}{Ta}{1,383}, \dstc{0}{-510}{\text{\bf{65}}}{Ta}{1,353}, \dstc{0}{-510}{\text{\bf{65}}}{Ta}{1,413}  to be compared with \dstc{0}{+530}{\text{\bf{65}}}{Ta}{1,396}, \dstc{0}{+530}{\text{\bf{65}}}{Ta}{1,366}, \dstc{0}{+530}{\text{\bf{65}}}{Ta}{1,426}  in \Fig{SpropIa}.  The mean slope obtained using linear regression on the center plots is +1.23$\times10^{-7}$, to be compared with the mean slope -1.45$\times10^{-7}$ for the \Ia=+0.53~A, $T$=0.065~K in \Tab{ratioTab}. The magnitudes of the slopes for the direct and reverse $\text{\Ia}$ are consistent\footnote{The 0~A data is the source of most of the noise.} with each other but of opposite signs. Hence, the electrization field reversed its direction when the  nuclear magnetization  was reversed, as expected for  a T-odd electric field.
\item In the second row, the middle figure plots \dstc{-510}{530}{\text{\bf{65}}}{Ta}{1,7}, \dstc{-510}{530}{\text{\bf{65}}}{Ta}{1,37}, \dstc{-510}{530}{\text{\bf{65}}}{Ta}{25,1}. Since the reverse field data is opposite to the direct field data, our criteria for the best offset is for the differenced datasets to minimize the tail when the datasets \stc{-510}{65}{Ta} and \stc{530}{65}{Ta} are {\it added} instead of subtracted. The best offsets according to that criterion was (1,7) and the corresponding linear regression model yielded a \gbar{1040}{65}{Ta}=$-2.57\cdot10^{-7}$V/s. Taking the ratio with \gbar{265}{65}{Ta} give
 \beq
	 \frac{ \text{\gbar{1040}{65}{Ta}} }{ \text{\gbar{265}{65}{Ta}} }=3.6\cong\frac{1040}{265}=3.9~.
 \eeq
Hence, the ratio is only 8\% below the theoretical value of 3.9 adding to the evidence that the field driving the current in the sample Ta wire is the electrization proportional to \Ia.
 \end{itemize}
Furthermore, the consistency of the reverse field data with the $1/T$ dependence of the electrization field is verified by dividing \gbar{1040}{65}{Ta} by $1040/265$ to obtain an equivalent point with \Ia=265~mA, and including the result at $T=65$~mK in  \Fig{gammaTTaMod}. Including the reverse current data as the sixth data point in the linear regression, but excluding the  \dstc{265}{530}{T}{Ta}{~\bar{m},\bar{n}} data point, yields for \gbar{265}{}{Ta}
\begin{align} \label{gammabarM2}
&\text{\gbar{265}{}{Ta}}=(-5.87\pm0.51)\cdot10^{-9}\frac{\text{V}\cdot\text{K}}{\text{s}}, ~p=0.00032 \\
&\text{\gbar{265}{}{Ta}}=(-5.87\pm5.22)\cdot10^{-9}\frac{\text{V}\cdot\text{K}}{\text{s}} \ne0,~99.95\text{\%}~\text{CL}.
\end{align}
Adding the reverse field data point decreases the $p$-value of proportionality to $1/T$ by a factor of 5. Since adding the reverse field data point was predicated on being allowed to divide \gbar{1040}{65}{Ta} by $1040/265$, the decrease of the $p$-value also adds to the evidence that the electrization field is proportional to \Isol.

This electrization reversal occurred in a context where the pill current \ips~ does not change sign and is in fact very similar to the \ips~ of the \Ia=+0.53A as seen in the third column plots of \Fig{RevB}. Comparing this to \Fig{SpropIa} for the \Ia=0~A data, the $\Delta$\ips~  ranges differ by about 1.5~mA. Comparing the $\Delta$\ips~ plot of the second row to \Fig{fixedTempCurr-T}, the ranges of \ditc{-510}{530}{\text{\bf{65}}}{Ta}{m,n} are similar to those of \ditc{265}{530}{\text{\bf{65}}}{Ta}{m,n}.  This is additional evidence that systematics from the time-dependent \ips~ magnetic field is not a significant source of noise, something that was also broadly concluded from the 0~A data in \Fig{fixedTempCurr-T}.

\begin{figure}
\resizebox{7 cm}{!}{\includegraphics*{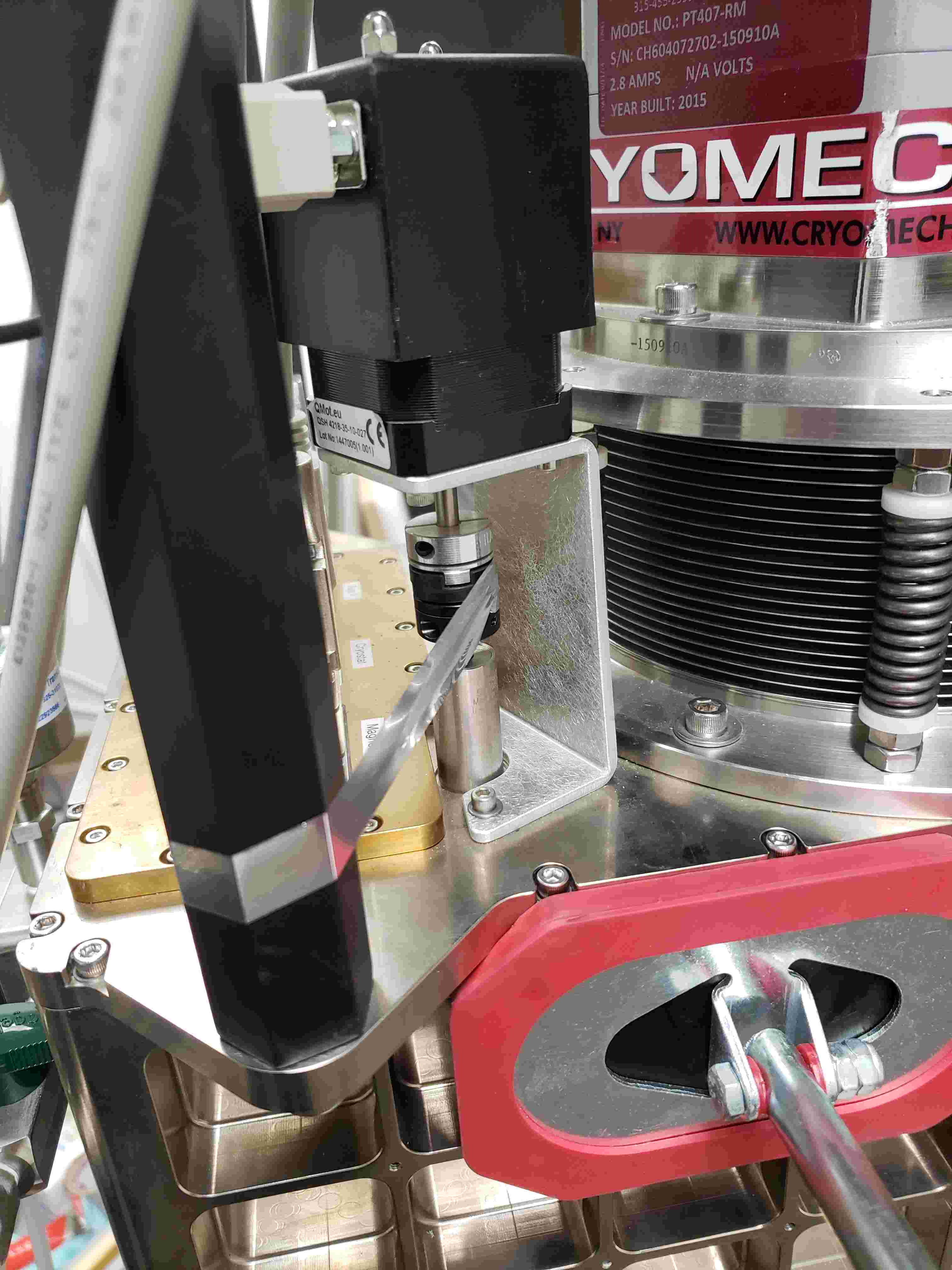}}
\caption{Mylar tape to increase the thermal contact between the pills and the 3~K stage heat sink. Also visible is one of the `third hands' to suppress vibrations generated by the compressor.}\label{mylar}
\end{figure}

\begin{center}
\begin{table}[t]
\begin{tabular}{|l|l|l|}
\hline
          &   52.5~mK   &      75~mK  \\
\hline
134~mA & 21.7 (781)     & 46.5 (1673)    \\
\hline
271~mA &  21.5 (773) &     45.3 (1629)   \\
\hline
542~mA &  21.8 (786)  &   45  (1618)    \\
\hline
\end{tabular}
\caption{Same as \Tab{TaDatasets} but for lead; a total of 7260 data points over 201.8 hours was collected at fixed temperatures 52.5~mK and 75~mK and three fixed currents. Compared to the number of hours listed in \Tab{TaDatasets} for 265~mA and 530~mA  at 52.5~mK, 75~mK, the hold times are significantly larger for Pb, consistent with the improved thermal contact of the FAA and GGG pills with the 3~K heat sink during REGEN thanks to a greater torque on the heat switch.}\label{PbDatasets}
\end{table}
\end{center}
\begin{figure*}
\centering
\resizebox{17 cm}{!}{
\begin{tabular}{ccc}
 \includegraphics*{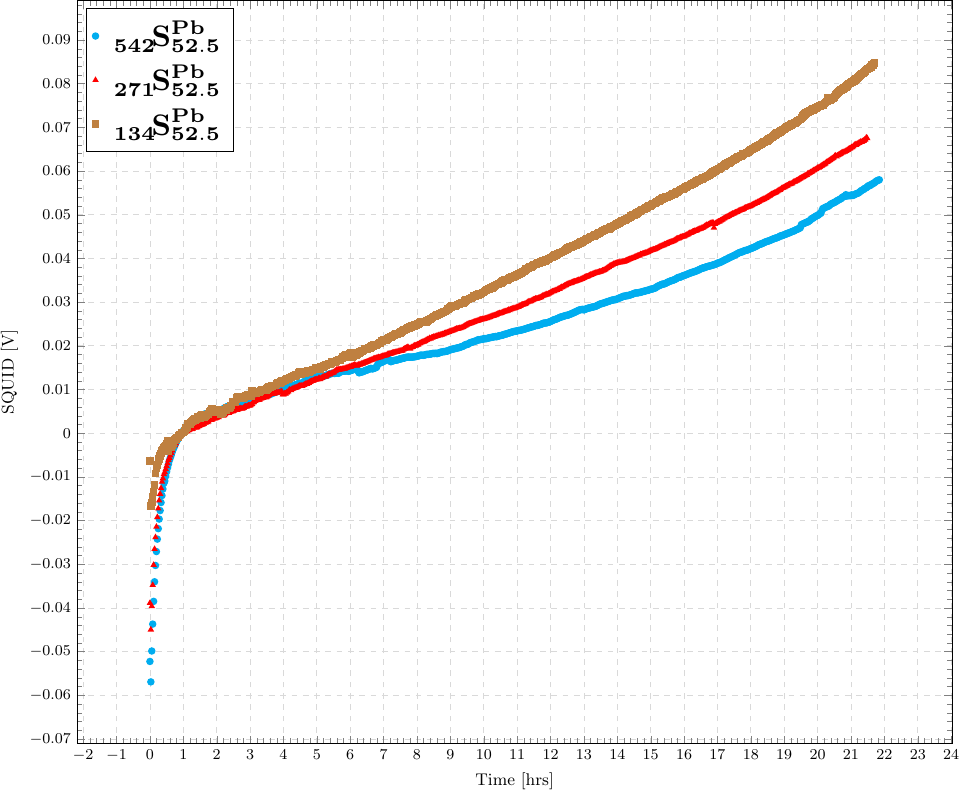}
&
 \includegraphics*{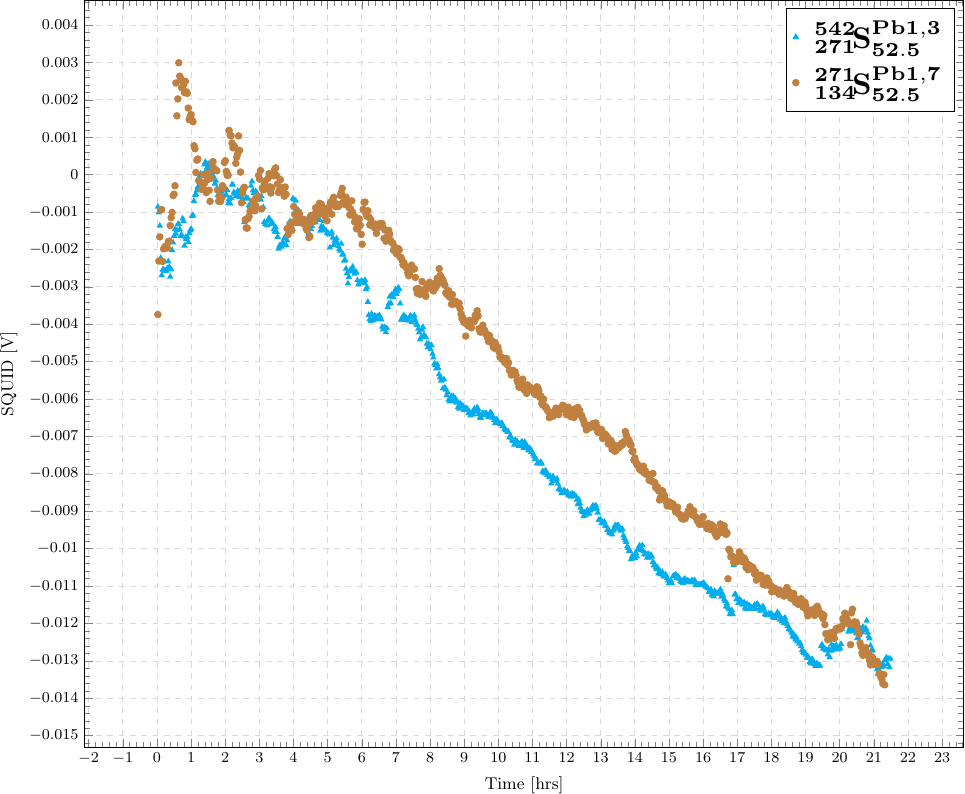}
&
 \includegraphics*{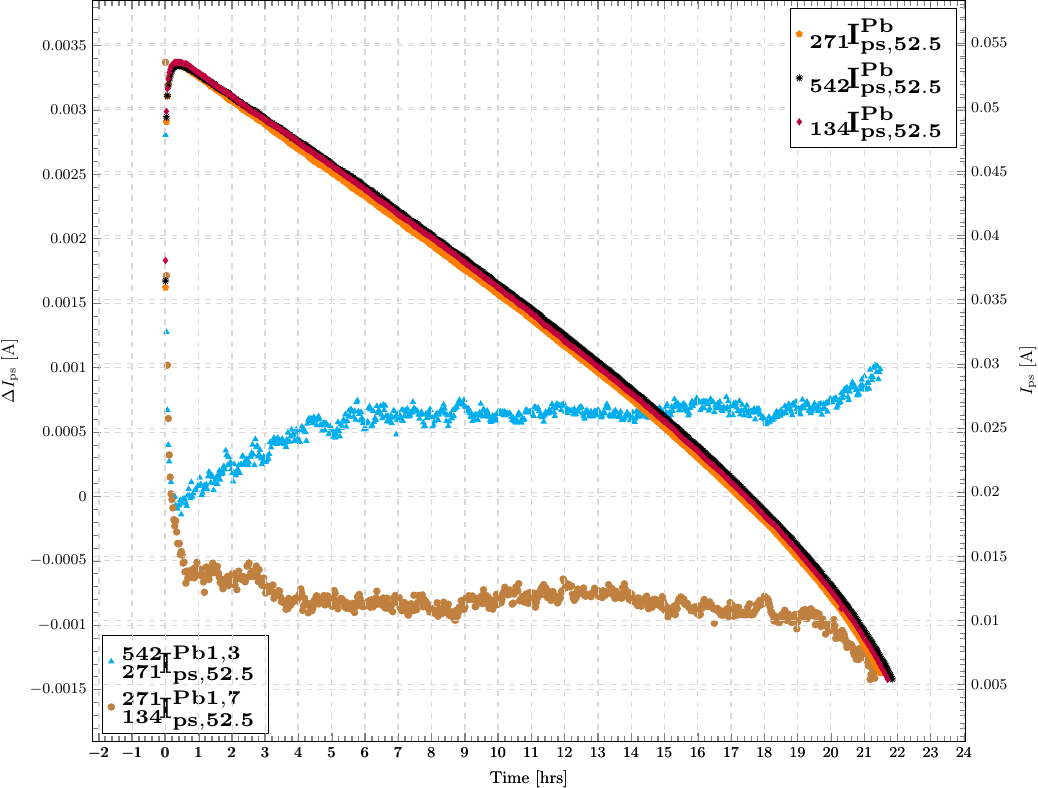}
\\
 \includegraphics*{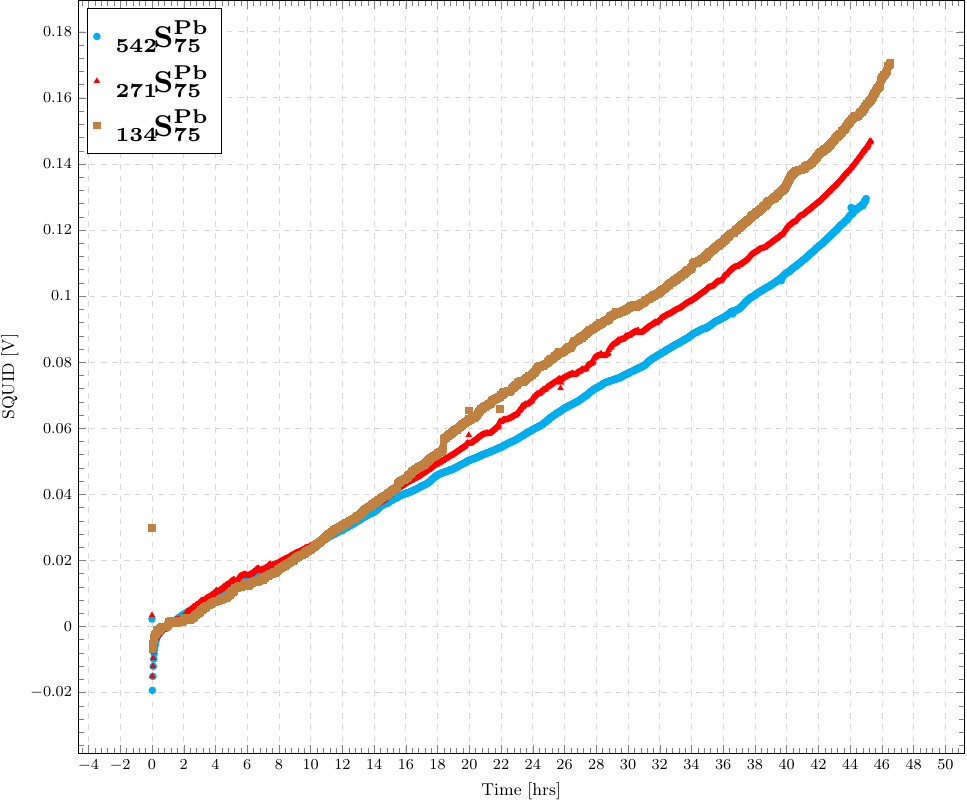}
&
 \includegraphics*{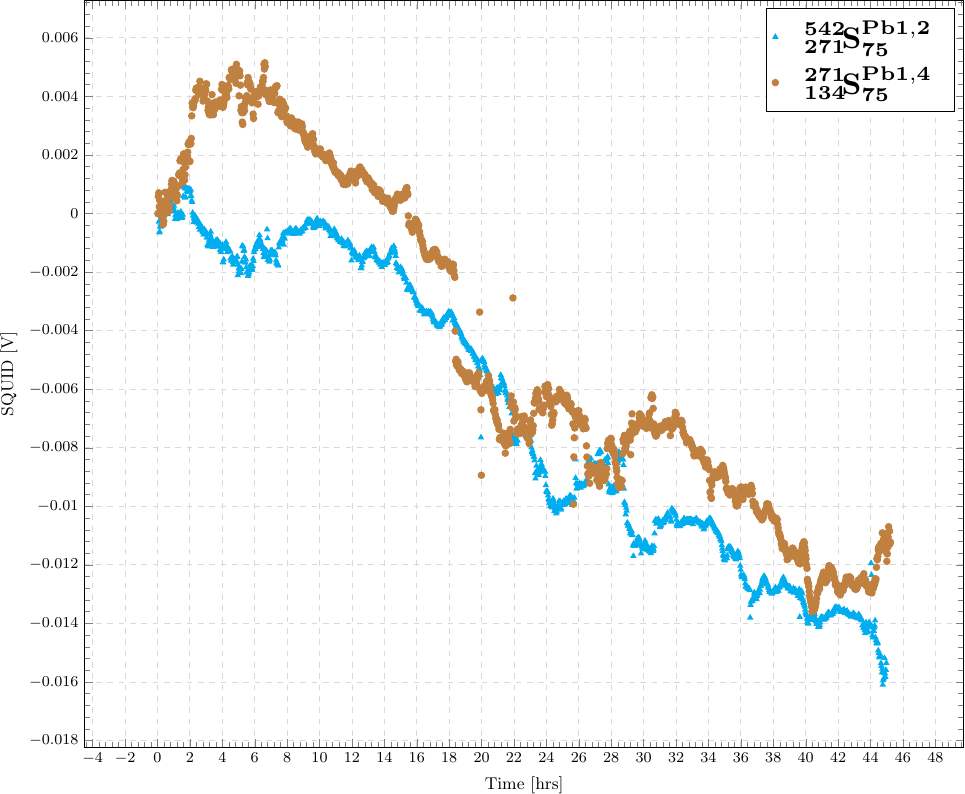}
&
 \includegraphics*{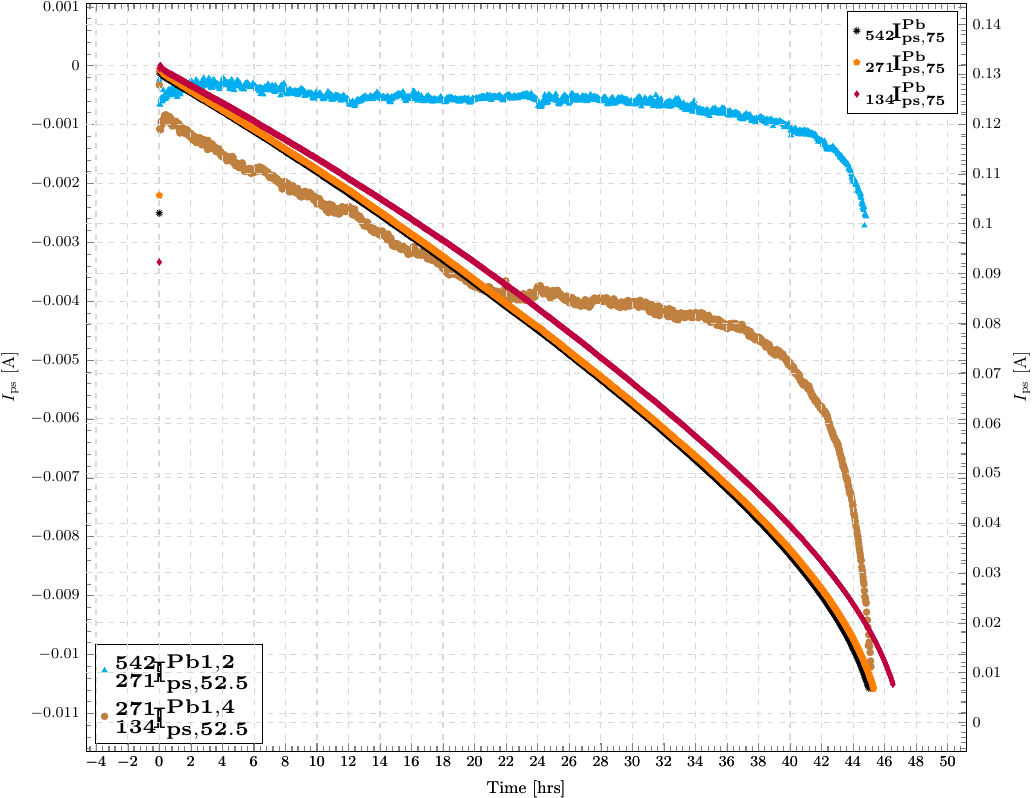}
\end{tabular}
}
%\vspace{-2.5cm}
\caption{Pb SQUID and PS I data at fixed temperature and current data. The first and second row data were collected at $T$=52.5~mK and 75~mK respectively.   The $1^\text{st}$ column displays the \stc{542}{T}{Pb}, \stc{271}{T}{Pb},  \stc{134}{T}{Pb} plots; the second column displays the \dstc{275}{542}{T}{Pb}{m,n} and \dstc{134}{271}{T}{Pb}{i,j} plots where $m,n,i,j$ have been adjusted to minimize the early time systematics as was done for Ta.  The $3^\text{rd}$ column displays the corresponding $\Delta$\ips=\dstc{275}{542}{T}{Pb}{m,n},  \dstc{134}{271}{T}{Pb}{i,j}  plotted on the left, while the \ips=\itc{I_a}{T}{Pb} for \Ia=134, 271, 542~mA are plotted on the right. It is noted that the \itc{I_a}{52.5}{Pb} plots coincide nearly exactly, while \itc{134}{75}{Pb} is slightly above \itc{271}{75}{Pb}and \itc{542}{75}{Pb} and has a correspondingly longer run time.}\label{fixedTempCurrPb}
\end{figure*}
\begin{center}
\begin{table}[t]
\begin{tabular}{|l|l|l|l|}
\hline
          &   \gtc{271}{T}{Pb}$^{m,n}$  &  \gtc{137}{T}{Pb}$^{i,j}$   &  $\rho^{\text{Pb}}_T$    \\
\hline
                            &    -1.98$\cdot10^{-7}$ (1,5)              &         -2.03$\cdot10^{-7}$ (1,13)   &     \\
52.5~mK &  -2.00$\cdot10^{-7}$ (1,3)              &         -2.07$\cdot10^{-7}$ (1,7) &        \\
                           &   -1.96$\cdot10^{-7}$ (1,1)              &         -2.01$\cdot10^{-7}$ (1,1) &       \\
\hline
 MEAN(\gtc{}{52.5}{Pb}) &       -2.00$\cdot10^{-7}$ & -2.03$\cdot10^{-7}$   &   0.98 \\
\hline
                            &    -1.08$\cdot10^{-7}$ (1,4)              &         -1.24$\cdot10^{-7}$ (1,7)   &      \\
75~mK  &  -1.08$\cdot10^{-7}$ (1,2)              &         -1.23$\cdot10^{-7}$ (1,4) &      \\
                           &   -1.08$\cdot10^{-7}$ (1,1)              &         -1.24$\cdot10^{-7}$ (1,1) &        \\
\hline
 MEAN(\gtc{}{75}{Pb}) &       -1.08$\cdot10^{-7}$ & -1.24$\cdot10^{-7}$   &   0.87 \\
\hline
%$\overline{ _{50}^{115}\text{Sn}+ _{50}^{117}\text{Sn} + _{50}^{119}\text{Sn}} $ & 4 & 3.71 & -1.023 & 1/2& 16.61 & 99.999 \\
%\hline
\end{tabular}
\caption{The slopes at $T\in$\Tset{Pb} for the different offsets  $m,n$ and $i,j$  given  in parentheses. The last column shows the mean ratio of the slopes. }\label{Pbslopes}
\end{table}
\end{center}
\subsubsection{Pb data}\label{sec:pb}
The lead data was taken as a control for tantalum because the ratio of free magnetizations $\Fedm{Ta}/\Fedm{Pb}=19$ is much smaller than 1. The temperatures, currents, number of data points and hours of data collection for lead are provided in \Tab{PbDatasets}. The Pb run times at fixed temperatures were increased by applying a torque on the heat switch during REGEN in order to improve the thermal contact between the pills and the 3~K heat sink. The heat switch can be opened and closed manually by turning a knob on the exterior. The torque was applied with mylar tape where one end was wrapped around the knob and  the other end was pulled and wrapped around one of the support rods of the ADR (see \Fig{mylar}).

The improved thermal contact was confirmed three ways: 1) the maximum temperature of the 1~K stage during RAMP UP of the \ips~ was about 0.5~K smaller than for Ta, 2) the minimum temperature achieved after REGEN in Pb was 38~mK compared with 44~mK in Ta, and 3) the run times were increased by about a third at 52.5~mK and about 14\% at 75~mK. This improved thermal contact during REGEN also produced nearly identical \ips~ time evolutions for different $\text{\Ia}$ at a fixed temperature as seen in the third column of \Fig{fixedTempCurrPb}. The nearly identical \ips~ at $T=52.5$~mK is also reflected in the nearly identical cardinalities of $|$\stc{134}{52.5}{Pb}$|$=781, $|$\stc{271}{52.5}{Pb}$|$=773, and $|$\stc{542}{52.5}{Pb}$|$=786. Hence, the systematics due to the varying \ips~ could be more efficiently subtracted out in the differentiated datasets \dstc{I_1}{I_2}{T}{Pb}{m,n}. This improved thermal contact during REGEN also yielded significantly smaller standard deviations about the fixed temperature during REG compared to the Ta standard deviations as seen in \Fig{TsetSDPb} compared to \Fig{SDplots}.

Complete datasets were created for three currents \Iset{Pb}=\{134~mA, 271~mA, 542~mA\} and two fixed temperatures \Tset{Pb}=\{52.5~mK, 75~mK\}. The \dstc{I_a}{}{T}{Pb}{}  are shown in  the first column of \Fig{fixedTempCurrPb}.\footnote{An overall minus sign on all the Pb data was removed for ease of comparison with the Ta data.} The \dstc{I_a}{}{T}{Pb}{} plots have an upward trend with the data collected at the largest \Ia=542~mA having the smallest uptrend, and the data collected at the smallest \Ia=134~mA having the largest. These features may be compared to the plots of \dstc{I_a}{}{75}{Ta}{} in   \Fig{fixedTempCurr-T} where at early times, the dataset with the smallest \Ia, \dstc{0}{}{75}{Ta}{}, shows the largest uptrend  while the dataset with the largest \Ia, \dstc{530}{}{75}{Ta}{}, has the smallest uptrend. The second column of \Fig{fixedTempCurrPb} shows plots of the differential datasets created to  minimize systematics. As in the case of tantalum, the $m,n$ in \dstc{I_{a,1}}{I_{a,2}}{T}{Pb}{m,n} were chosen to minimize the early times systematics.

Since \dstc{134}{542}{T}{Pb}{1,1}=\dstc{271}{542}{T}{Pb}{1,1}+\dstc{134}{271}{T}{Pb}{1,1} for the first $N_\text{min}$ elements where $N_\text{min}$ is the smallest of $|$\dstc{134}{542}{T}{Pb}{1,1}$|$, $|$\dstc{271}{542}{T}{Pb}{1,1}$|$, and $|$\dstc{134}{271}{T}{Pb}{1,1}$|$, only \dstc{275}{542}{T}{Pb}{m,n} and \dstc{134}{271}{T}{Pb}{i,j} need be considered. \Fig{fixedTempCurrPb} plots the \dstc{275}{542}{T}{Pb}{m,n} and \dstc{134}{271}{T}{Pb}{i,j}  for $m,n,i,j$ that minimize the early times systematics. The linear regression slopes are provided in \Tab{Pbslopes}. The slopes at $T=52.5$~mK are about twice as big as the slopes at $T=75$~mK to be compared with the expected value of $75/52.5=1.43$.

The result on the linear regression slopes may be a statistical fluctuation, but the lack of dependence on $\text{\Isol}$ is even more problematic as the mean ratios of \Tab{Pbslopes} satisfy
\beq
\rho^{\text{Pb}}_T<1 ~~\text{compared to}~ \rho^{\text{Pb}}_T=\frac{271}{134}=2.02~~\forall T\in\text{\Tset{Pb}}
\eeq
The ratios $\rho^{\text{Pb}}_T$ are  inconsistent with the expected ratio of \Ia currents of 2.02 for the detection of a nuclear EDM signal. The small $\rho^{\text{Pb}}_T$ likely results from  the fact that  $\Fedm{Pb}/\Fedm{Ta}\ll1$ implying that much longer run times are required for Pb compared to Ta to confirm a NEDM signal. The results obtained for Pb are similar to what would have been obtained for Ta  if that data had been collected over an hour or so: the slopes for would not have been consistent with a NEDM signal. To properly confirm a NEDM signal for Pb at 52.5~mK at \Ia=271~mA would likely require over 100 hours of data collection.

\begin{center}
\begin{table}[t]
\begin{tabular}{|c|c|c|}
\hline
  \Ia (mA)        &   Ta     &    Pb      \\
\hline
265/271 & 52.0 (1868) $[45,151]$    & 63.7 (2289) $[39,213]$    \\
\hline
530/542 & 51.9 (1865) $[46,152]$   &  70.2 (2524)  $[39,285]$  \\
\hline
\end{tabular}
\caption{Fixed current at zero-field datasets. The numbers at the intersection of the $\text{\Ia}$ (rows) and the elements (columns) give the total number of hours of data and, in parenthesis, data points collected at $\text{\Ia}$ over the given temperature range in mK for Ta or Pb. For example, 52 hours (1868 data points) were collected for tantalum at an applied current of 265~mA over a temperature range 45~mK < $T$ < 151~mK.}\label{TaZeroFieldDatasets}
\end{table}
\end{center}
\begin{figure*}
\centering
\resizebox{17 cm}{!}{
\begin{tabular}{cc}
 \includegraphics*{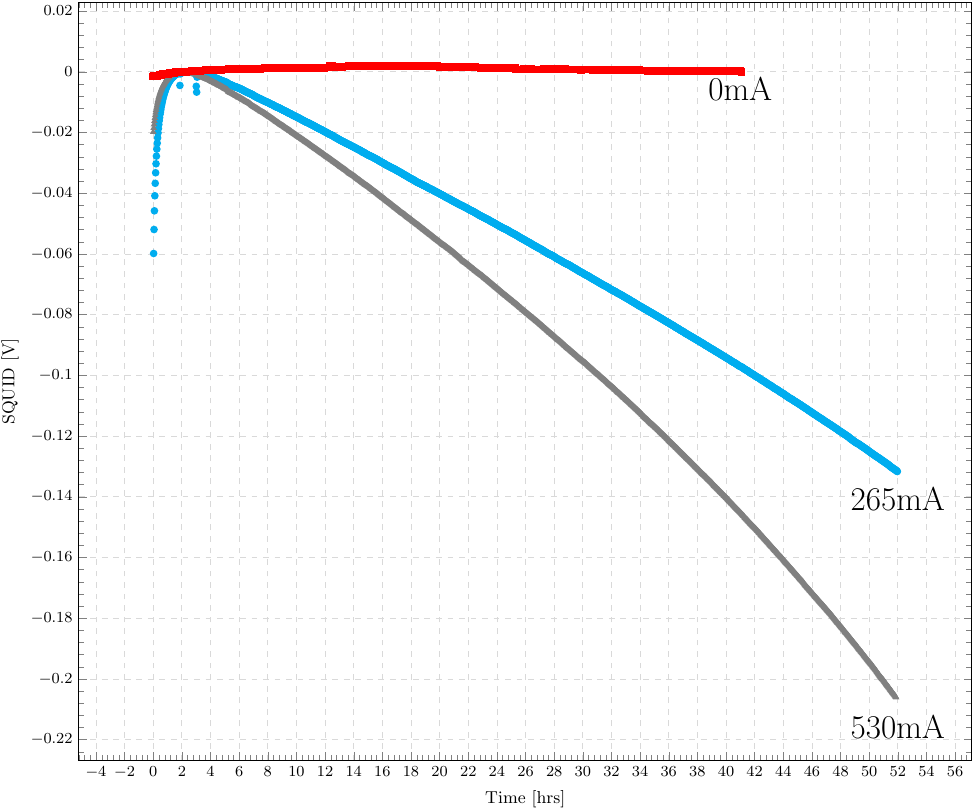}
&
 \includegraphics*{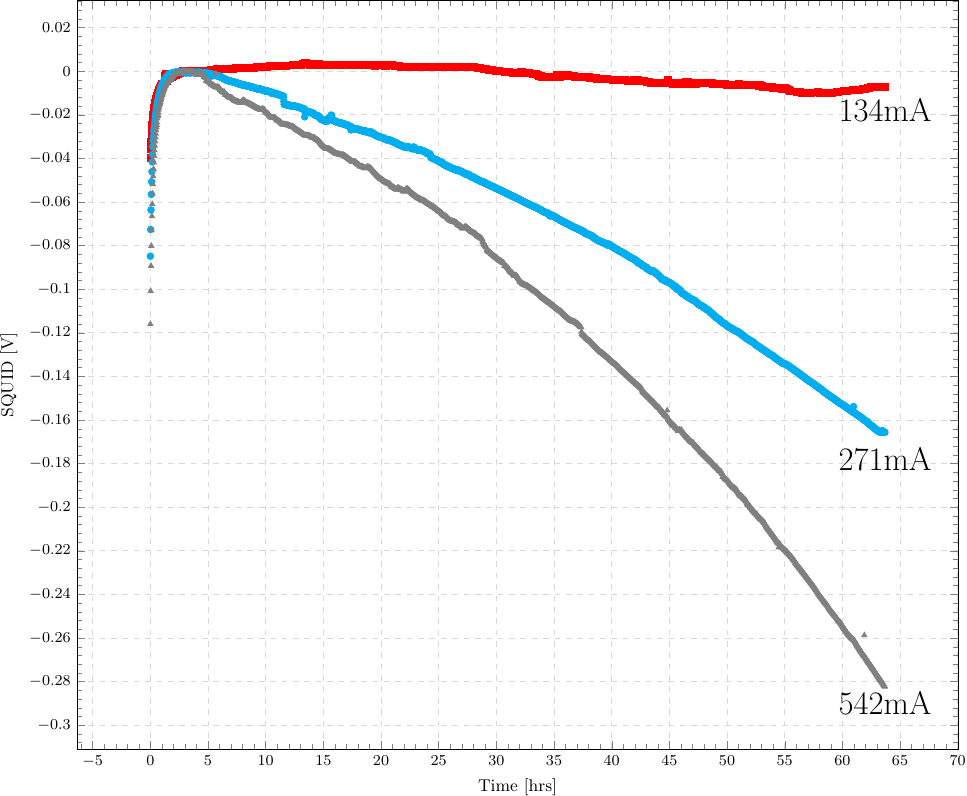}
\end{tabular}
}
%\vspace{-2.5cm}
\caption{Zero-field data for Ta and Pb where \Iset{Ta}=\{0~mA,265mA,530mA\}, \Iset{Pb}=\{134~mA,271mA,542mA\}. It is seen that the SQUID voltage depends on \Ia and effectively vanishes as \Ia$\rightarrow0$.}\label{TaPbsubKall}
\end{figure*}

\subsection{Zero-field data}\label{sec:zfd}

Zero-field data refers to data collected in the absence of a \ips-generated external magnetic field. Therefore, the sample assembly temperature is not fixed and increases through ambient heat absorption. The zero-field datasets were collected at \Iset{Ta}=\{0mA,265mA,530mA\}, \Iset{Pb}=\{134mA,271mA,542mA\} for tantalum and lead respectively, and are plotted in \Fig{TaPbsubKall} where it is observed that the variations of the SQUID voltage depend on the temperature (see \Fig{tempTa}) and vanish as \Isol$\rightarrow$0. This dependence can be written as
$I_\text{sol}P(I_\text{sol})f(T)$ where $T=T(t)$ where $t$ is time.
Although this dependence is relatively complex, zero-field data can still be used to support the hypothesis of an electrization field by using \Eq{intdidt} with the lead data as a control and using the main results obtained summarized thus
\begin{itemize}
\item From \Eq{gammabarM}, $ _{265}\!\bar{\gamma}^{\text{Ta}}=-6.50\cdot10^{-9}$.
\item $ _{265}\!\bar{\gamma}^{\text{Ta}}$ is proportional to $\text{\Isol}$ and therefore to \Ia.
\item Pb data does not have an explicit time dependence stemming from the bulk NEDM electric field and can only depend on $T$ and \Ia.
\end{itemize}
The Pb dependence on $T$ and $\text{\Ia}$ can come from a number of external sources, for example, the fact that $\text{\Isol}$ can be affected by changes in the nuclear magnetization (and corresponding surface current) of the Al booster solenoid core penetration depth or, possibly, the magnetization of the salt pills. To experimentally quantify the consistency of the zero-field data with our main results, a comparison between the zero-field data of Pb and Ta was performed. Define the dataset
\begi
\text{\stc{I_a}{\text{\bf{SubK}}}{C}}=\{($s_i,t_i,T_i$), $i$=1,2,...,$N^\text{C}_{I_a}$\} where $s_i$ is the SQUID data point collected at time $t_i$, temperature $T_i$ for a sample wire made of element C=Ta or Pb and a $\text{\Isol}$ induced from a main solenoid current \Ia. Each sub-Kelvin dataset is  a line in the 3-dimensional space of SQUID voltage, time and temperature.
\ei
Using this notation, datasets that depend only on temperature and $\text{\Ia}$ can be defined using our main results
\begin{align}\label{subk530a}
& \text{\stcbar{530}{\text{\bf{SubK}}}{Ta}}=\{s_i~|~ s_i = s_i^\prime-2~ _{265}\!\bar{\gamma}^{\text{Ta}}\int_{0}^{t_i}\frac{\text{d}t}{T},~\nonumber\\
&~~~~~~~~~~~~~~~~~~~~~~~~~~~~~~~~~~~~\forall~s_i^\prime\in \text{\stc{530}{\text{\bf{SubK}}}{Ta}}   \}   \\ \label{subk265a}
 &\text{\stcbar{265}{\text{\bf{SubK}}}{Ta}}=\{s_i~|~ s_i = s_i^\prime-~ _{265}\!\bar{\gamma}^{\text{Ta}}\int_{0}^{t_i}\frac{\text{d}t}{T},\nonumber\\
 &~~~~~~~~~~~~~~~~~~~~~~~~~~~~~~~~~~~~\forall~s_i^\prime\in \text{\stc{265}{\text{\bf{SubK}}}{Ta}}   \}   \\ \label{subk530b}
 &\text{\stcbar{530}{\text{\bf{SubK}}}{Pb}}\subset \text{\stc{530}{\text{\bf{SubK}}}{Pb}} \\ \label{subk265b}
 &\text{\stcbar{265}{\text{\bf{SubK}}}{Pb}}\subset \text{\stc{265}{\text{\bf{SubK}}}{Pb}} \\ \label{intnEDM}
& \int_{0}^{t_i}\frac{\text{d}t}{T}\equiv\sum_{j=1}^{i}\frac{t_{j+1}-t_j}{T_j}
 \end{align}
where $t_j,~T_j\in$\stc{I_a}{\text{\bf{SubK}}}{Ta} and the datasets with a bar, \stcbar{I_a}{\text{\bf{SubK}}}{C}, depend on $T$, $\text{\Ia}$ and the sample wire element C only\footnote{Note that the same $\text{\Ia}$ labels are used for Pb even though the $\text{\Ia}$ used for the Pb runs were about 2\% higher: 271~mA vs 265~mA and 542~mA vs 530~mA.}. Each Pb data point  in the subset \stcbar{I_a}{\text{\bf{SubK}}}{Pb}} was selected by finding the $s_i\in$\stc{I_a}{\text{\bf{SubK}}}{Pb} collected at the  temperature closest to each temperature $T_i$ in $(s_i,t_i,T_i)\in$\stc{I_a}{\text{\bf{SubK}}}{Ta}. Hence, 
\beq
|\text{\stcbar{I_a}{\text{\bf{SubK}}}{Pb}}|=|\text{\stcbar{I_a}{\text{\bf{SubK}}}{Ta}}|=|\text{\stc{I_a}{\text{\bf{SubK}}}{Ta}}|~.
\eeq
The actual $t_i$ corresponding to a particular $s_i$ in \stc{I_a}{\text{\bf{SubK}}}{Pb} is irrelevant by our third main result, and the $t_i$ that corresponds to the $T_i$ from the Ta datasets are assigned to the $s_i\in$\stcbar{I_a}{\text{\bf{SubK}}}{Pb} as well. \Eq{subk530a} uses the second main result that $_{530}\!\bar{\gamma}^{\text{Ta}}=2~_{265}\!\bar{\gamma}^{\text{Ta}}$. \Eq{subk530a} and \Eq{subk265a} use the first main result, the actual value of $_{265}\!\bar{\gamma}^{\text{Ta}}$ given by \Eq{gammabarM}. The third main result that the Pb datasets depend only on temperature and $\text{\Ia}$ is expressed in \Eq{subk530b} and \Eq{subk265b}.

Since \stcbar{I_a}{\text{\bf{SubK}}}{C} depends on $\text{\Ia}$, the subtraction approach  used for the fixed temperature datasets where the \iitc was independent of $\text{\Ia}$ cannot be used since a subtraction of \stcbar{I_a}{\text{\bf{SubK}}}{C} collected at a specific  $\text{\Ia}$ from \stcbar{2I_a}{\text{\bf{SubK}}}{C} will not remove the dependence on $\text{\Ia}$ requiring a new approach.

In light of the fact that the same booster and sample solenoids were used for both the Pb and Ta runs, the main difference between \stcbar{I_a}{\text{\bf{SubK}}}{Pb} and \stcbar{I_a}{\text{\bf{SubK}}}{Ta} will be in the mutual inductance of the wire with the  SQUID\footnote{The differences in $\text{\Ia}$ of about two percent between the Ta and Pb run were experimentally insignificant. As for the self-inductance, it is nearly  identical across the different runs since most of the self-inductance comes from the shape of the sample solenoid and the Pb and Ta sample wires have the same length.}. If $M_\text{C-PC}$ is the mutual inductance between the SQUID pickup coil and the sample wire made of C=Ta or Pb,  and if $M_\text{Ta-PC}=\bar{m}M_\text{Pb}$, then the relationship between \stcbar{I_a}{\text{\bf{SubK}}}{Pb} and \stcbar{I_a}{\text{\bf{SubK}}}{Ta} will be \stcbar{I_a}{\text{\bf{SubK}}}{Ta}=$\bar{m}$~\stcbar{I_a}{\text{\bf{SubK}}}{Pb} for identical, noiseless experimental conditions and setting any additive SQUID voltage constant to zero\footnote{Something that can always be done as SQUIDs do not measure absolute magnitudes of magnetic fields.}. However, since 1) the temperature time-dependence collected during the Pb experimental runs was significantly different from that of the  Ta runs, and 2) the datasets are not noiseless,  \stcbar{I_a}{\text{\bf{SubK}}}{Pb} and \stcbar{I_a}{\text{\bf{SubK}}}{Ta} will have a constant ratio $\bar{m}_{I_a}$ that  depends on the actual experimental noise and conditions of the run which is labeled by the subscript \Ia=530 or 265. In order to conclude that the zero-field data is consistent with the fixed temperature results, the relationship to verify is
\beq\label{sbareq}
\text{ \stcbar{I_a}{\text{\bf{SubK}}}{Ta}}= \bar{m}_{I_a}\text{ \stcbar{I_a}{\text{\bf{SubK}}}{Pb}}+\bar{b}_{I_a}~.
\eeq
where $\bar{m}_{I_a},~\bar{b}_{I_a}$ are found using linear regression and \Eq{sbareq} integrates all of our main results. The validity of \Eq{sbareq} will be assessed by comparing it to
\beq\label{scomp}
\text{ \stc{I_a}{\text{\bf{SubK}}}{Ta}}= {m}_{I_a}\text{ \stcbar{I_a}{\text{\bf{SubK}}}{Pb}}+{b}_{I_a}~.
\eeq
where $m_{I_a},~b_{I_a}$ are also found using linear regression. It is expected that \stc{I_a}{\text{\bf{SubK}}}{Ta} and  \stcbar{I_a}{\text{\bf{SubK}}}{Pb} are highly correlated since their data points were collected using the same solenoid assembly under similar conditions. However, if the zero-field data are consistent with the theory presented in this work, one expects that removing the time-dependent contribution proportional to $_{265}\!\bar{\gamma}^{\text{Ta}}$ from \stc{I_a}{\text{\bf{SubK}}}{Ta}, i.e. using \stcbar{I_a}{\text{\bf{SubK}}}{Ta},   leads to an even better correlation with \stcbar{I_a}{\text{\bf{SubK}}}{pb} which does not have that contribution.

The linear regression model datasets implementing this approach are defined
\begin{align}\label{mod}
&\text{\mtcbar{I_a}{\text{\bf{SubK}}}{Ta}}\!=\!\{s_i| s_i=\bar{m}_{I_a}~s_i+\bar{b}_{I_a} ~\forall~s_i\in \text{\stcbar{I_a}{\text{\bf{SubK}}}{Pb}}  \} \\
&\text{\mtc{I_a}{\text{\bf{SubK}}}{Ta}}\!=\!\{ s_i| s_i = m_{I_a}~s_i+b_{I_a} ~\forall~s_i\in \text{\stcbar{I_a}{\text{\bf{SubK}}}{Pb}}  \} 
\end{align}
where $\bar{m}_{I_a},~\bar{b}_{I_a},~m_{I_a},~b_{I_a}$ are the linear regression parameters on the Pb datasets defined in \Eq{sbareq} and \Eq{scomp} respectively.
\begin{center}
\begin{table}[t]
\begin{tabular}{|c|c|c|c|c|}
\hline
  \Ia   &  1-$R^2$  \mtc{I_a}{\text{\bf{SubK}}}{Ta} &   1-$R^2$  \mtcbar{I_a}{\text{\bf{SubK}}}{Ta} & dof  & $F$  \\
\hline
265~mA  & $12.2\cdot10^{-3}$  &  7.3$\cdot10^{-3}$ & 1836 &  2.15\\
\hline
530~mA  & 7.5$\cdot10^{-3}$  & 2.3$\cdot10^{-3}$& 1834  & 4.49  \\
\hline
\end{tabular}
\caption{}\label{r2values}
\end{table}
\end{center}
The first row of \Fig{TaPbPlots} displays the \Ia=530~mA  plots while the \Ia=265~mA plots are in the second row. In order to make visual comparisons easier, the time domains of all  plots are identical and the maximum value of the ordinate is 0.24~V above the minimum value of the ordinate in all plots.  It is visually  clear that the \Ia=265~mA zero-field datasets are flatter than the \Ia=530~mA datasets verifying the dependence on \Ia. In the first column, the plots of  \stc{I_a}{\text{\bf{SubK}}}{Ta},  \stcbar{I_a}{\text{\bf{SubK}}}{Ta} and  \stcbar{I_a}{\text{\bf{SubK}}}{Pb} are shown and the \stcbar{I_a}{\text{\bf{SubK}}}{Ta} are always flatter than the  \stc{I_a}{\text{\bf{SubK}}}{Ta} indicating that the subtraction of \Eq{intnEDM} flattens the tantalum datasets.  The second column shows the plots of \stc{I_a}{\text{\bf{SubK}}}{Ta} and  \mtc{I_a}{\text{\bf{SubK}}}{Ta}, while the third column shows the plots of \stcbar{I_a}{\text{\bf{SubK}}}{Ta} and  \mtcbar{I_a}{\text{\bf{SubK}}}{Ta}.

It is visually apparent from \Fig{TaPbPlots} that \stcbar{I_a}{\text{\bf{SubK}}}{Ta} is a  better linear fit to \stcbar{I_a}{\text{\bf{SubK}}}{Pb} than \stc{I_a}{\text{\bf{SubK}}}{Ta}. From the $1-R^2$ values of \Tab{r2values}, the unexplained variance  of  the linear fit of  \stcbar{530}{\text{\bf{SubK}}}{Ta} to \stcbar{530}{\text{\bf{SubK}}}{Pb} is less than a third of the unexplained variance from the  \stc{530}{\text{\bf{SubK}}}{Ta} fit, while the unexplained variance of the \stcbar{265}{\text{\bf{SubK}}}{Ta} fit is 0.6 of the unexplained variance of the \stc{265}{\text{\bf{SubK}}}{Ta} fit.

Taking as the null hypothesis that  \stcbar{I_a}{\text{\bf{SubK}}}{Ta} is no better at explaining the Pb zero-field data than \stc{I_a}{\text{\bf{SubK}}}{Ta}, the number of degrees of freedom and the $F$ statistics in \Tab{r2values} yield a probability of zero that the null hypothesis is true. Hence, we accept the alternative hypothesis that \stcbar{I_a}{\text{\bf{SubK}}}{Ta} explains significantly more of the Pb data, \stcbar{I_a}{\text{\bf{SubK}}}{Pb}. This is entirely consistent with the theory that the electrization field is inversely proportional to $1/T$.

These zero-field results are consistent with our three, main, fixed-temperature results reiterated here: 1) $ _{265}\!\bar{\gamma}^{\text{Ta}}=-6.50\cdot10^{-9}$, 2) $ _{265}\!\bar{\gamma}^{\text{Ta}}$ is proportional to $\text{\Ia}$ with the result $ _{530}\!\bar{\gamma}^{\text{Ta}}=2~ _{265}\!\bar{\gamma}^{\text{Ta}}$ and 3) Pb data does not have a significant time-dependence stemming from the bulk NEDM electric field and can only depend on $T$ and \Ia.

It is remarkable that the fixed temperature results used to determine the  time derivative of \Is, \Eq{didt}, are consistent with the zero-field data results which represent the time integration of \Eq{didt} over thousands of data points collected over 51 hours during which the temperature increased in a non-linear fashion (see \Fig{tempTa}).
\begin{figure*}
\centering
\resizebox{17 cm}{!}{
\begin{tabular}{ccc}
 \includegraphics*{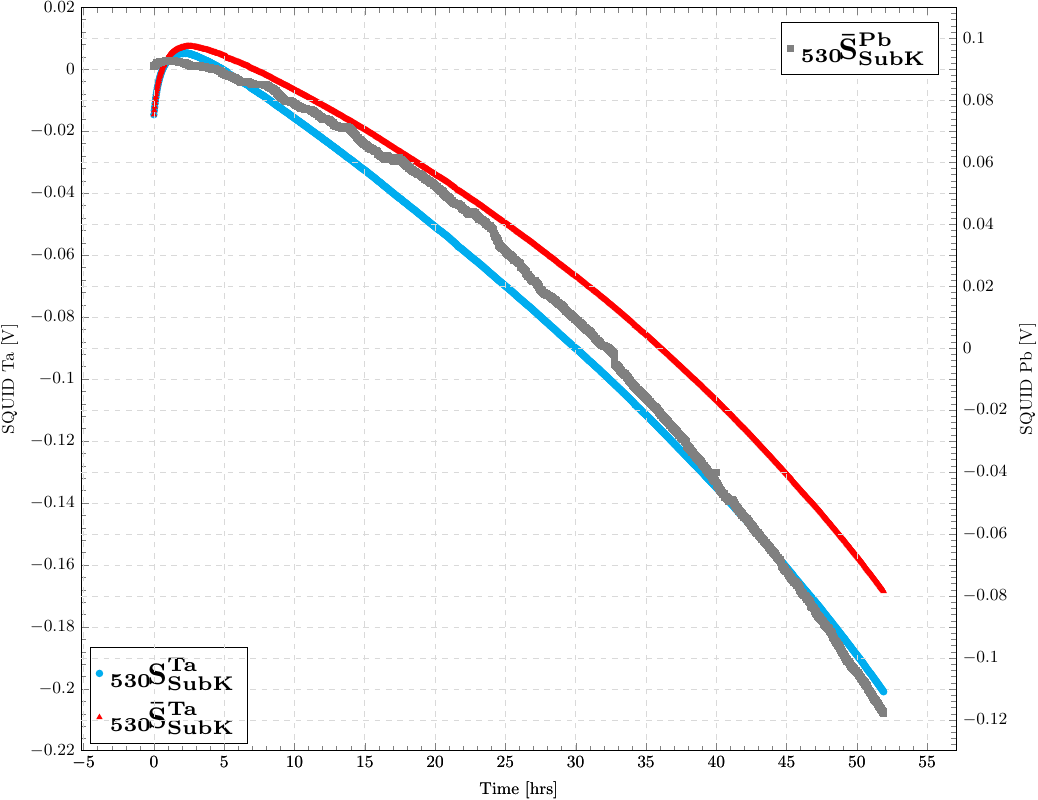}
&
 \includegraphics*{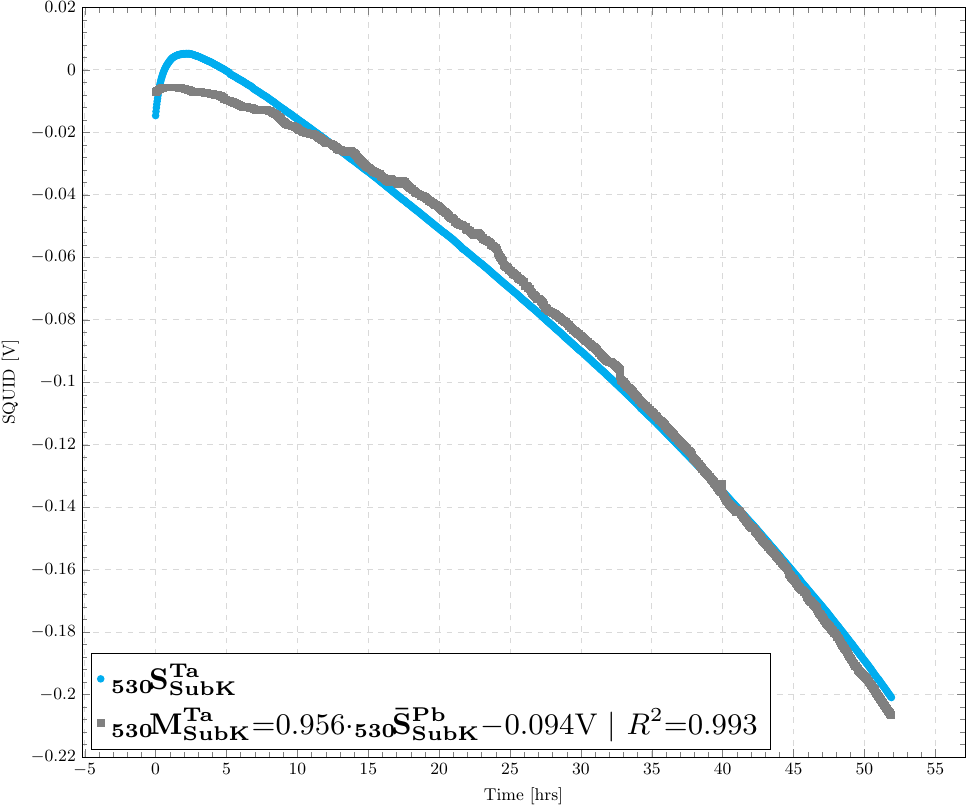}
&
 \includegraphics*{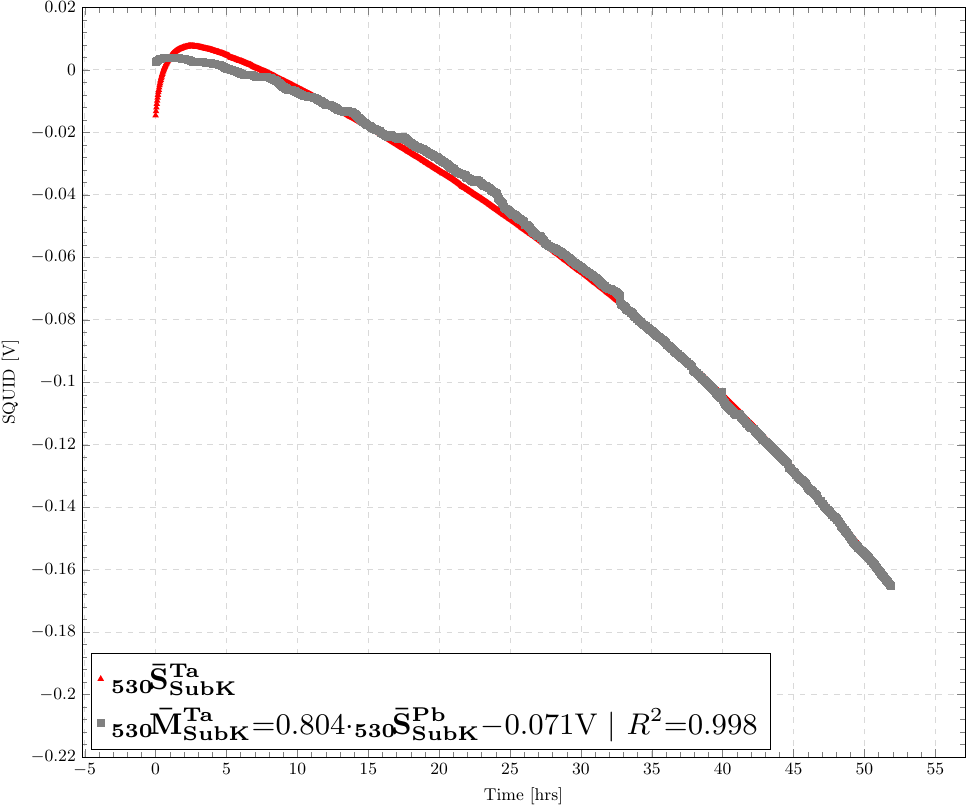}
\\
 \includegraphics*{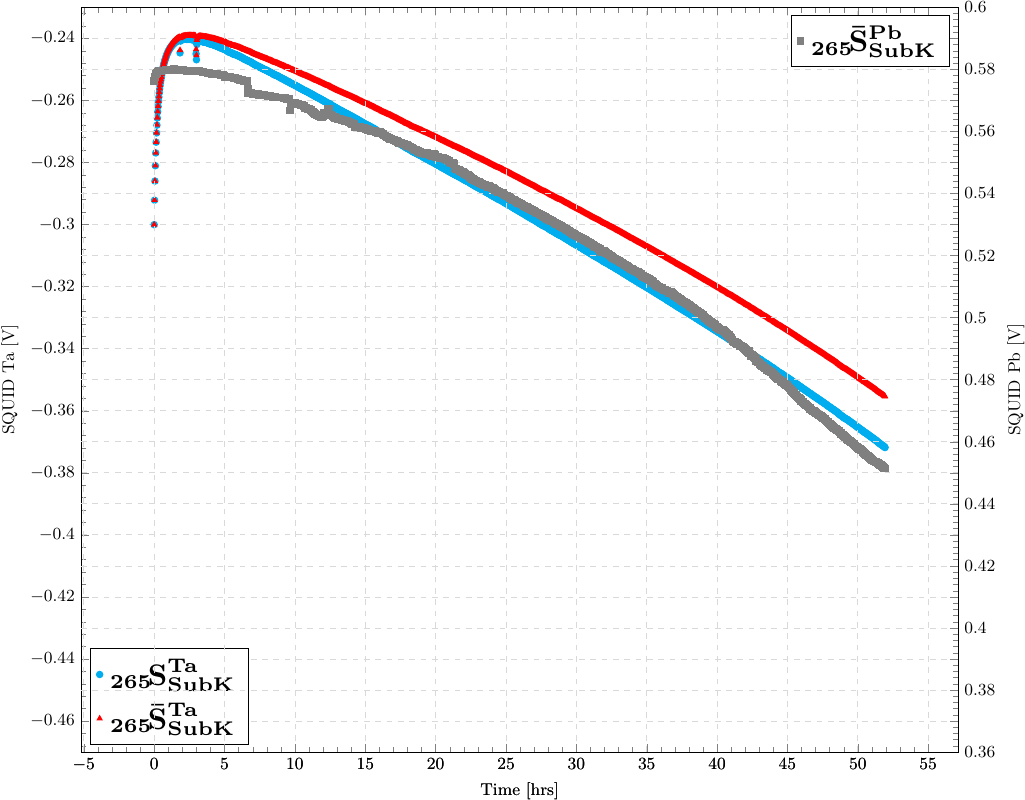}
&
 \includegraphics*{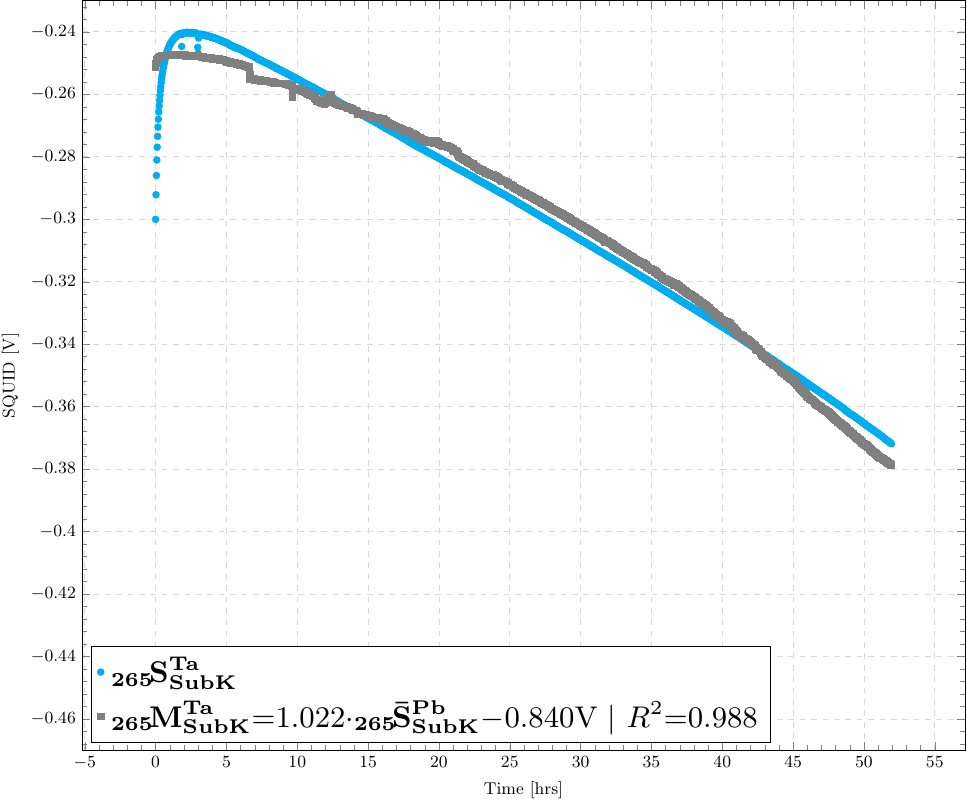}
&
 \includegraphics*{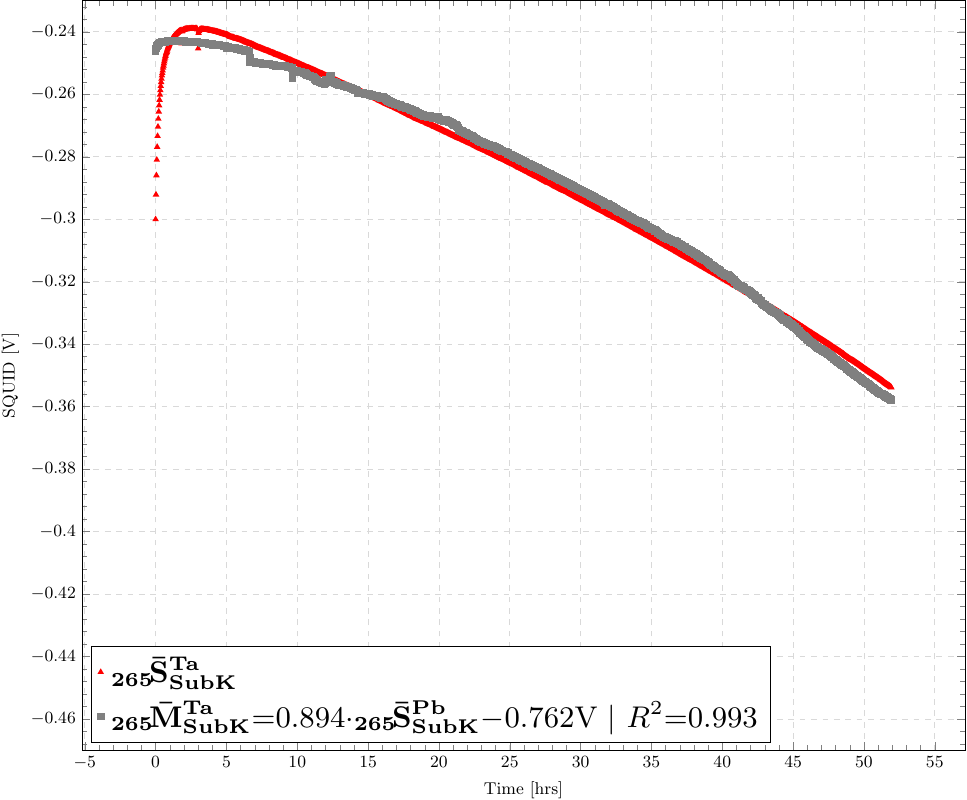}
\end{tabular}
}
%\vspace{-2.5cm}
\caption{Zero-field data for Ta and selected data points from Pb. The difference between the maximum and  minimum SQUID voltages, 0.24, is the same across all plots to make salient the dependence on \Ia: the smaller $\text{\Ia}$ has a smaller range of values for the same time domain. The blue and gray plots of the first column represent the zero-field data of Ta and Pb respectively, with the right-side ordinates providing the SQUID voltages for the Pb zero-field truncated datasets, \stcbar{I_a}{\text{\bf{SubK}}}{Pb}. The red plots of the first column correspond to the Ta sub-Kelvin data with the NEDM contribution, \Eq{intnEDM}, subtracted from each data point, i.e., \stcbar{I_a}{\text{\bf{SubK}}}{Ta}. The second column plots the linear regression model of the Ta data on the Pb data. The third column plots the linear regression model of the Ta-NEDM data on the Pb data.}\label{TaPbPlots}
\end{figure*}

\begin{figure}
\resizebox{8 cm}{!}{
\begin{tabular}{c}
  \includegraphics*{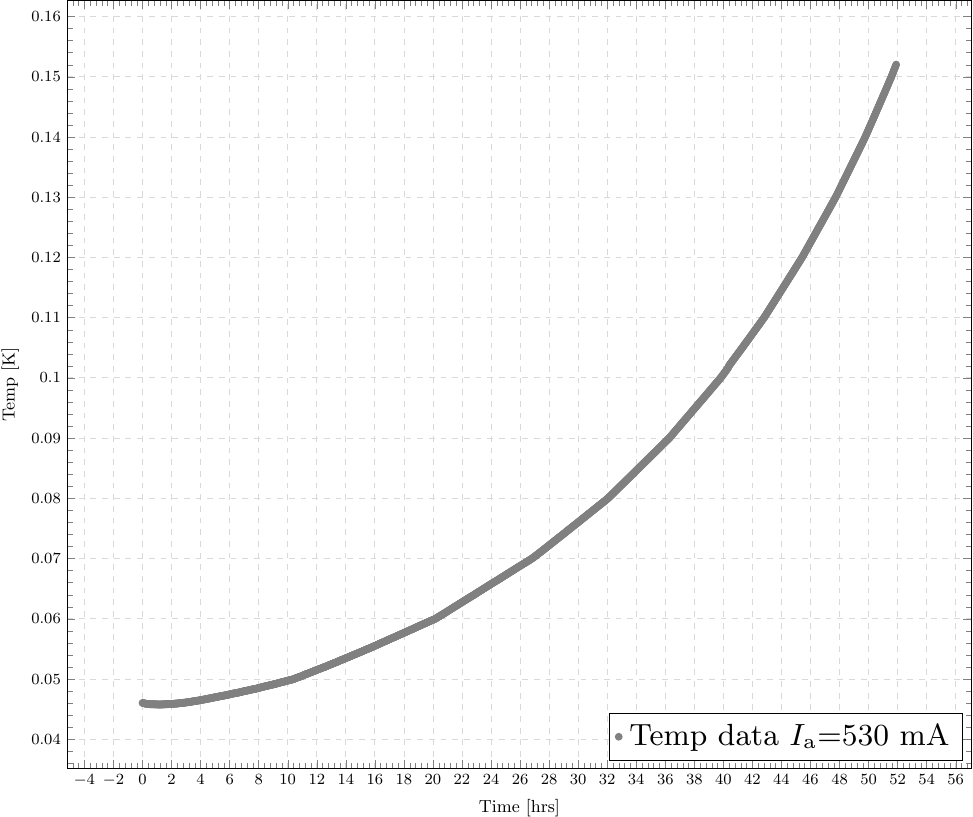}
\\
  \includegraphics*{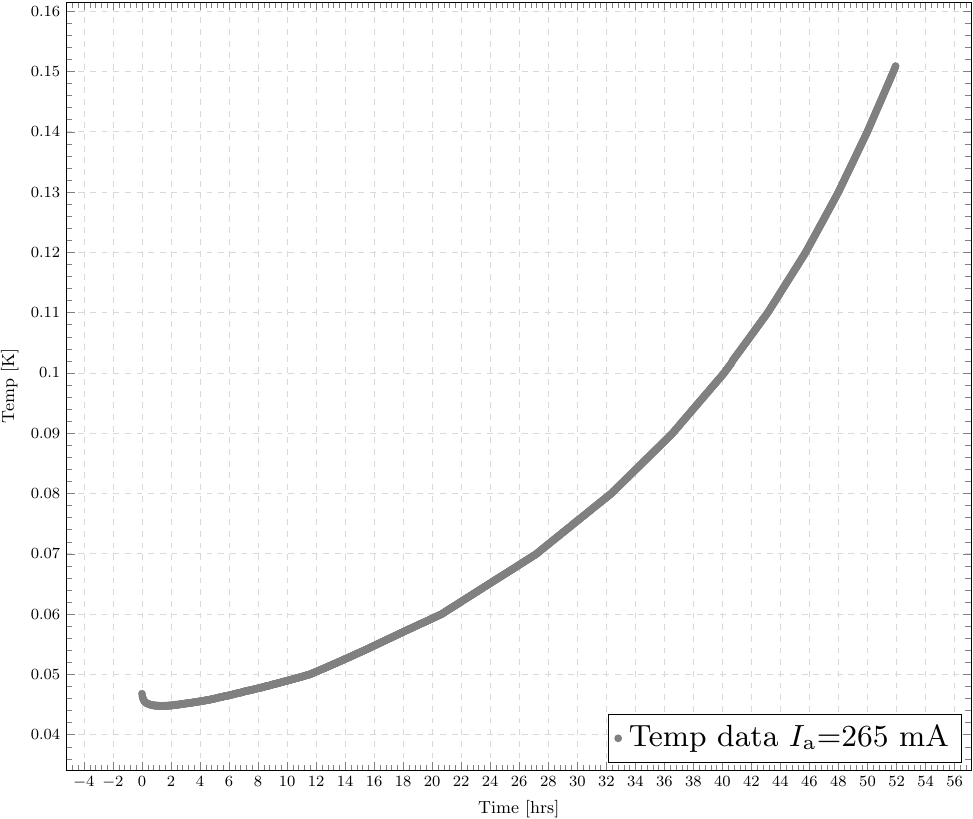}
\end{tabular}
}
%\vspace{-2.5cm}
\caption{Tantalum temperature data for the zero-field experimental runs at \Ia=530~mA, 265~mA.}\label{tempTa}
\end{figure}
\subsection{Systematics discussion}\label{sec:systematics}

The \Ia=0~A data in the first column of \Fig{fixedTempCurr-T} indicated the existence of correlated noise that generally trends up with time and may be mostly due to the decreasing pill solenoid current, \ips. However, comparison with the $\text{\Ia}\ne0$~A data showed  that this correlated noise was sub-dominant at all temperatures, and that the  $\text{\Ia}\ne0$ data would always eventually trend down with time, overcoming the 0~A correlated noise. This is particularly clear for the $T=52.5$~mK data where complete datasets exist for $\forall\text{\Ia}\in\text{\Iset{Ta}}$. Furthermore, the plots of the middle column of \Fig{fixedTempCurr-T} showed that the subtraction of the \stc{265}{T}{Ta} datasets from the \stc{530}{T}{Ta} datasets produced highly linear plots providing some confidence that  the correlated noise in the fixed temperature, differenced datasets had been suppressed.

At fixed temperatures, the \stc{265}{T}{Ta} and  \stc{530}{T}{Ta} datasets were collected at different times and had different cardinalities. This led to our criterion for choosing an offset to apply when subtracting a point from $a_i\in\text{\stc{265}{T}{Ta}}$ from a point $b_i\in\text{\stc{530}{T}{Ta}}$. However, \Fig{gammaTTaMod} shows that changing the offset by the equivalent of 3000 seconds results in all the points in \Fig{gammaTTaMod} moving up if the $\gamma_\text{max}$ slopes are used or down with the $\gamma_\text{min}$ slopes, while \gbar{I_\text{a}}{}{Ta} is relatively unaffected. Indeed, using $\gamma_\text{max}$ yields a \gbar{I_\text{a}}{}{Ta}=$-5.62\cdot10^{-9}$~VK/s while using $\gamma_\text{min}$ yields a \gbar{I_\text{a}}{}{Ta}=$-6.24\cdot10^{-9}$~VK/s both of which are well within a single standard deviation of result quoted in \Eq{gammabar}. Specifically, over  a range of time of 6000~s ($\pm3000$~s), the slope changed by $\pm0.31$~V$\cdot$K/s. This is a test of the robustness of the results.

A possible signal correlated with time is resistivity in the BiPb weld of the booster/sample solenoid, a possibility that can be taken as the null hypothesis for explaining the data. However, that possibility is rejected by the zero-field data. Indeed, energy dissipation in the solenoid weld would have occurred for both the Ta and Pb zero-field data and implies that \stcbar{I_a}{\text{\bf{SubK}}}{Ta} should be no better at explaining the \stcbar{I_a}{\text{\bf{SubK}}}{Pb} data than \stc{I_a}{\text{\bf{SubK}}}{Ta}. That hypothesis was found to have a probability of zero of being true. Removing $\int\text{\gbar{I_\text{a}}{}{Ta}}/T(t)\text{d}t$ from the Ta zero-field data significantly improved the correlation with the Pb zero-field data, in agreement with the NEDM theory presented at the beginning of this article.
The resistivity hypothesis is also inconsistent with the following fixed temperature data results:
\begi
The resistivity hypothesis predicts a linear dependence on $\text{\Ia}$ to leading order for both the Pb and Ta fixed temperature data. Only for the Ta data was this true in agreement with the NEDM prediction which says that the effect is proportional to the free magnetization.
The resistivity hypothesis predicts a linear dependence on the temperature to leading order\footnote{It is assumed that the resistivity function can be approximated by a Taylor expansion in power of $T$.} for small temperature variations. To test this, a linear regression on $T$ of the seven slopes of \Fig{gammaTTaMod} was performed and produced an inferior agreement with the data compared to the linear regression on $1/T$. Calculating the F-statistic by evaluating the ratio of the residual sum of squares of the $T$ and $1/T$ hypotheses, leads to the rejection of the $T$ hypothesis with a probability of 0.90. This is consistent with the zero-field data, which, to emphasize, is a much stronger test of the $1/T$ hypothesis since it involves a time integration over thousands of data points. These arguments apply to the possibility of resistivity in the sample wire welds.
\ei

Resistivity in the solenoid and sample wire welds having been rejected by the data, and an external signal during the Ta runs having been shown to be sub-dominant and suppressed in the differenced data \dstc{I_{\text{a}1}}{I_{\text{a}2}}{T}{Ta}{~m,n}, it is concluded that only intrinsic properties of tantalum and lead can explain the 1000+ hours of fixed temperature and sub-Kelvin data simultaneously. The data fits the theory presented in this paper to a high degree of confidence, and that intrinsic property must be an electrization field induced by the tantalum nuclear electric dipole moment, but suppressed in lead because of its comparatively small free magnetization. Using \Eq{gammabarM}, the tantalum nuclear EDM is (assuming a value \Isol=(1.8+2.0)/2=1.9~A):
\begin{align}
&|d_\text{e}^\text{Ta}|\!=\!(3.39\pm0.31_\text{stat})\cdot10^{-32}e\cdot\text{cm}\\
&|d_\text{e}^\text{Ta}|\!=\!(3.39\pm3.18)\cdot10^{-32}e\cdot\text{cm}>0,~\text{99.985\%CL}
\end{align}
In addition to the linear regression statistical uncertainty, there's also the fact that the Lakeshore RuOx thermometers (R102A-AA) installed in the ADR are interchangeable,\footnote{Interchangeable refers to the fact that the thermometer can be interchanged with another thermometer that follows the standard curve. The temperature is calculated from the resistance of the RuOx using Chebyshev polynomials up to 8th order.} with the resistance varying smoothly with temperature; the RMS deviation between the data and the standard curve fit is 0.5~mK for 0.05~mK$<T<$0.95~K. Hence, there may be a systematic uncertainty from the thermometer whose standard curve is within a band of $\pm0.01$~K\footnote{This uncertainty applies to the range 50mK$<T<$300mK. That uncertainty is likely a bias systematic because the RMS deviation with the standard curve is 0.5~mK and the range of temperature for our fixed temperature data is very small at 0.0525$\le T\le$0.105~K. If the reading of the thermometer at a specific temperature could be anything within the range $\pm$10~mK according to some probability density function, the RMS deviations of the data with the standard curve would likely be of that magnitude as well.}, as well as uncertainties on \Isol, the self-inductance of the sample wire and the mutual inductance between the sample wire and the SQUID pickup coil:
\begin{itemize}
\item Thermometer: The reported RMS deviations between test temperature data and the RuOx standard curve is $\pm0.0005$~K indicating that the RuOx resistance varies smoothly with temperature as confirmed in \cref{tempTa}. Fixed temperature data was collected for $0.0525\text{K}\le T\le0.105\text{K}$ and the smoothness of \cref{tempTa} combined with the tiny RMS imply that the uncertainty on the thermometer readings is likely systematic. Below are 4 extreme scenarios to add or subtract 0.01~K to the temperature and the resulting CL for a non-zero Ta nuclear EDM:
\begin{enumerate}
\item 0.01~K is added to all temperatures:  $|d_\text{e}^\text{Ta}|=(4.82\pm0.67_\text{stat})\cdot10^{-32}e\cdot\text{cm}$ and $(4.82\pm4.21)\cdot10^{-32}e\cdot\text{cm}>0$ at 99.85\%CL
\item 0.01~K is subtracted from all temperatures: $|d_\text{e}^\text{Ta}|=(2.69\pm0.34_\text{stat})\cdot10^{-32}e\cdot\text{cm}$ and $(2.69\pm2.12)\cdot10^{-32}e\cdot\text{cm}>0$ at 99.85\%CL
\item the steepest slope defined by compressing the temperature range according to the formula $T^\prime=T+0.01\text{K}(0.105+0.0525-2\cdot T^\prime)/0.0525$: $|d_\text{e}^\text{Ta}|=(6.44\pm1.01_\text{stat})\cdot10^{-32}e\cdot\text{cm}$ and $(6.44\pm6.33)\cdot10^{-32}e\cdot\text{cm}>0$ at 99.85\%CL
\item the shallowest slope defined by stretching the temperature range according to the formula $T^\prime=T-0.01\text{K}(0.105+0.0525-2\cdot T^\prime)/0.0525$:  $|d_\text{e}^\text{Ta}|=(2.33\pm0.28_\text{stat})\cdot10^{-32}e\cdot\text{cm}$ and $(2.33\pm1.73)\cdot10^{-32}e\cdot\text{cm}>0$ at 99.85\%CL
\end{enumerate}
Hence, regardless of how we add or subtract 0.01~K, the Ta nuclear EDM is always $>0$ to at least the 99.85\%CL. 
\item The $\text{\Isol}$ range when averaging the numerical and experimental values is $\text{\Isol}=1.9\pm0.1$A leading to an uncertainty of $\pm5$\% on $|d_\text{e}^\text{Ta}|$
\item The uncertainty on $M_\text{Ta-PC}$ is $\pm$15\% when moving the sample wire down $-0.5$~mm so that it touches the back of the SQUID or up +0.5~mm.
\item The uncertainty on \Lloop ~is $\pm$4\% due to the unknown orientation of the sample wire segment ``hook'' against the SQUID.
\end{itemize}

Assuming all nuclei/atoms have electric dipole moments of the same order of magnitude, this result is consistent with the upper limit obtained in Ref.~\cite{Graner2016} for $^{199}$Hg: $|d_\text{e}^\text{Ta}|<7.4\cdot10^{-30}e\cdot\text{cm}$.

One can also put an upper-bound on the $|d_\text{e}^\text{Pb}|$ by noting that such a limit should be proportional to the sensitivity of our experiment and to the ratio of the free magnetization of tantalum over lead. Using the statistical uncertainty on $|d_\text{e}^\text{Ta}|$ as representative of the sensitivity, we estimate for an upper-bound at the 95\% confidence level 
\beq
|d_\text{e}^\text{Pb}| \lesssim t_{95}\cdot0.31\cdot10^{-32}\frac{\Fedm{Ta}}{\Fedm{Pb}}=1.4\cdot10^{-31}
\eeq

\section{Discussion and future applications}\label{sec:disc}

\begin{center}
\begin{table}[t]
\begin{tabular}{|c|c|c|}
\hline
Isotope    & $T_\text{c}$(K)    &Compound ($T_\text{c}$)     \\
\hline
$^{1,2}$H & -- &  Hydrides, e.g. H$_3$S \\
& & ($\ge7$K, $\ge50$GPa)~\cite{Eremets2008,https://doi.org/10.48550/arxiv.1412.0460}    \\
\hline
$^{6,7}$Li & 4$\cdot10^{-4}$~\cite{Tuoriniemi2007} & C$_2$Li (1.9K)~\cite{Belash1989}    \\
\hline
$^9$Be &  0.026~\cite{Falge1967}  & --    \\
\hline
$^{10,11}$B & -- & YB$_6$ (8.4K)~\cite{Fisk1976,Szab2007}, \\
& & ZrB$_{12}$~\cite{Matthias1968}, MgB$_2$~\cite{Nagamatsu2001}    \\
\hline
$^{13}$C & -- & SiC:B (1.5K), SiC:Al (1.4K)~\cite{Muranaka2008}    \\
\hline
$^{15}$N & -- &  TiN (5.6K), ZrN (10K), NbN (16K) \\
& & ~Tab.V~in~\cite{Matthias1963}     \\
\hline
$^{17}$O & -- & TiO$_x$ ($\sim$1K), NbO$_x$ ($\sim$1K)~\cite{Hulm1972}    \\
\hline
$^{39}$K & -- & C$_3$K (3K), C$_6$K (1.5K)~\cite{Belash1989}    \\
\hline
$^{207}$Pb & 7.19 & --    \\
\hline
$^{209}$Bi & 5.3$\cdot10^{-4}$ & Bi$_{1-x}$Pb$_x$ ($\cong$9K for $x\cong0.5$)    \\
\hline
\end{tabular}
\caption{Nuclei where magnetic moments have been calculated in EFT and are superconducting either in their pure form or in a compound. Only stable, naturally occurring isotopes are included.}
\label{nucleiSuper}
\end{table}
\end{center}

In \cref{sec:theory}, \cref{Ebar,didt,betaSlope} were derived for the long range electrization field induced by the point nuclear EDMs in a superconductor. To experimentally verify this theory, three pieces of evidence were used
\begin{enumerate}
\item The current \Is~ in a sample Ta wire must be inversely proportional to the temperature.
\item \Is~ must be proportional to the solenoid magnetic field, i.e., it must be proportional to \Isol$\propto$\Ia.
\item The electrization field is proportional to the free magnetization, $\Fedm{ele}$, an element-dependent parameter that describes the amount of magnetization available to produce an electrization field. Hence, the electrization field must be suppressed in a metal with a suppressed free magnetization like Pb.
\item When performing a linear regression of Ta on Pb, removing the electrization field contribution from tantalum before performing the linear regression leads to a significantly better correlation with Pb then the linear regression where the electrization field contribution in tantalum is left intact.
\item Reversing the sample solenoid magnetic field also reverses the electrization field, a point that is related to the proportionality to $\text{\Ia}$ and to the fact that the electrization field is proportional to the nuclear magnetization.
\end{enumerate}
All of these benchmarks were experimentally shown to be true. Furthermore, it was also shown in \cref{sec:systematics} that neither external systematics nor resistivity could account for the experimental observations. Hence, a determination was made that the tantalum electrization field had been measured. Combining that measurement with \Eq{gammabarM} then provides an experimental value for the tantalum nuclear EDM. With this result, it is found that if nuclear EDM's are entirely due to the QCD $\tqcd$ term and that nuclear EDM's are of the same order of magnitude as those of the nucleons as suggested from lattice simulations of light nuclei~\cite{Dragos2021}, one obtains
\beq
\tqcd&\sim& 10^{-16}~\text{from Ta}\\
\tqcd&\lesssim& 10^{-15}~\text{from Pb}~.
\eeq
The $\tqcd$ upper-bound stemming from the nearly spherical $^{207}$Pb result may be more in line with the nucleon EDMs with the caveat that nuclear EDMs have contributions from CP-violating interactions like $\pi NN$.

Implications for supersymmetry have been obtained from the experimental constraints on the nuclear EDM of $^{199}$Hg~\cite{Graner2016,Nakai2017}. Extracting constraints from our tantalum nuclear EDM is not as simple as transferring those stemming from the experimental constraints on $^{199}$Hg~\cite{Graner2016} to the measured value of the tantalum nuclear EDM. Indeed, the strongest contribution to the $^{199}$Hg EDM is from the Schiff moment, specifically the CP-violating pion-nucleon interaction~\cite{Nakai2017}. This does not apply in a superconductor since the core electrons bound to the lattice sites no longer completely screen the nuclear dipole moment, even for pointlike nuclei. Implications for supersymmetry from the tantalum result are unfortunately beyond the scope of this paper but the supersymmetry model space is likely to be found further constrained in a future analysis.

\subsection{Large nuclear EDM dataset and CP-odd parameter constraints}

Near term plans include increasing the hold time of fixed temperature runs by improving the heat sinking at the 3~K and 1~K stages, followed by the measurement of the nuclear EDMs of indium, tin and aluminum. In addition, the sample assembly will be manufactured according to a more compact design in order to reduce its heat load, minimize uncertainties on the inductances, improve the magnetic shielding and the thermal contact with the 50~mK stage.  Depending on funding, the RuOx thermometer will also be matched.

In the longer term, the electrization field makes it possible to create a large dataset of nuclear electric dipole moments within a few years. Such a nuclear EDM dataset could then be used to over-constrain the free parameters of effective field theories (EFTs)~\cite{https://doi.org/10.48550/arxiv.2404.00516, Buchalla2014} where CP-odd operators are ordered in a chiral power counting scheme. The target nuclei for this project could be the same nuclei where EFT has been successfully used to calculate nuclear magnetic moments, typically light nuclei ($A\le10$~\cite{ChambersWall2024,weng2025}, $A<20$~\cite{Martin2023}) and doubly-closed nuclei $\pm$~one nucleon~\cite{Li2018}. A non-exhaustive list of these nuclei that exhibit superconductivity in pure form or in compounds is provided in \Tab{nucleiSuper}.

Some of these nuclei are superconducting only within a compound. However, if there are $n_\text{c}$ elements in the compound, and each of these element occupies a lattice site in the crystalline structure, one can think of the total electrization field of the compound as being the sum of the electrization fields induced by $n_\text{c}$ lattices. \Eq{betaSlope} can then be rewritten as
\beq\label{didtgen}
\EL=S_\text{edm} \sum_{i=1}^{n_\text{c}} F_\text{edm}^i  \bar{d}_\text{e}^i
\eeq
where the sum is over all the elements constituting the compound. The main theoretical requirement is knowing the average $N_{\text{c},i}$ for each element $i$ forming the compound. If all of the elements constituting the compound have a known EDM or no nuclear magnetic moment except for the target nuclei, then that EDM may be determined. For example, only 1.07\% of the natural abundance of carbon has a magnetic moment unlike potassium where 93.2\% of the natural abundance is $^{39}$K (see \cref{nucData}); it follows that C$_3$K on its own could provide an excellent measurement of $\bar{d}_\text{e}^\text{K}$. In combination with a measurement using the C$_6$K compound, both $\bar{d}_\text{e}^\text{K}$ and $\bar{d}_\text{e}^\text{C}$ could be determined. \Eq{didtgen} can also be used for a single element with different proportions of its isotopes to determine the EDM of each isotope as would be the case for lithium, to take one example.

Another example from \Tab{nucleiSuper} shows that boron, oxygen and nitrogen can be determined from compounds with niobium and zirconium. Both Nb and Zr are superconducting in their pure forms with critical temperature of 9.26~K and 0.55~K respectively. Thus, Nb and Zr have nuclear EDMs that are relatively easy to measure. Using ZrB$_{12}$ to measure the nuclear EDM of boron, one can then determine the nuclear EDM of magnesium from MgB$_2$ even though neither element is superconducting in its pure form.

$^1$H could provide an excellent measurement of the proton EDM keeping in mind the possibility of CP-violating electron-proton interactions, but at this stage, its compounds like H$_3$S become superconducting only under high pressure.\footnote{It is worth mentioning that only 0.76\% of sulfur has a non-zero nuclear magnetic moment.} The challenge of measuring the nuclear EDM of $^1$H using a sample assembly that includes a sample solenoid is that the sample solenoid would also be under high pressure. That challenge may be overcome if, instead of using an external magnetic field generated by a sample solenoid, the oscillations described in Appendix~A are used to determine the nuclear EDM, assuming their existence is confirmed.

\subsection{Measuring the electron EDM}

Cooper pairs with total spin $S=1,~L=1$ have their constituent electron spins aligned; therefore, their EDM's must also be aligned. Superconductors where the Cooper pairs have $S=1,~L=1$ quantum numbers are called $p$-wave superconductors, and the first strong candidate discovered was Sr$_2$RuO$_4$~\cite{Maeno1994,Mackenzie2003} with a $T_\text{c}=0.93$~K and a critical field of 0.067~T~\cite{Jerzembeck2023}, similar to tantalum. However, there exists a debate over whether Sr$_2$RuO$_4$ might in fact be a $d$-wave ($S=0,~L=2$) superconductor~\cite{Hassinger2017}. In 2019, the iron-based superconductor NdFeAs(O,F) was identified as a bulk $p$-wave superconductor with a $T_\text{c}=40.5$~K~\cite{Talantsev2019}. Other iron-based superconductors were also identified as likely $p$-wave superconductors confirming their existence further.

Each electron in a Cooper pair may have a non-zero pointlike EDM that generates contact and long range electric fields, $\mathbf{E}^\text{\bf e}_{\vlr}(\bm{r})$ and $\mathbf{E}^\text{\bf e}_{\bm{\delta}}(\bm{r})$ respectively, where $\bm{r}$ is the current location of the electron. Electrons cannot have contact interactions with lattice sites in a superconductor, implying that the lattice sites can never interact with $\mathbf{E}^\text{\bf e}_{\bm{\delta}}(\bm{r})$. On the other hand, an ion at a lattice site with charge $N_\text{c}e$ is subjected to the electric force due to $\mathbf{E}^\text{\bf e}_{\vlr}(\bm{r})$. By Newton's third law, the electron will also be subjected to a force of equal magnitude and opposite direction from the lattice site ion.
In a sample assembly where a solenoid carrying a current $\text{\Isol}$ generates a magnetic field in the penetration depth of a $p$-wave superconductor, an electrization field can appear with the form
\begin{align}\label{EDMcp}
&\EL=d_\text{e}\frac{N_\text{c}N_\text{CP}\sinh\left(\eta\right)\sinh\left(\frac{1}{2}\eta\right)}{3\epsilon_0\sinh\left(\frac{3}{2}\eta\right)} \\
&\eta\equiv \frac{(g_\text{e}-1)\mu_\text{B}\mu_0N_\text{sol}\text{\Isol}}{k_\text{B}T}
\end{align}
where $g_\text{e}$ is the electron gyromagnetic factor, $\mu_\text{B}$ is the Bohr magneton, $d_\text{e}$ is the electron electric dipole moments, $N_\text{CP}$ is the number density of Cooper pairs and the other parameters are as before. \Eq{EDMcp} is derived in \cref{app:avrMagMom}. Note that Curie's law no longer applies for the parameters used in this experiment, $N_\text{sol}$=1754, $T\sim0.05$~K and \Isol$\sim2.5$~A, because the magnetic moment of the electron is two thousand times larger than a Bohr magneton. In fact, for those parameters, the hyperbolic tangent equals 1, and in order to verify the existence of an electrization field, the experiment would have to be performed at lower magnetic fields $\sim0.001$ or higher temperatures, $T\sim0.5$~K. As before, the energy for the work performed on the supercurrent by the electrization field ultimately comes from \Isol.

The Standard Model of particle physics predicts an electron EDM of the order of $10^{-38}~e\cdot$cm and the current upper bound on the magnitude of the electron EDM is $4.1\cdot10^{-30}~e\cdot$cm~\cite{Roussy2023} obtained using electrons subjected to the intra-molecular electric fields of an HfF$^+$ ion trap.

The work presented in this paper demonstrated a nuclear EDM sensitivity of at least $10^{-32}~e\cdot$cm. The corresponding sensitivity for an electron EDM is of the order of $10^{-35}~e\cdot$cm because the fraction of electrons aligned with the magnetic field will be of the order of $10^3$ times larger from the ratio of the electron magnetic moment to the Bohr magneton. A measurement at that sensitivity would severely constrain the parameter space for the Higgsino and CP-violating phases~\cite{Cesarotti2019,Kaneta2023}. In practice, the standard model limit could be reached through further noise reduction and longer run times. In order to use NdFeAs(O,F) to measure the electron EDM, $d_\text{e}^\text{Fe}$, $d_\text{e}^\text{As}$, and $d_\text{e}^\text{Nd}$ could be determined by measuring the electrization field of other iron-based superconductors to over-constrain all the necessary parameters, including the electrization field of the $S=1$ Cooper pairs\footnote{It's important to note that arsenic is a semiconductor, not a metal and that its free magnetization will be more suppressed due to Schiff screening. Hence, a measurement of $d_\text{e}^\text{As}$ would require significantly higher sensitivity.}. Alternatively, if FeSe$_{0.5}$Te$_{0.5}$ is a $p$-wave superconductor, it could be used `out of the box' since the free magnetization of Fe, Se and Te are all very suppressed leaving only the electrization field of the electrons. Indeed, $^{57}$Fe, $^{77}$Se, $^{123}$Te and $^{125}$Te are the only isotopes with a nuclear magnetic moment, and their $\Fedm{Fe,Se,Te}$ are all suppressed by their abundances ($<8$\%) and their magnetic moments ($<0.9\mun$). In addition, selenium and tellurium are semiconductors and are further suppressed by Schiff screening with $N_\text{c}<1$. Ultimately, it may be necessary to use elements where the isotopes with $\mun\ne0$ are further suppressed to reach the required sensitivity to measure the electron EDM.

\subsection{Finite size nuclei}

In this work, the nucleus was treated as pointlike and the potential impact of spatially extended nuclei with near degenerate states of opposite parity within a relativistic context was not considered. Since valence electrons in a superconductor are no longer bound to lattice sites and the atomic electrons can no longer completely screen a pointlike nuclear electric dipole, the impact of Schiff moments may be sub-leading. However, Schiff moments are still present when $N_\text{c}\ne0$ which brings up the possibility that octupole enhancements could be detectable using the electrization field. In Ref.~\cite{Engel2000}, it was found that nuclei with strong octupole vibrations could have enhanced Schiff moments; as such, long-lived actinides would make good candidates to detect them~\cite{Flambaum2020}. In particular, $^{235}$U and $^{231}$Pa have non-zero natural abundances and are superconducting in their pure form and/or in compounds, with $T_\text{c}\sim$1~K. Hence, the method outlined in this current work could be used to measure atomic electric dipole moments for comparison with theoretical expectations. One challenge is that these isotopes are rare, regulated, toxic elements leading to additional expense as well as non-trivial technical, administrative and safety challenges for fashioning of appropriate sample assemblies.

However, if future experiments have the sensitivity to measure the atomic EDMs of insulators and semiconductors that are part of superconducting compounds and molecules, that data could be critical to understand how Schiff screening works theoretically.

\subsection{Implications for axion searches}\label{sec:imp}

The current work's main purpose was to describe the theoretical foundations of the electrization field and experimentally demonstrate that supercurrents are highly sensitive to nuclear EDMs. That sensitivity leads to the question of whether a supercurrent under the influence of an electrization field can be used to detect axion-like particles (ALPs).

The experimental results presented in this work prove that the supercurrent is extremely sensitive to the electrization field whose strength is determined by the nuclear EDM. In particular, the electromotive force of \Eq{didtgen} may have a contribution from the axion background field superimposed on the electromotive force due to the nuclear EDM. The axion-induced EDM electromotive force would have a time dependence from two sources: one that comes from the axion-gluon coupling~\cite{Pospelov1999,Abel2017} and another that comes from variations of the local axion density as Earth rotates on its axis and orbits the sun~\cite{Sikivie2002,Przeau2015}.

The background axion field will oscillate at a frequency of $m_\text{a}c^2/\hbar$ where $m_\text{a}$ is the axion mass. The oscillation amplitude will be proportional to the square root of the local dark matter density, the axion coupling constant with gluons, $C_\text{G}$, and inversely proportional to the axion decay constant, $f_\text{a}$. In Ref.~\cite{Abel2017}, the authors were able to impose constraints on $C_\text{G}/f_\text{a}$ from the cold neutron experiments at Sussex–RAL–ILL~\cite{Baker2014} and the follow-up experiment at PSI~\cite{Baker2011} that had a sensitivity to the neutron EDM down to approximately $10^{-26}e\cdot\text{cm}$ by searching for shifts in the Larmor frequency for opposite orientations of an applied electric field. The authors of Ref.~\cite{Abel2017} were able to impose their constraint on $C_\text{G}/f_\text{a}$ by obtaining a null result on finding an induced EDM in the Sussex–RAL–ILL~\cite{Baker2014} and  PSI  Larmor frequency data. 

However, the supercurrent driven by the nuclear EDM electrization field is orders of magnitude more sensitive to nuclear EDMs than other existing methods. This sensitivity opens up the possibility of much greater constraints on axion parameters if not outright discovery. Possible experimental concepts for the search of axions or ALPs include:
\begin{enumerate}
\item Searching for an unexplained frequency in the power spectrum of the supercurrent data observed by different groups at different locations. If that frequency corresponds to an ALP mass that has not been excluded by other data, it may be an indication of a cosmic ALP background.
\item A search for an electrization field due to an ALP flux stemming from the galactic wind or the sun. This flux could be detected using two sample assemblies with opposite sample solenoid magnetic fields that are aligned and anti-aligned with the purported ALP flux: any positive detection in the differenced data between the two assemblies would be proportional to the ALP flux.
\item A search for density fluctuations of the ALP background due to the gravitational concentration that can produce dense dark matter hairs~\cite{Przeau2015}. The right ascension and declination of the galactic axion dark matter flux relative to the solar system are approximately 138$^\circ$ and $-48^\circ$ respectively. Hence, the azimuths of experiments located near latitudes 48$^\circ$N/S could coincide once daily with the axis parallel to the background ALP wind where one is most likely to find dark matter hairs at the surface of the Earth. Experiments would search for a daily periodic ``bump'' in the SQUID data that would occur at the local sidereal time corresponding to the right ascension of the axion wind, around 9h12m in the northern hemisphere with a yearly modulation accounting for the Earth's orbit around the sun. Collecting data for a sample wire sensitive to axions simultaneously with a sample wire whose sensitivity to axions is suppressed as a control could help confirm evidence for ALPs.
\end{enumerate}

\section{Summary and conclusions}

Over a thousand hours of data was collected to make a first non-zero detection of a permanent nuclear electric dipole moment, $|d_\text{e}^\text{Ta}|=3.39\cdot10^{-32}e\cdot\text{cm}$, as well as put an upper bound on the control metal electric dipole moment of $^{207}$Pb, $|d_\text{e}^\text{Pb}| \lesssim 1.2\cdot10^{-31}$. This was achieved in a superconducting sample assembly where an electromotive force in tantalum was shown to follow Curie's law but not in the control sample made of lead. Furthermore, it was shown that when the integrated temperature-dependent effect was removed from the tantalum zero-field data points, they now correlated very closely with the lead zero-field data points. External fields as well as a possible infinitesimal, residual resistance in the superconducting welds were shown to be sub-dominant or contradict the observed data.

In conclusion, it was found that the electrization field is able to achieve a sensitivity to the CP-odd nuclear EDM several orders of magnitude beyond current techniques, and seems poised to contribute decisively to the discovery of a vast number of CP-odd phenomena.

\section{Acknowledgments}

I am forever grateful to my wife Kerry, and our children, Marcel and Edith, for their continuous support, patience and understanding of the many weekends I spent at the lab over the last five years. I dedicate this paper to them, and to the hundreds of millions of individuals currently lost to science because of the systemic forces that steer them away from STEM fields or from education altogether.

\begin{appendices}

\section{Potential of electrization field and the Schr\"{o}dinger equation}\label{app:potential}

In the main text, the discussion was in terms of the electric field and its long range component referred to as the electrization field. However, the same conclusions can be drawn from the potential itself, including that the total potential can be split into a contact component, $V_\delta(\theta)$ and a long range component, $V_\lambda(\theta)$ that cancel. Working in 1-dimension with a circular conductor of radius $R$, we must have
\beq
V(\theta)=V_\delta(\theta)+V_\lambda(\theta)=0\Rightarrow V_\lambda(\theta)=-V_\delta(\theta)
\eeq
$V_\delta(\theta)$ can be calculated from a line integral of the contact electric field along a circular arc $\Delta\theta$. Including a screening factor $\sigma_\text{sf}$, a number density of nuclei $N_\text{N}/V$, and setting $V_\lambda(\theta)=0$ at $\theta=0$ gives
\begin{align}
& \mathbf{E}_{\bm{\delta}}\!=\!-\frac{N_\text{N}}{3\epsilon_0V}\sigma_\text{sf}\den\!\sum_i\delta(\theta)\bm{\hat{\theta}} \\
&V_\lambda(\theta)\!=\! -V_\delta(\theta)\!=\!-\int_0^\theta \mathbf{E}_{\bm{\delta}}\cdot\bm{\hat{\theta}}\text{d}\theta\!=\!\frac{N_\text{N}}{3\epsilon_0V} \sigma_\text{sf}\den R\theta~.
\end{align}
This is the RHS of \Eq{betaSlope} after including the factors from Curie's law and dividing by \Lloop. Hence, in a superconductor, only $V_\lambda(\theta)$ is perceived by the Cooper pairs and the long range potential is proportional to $\theta$.

The 1-d Schr\"{o}dinger equation for a particle of mass $m$ and charge $q$ evolving in $V_\lambda(\theta)$ is
\beq\label{schrodinger}
& &-\frac{\hbar^2}{2mR^2}\frac{\text{d}^2\psi(\theta)}{\text{d}\theta^2}+\kappa\theta\psi(\theta)=E\psi(\theta) \\
& &\kappa\equiv \frac{N_\text{N}}{3\epsilon_0V} \sigma_\text{sf}q \den R~.
\eeq
Making the change of variables $\theta=\theta_0\theta^\prime$ where $\theta_0=[\hbar^2/(2mR^2\kappa)]^{1/3}$ and $\epsilon=(2mR^2\theta_0^2/\hbar^2)E$
\beq
-\frac{\text{d}^2\psi}{\text{d}\theta^{\prime^2}}+(\theta^\prime-\epsilon)\psi=0
\eeq
With the final change of variable $x=\theta^\prime-\epsilon$, this is the Airy equation with known transcendental solutions. This problem is slightly different from the real physical system since the magnetization and electrization fields are  postulated to have an fixed orientation and the potential is time independent. In the real problem, the magnetization's direction is determined by a magnetic field generated by a solenoid supercurrent that also provides the energy of the sample wire supercurrent and where the real electrization potential depends on time. In this toy model where the electrization potential is unchanging, the total mechanical energy is a constant of motion and the stationary solutions of \Eq{schrodinger} can be used to build wave packets to model the real problem on a time scale where the real electrization potential remains constant according to a predetermined criterion.

The Airy solution $\text{Bi}(\theta)$ is divergent leaving as the only valid solution $\psi(\theta)=a\text{Ai}(\theta/\theta_0-\epsilon)$ where $a$ is a normalization constant. Since the potential is not periodic in $\theta$ neither are the stationary solutions implying that the wave function is not single-valued at a specific $\theta$ limited to the range $0\le\theta<2\pi$. The potential energy of the charged particle is not single valued since it depends on the path it took to arrive at angle $\theta$, whether it went around the circuit once or ten times. This problem is exactly analogous to an infinite plate with a surface charge density that is located at $z\rightarrow-\infty$ and produces a constant electric field that fills all space, and a corresponding potential that is proportional to $z$ with $V(z)=0$ at $z=0$. The probability density for finding the particle at an angle $\theta$ is then given by the sum over all the possible paths that could have led it to that angle $\theta$
\beq
P(\theta)=\sum_{n=-\infty}^\infty |\psi(\theta+2\pi n)|^2, ~n\in \mathbb{Z},~0\le\theta<2\pi
\eeq
Boundary conditions can quantize the energy. For example, imposing the momentum initial value $\psi^\prime(0)=0$ leads to the quantization of $\epsilon$ since only those values satisfying $\text{Ai}^\prime(-\epsilon$)=0 are allowed.

\section{\scalebox{1.25}{\Isol}=0 oscillating solutions}\label{app:osc}

Consider the sample current vector in cylindrical coordinates ($\rho,\phi,z$) where $\rho$ is the distance from the wire axis, $\bm{\hat{z}}$ is the direction co-axial with the wire axis, and $\bm{\hat{\phi}}$ is the azimuthal angular direction
\beq
\mathbf{I}_\text{\bf{s}}=I_\phi\bm{\hat{\phi}}+I_\text{z}\mathbf{\hat{z}}
\eeq
where $I_\text{z}$ is what we previously wrote as \Is, the current coaxial to the sample wire axis, and $I_\phi$ is the current in the $\phi$-direction that . \Eq{energyCons} is then rewritten as
\beq\label{energyCons2}
\frac{\text{d}~}{\text{d}t}\left(\frac{1}{2}L_\text{z}I_\text{z}^2+\frac{1}{2}L_\phi I_\phi^2\right)=-\frac{\text{d}~}{\text{d}t}\left( \frac{1}{2}\text{\Lsol}I_\text{sol}^2 \right)
\eeq
where $L_\phi$ is the self inductance corresponding to the flux in the penetration depth due to $I_\phi$ and $L_\text{z}$ is the self-inductance for the circuit loop. In the absence of an external magnetic field, we can set \Isol=0. In the absence of an external magnetic field, the electrization field can not do work on $\mathbf{I}_\text{\bf{s}}$ as the direction of the field is always perpendicular to the direction of the sample wire current. The solution to \Eq{energyCons2} is an ellipse where the integration constant is the total field energy. An estimate of $L_\phi$ can be obtained in a London-inspired phenomenological model where the magnetic field depends only on $\rho$ and is independent of $\phi$ and $z$ 
\beq
& &\frac{1}{2}L_\phi I_\phi^2 = \frac{1}{2\mu_0}\int_0^{R_\text{w}} B^2(\rho)2\pi l_\text{w}\rho\text{d}\rho\\
& &B(\rho)=\mu_0 e^{-(R_\text{w}-\rho)/\lambda_\text{L}} \frac{I_\phi}{l_\text{w}} \\
& &\Rightarrow L_\phi=\frac{u_0\pi R_\text{w}\lambda_\text{L}}{l_\text{w}}
\eeq
where $R_\text{w}$ is the sample wire radius, $\lambda_\text{L}$ is the penetration depth and $l_\text{w}$ is the length of the  wire. Taking as initial values $\mathbf{I}_\text{\bf{s}}=I_0\hat{\text{z}}$ the solutions are
\begin{align}
&I_\text{z}=I_0\cos(\omega_\text{EDM}^0t)\\
&I_\phi=I_0\sqrt{\frac{L_\text{z}}{L_\phi}}\sin(\omega_\text{EDM}^0t)\\ \label{omegaedm}
&\omega_\text{EDM}^0\equiv \frac{\denbar \mu_0^2c^2}{\sqrt{L_\text{z}L_\phi}}    \left( \frac{N_\text{c}}{Z} \right)\frac{(I_\text{N}+1)}{I_\text{N}} \frac{a_\text{b}N_\text{N}}{9} \frac{\mun}{k_{\text{B}}T} 
\end{align}
As long as $|\mathbf{I}_\text{\bf{s}}|$ does not exceed the critical current of the superconductor, such oscillations in the currents could be detectable.

Hence, consider the persistent current data  in \cite{File1963} with the objective to check if a nuclear EDM electrization field could explain the variation in their data. For their experiment, they built a solenoid of 984 turns with a diameter of 0.1016~m, a length of 0.254~m using a Nb-25\%Zr wire with a diameter of 0.5~mm. The inductance of their solenoid was not provided in their paper, but was numerically calculated to be $L_\text{z}$=0.0326~H. The initial magnetic field inside the center of their solenoid was 0.210488~T. They performed a linear regression fit of their data and measured the following slopes for two runs:
\beq
&m_{21}\!=\!(1.674\pm0.19)\cdot 10^{-10}~\text{T/hr} ~\text{21d}~(504\text{hrs})~~~\\
&m_{37}\!=\!(2.022\pm0.15)\cdot 10^{-10}~\text{T/hr} ~\text{37d}~(888\text{hrs})~~~\\
&B_0\!=\!0.210488~\text{T}
\eeq 
From the data of their first run and  assuming a temperature of 4~K, the corresponding frequency $\omega_\text{EDM}^0$ is  found to be
\begin{align}
&\omega_\text{EDM}^0\!=\!\frac{\text{arccos}(1-m_{21}504\text{s}/B_0)}{540\cdot3600\text{s}}\!=\!4.9\cdot10^{-10}~\text{Hz} \\
&\bar{d}_e^\text{Nb}\!=\!3.5\cdot10^{-28}e\cdot\text{cm}
\end{align}
where the Nb nuclear EDM was calculated using \Eq{omegaedm}, the Nb values of \Tab{nucData} and $L_\phi=1.26\cdot10^{-19}$. $\bar{d}_e^\text{Nb}$ is 4 orders of magnitude larger than the Ta nuclear EDM making the explanations provided by the authors far more likely\footnote{The authors of \cite{File1963} speculated that this observation may be due to resistance of the spot weld or mechanical expansion of the wire due to magnetic radial pressure.}.

It is also worthwhile estimating the impact of $\omega_\text{EDM}^0$ on the Ta data. For our experiment we have $L_\phi=7.8\cdot10^{-17}$ and $L_\text{z}$=\Lloop. Using our measured value for the Ta nuclear EDM of $3.1\cdot10^{-32}e\cdot\text{cm}$,  $\Delta t$=48~hrs,   \Is=1$\mu$A, and $T=0.1$~K yields
\begin{align}
&\omega_\text{EDM}^0\!=\!1.4\cdot10^{-11}~\text{Hz} \\
&\frac{\Delta\text{\Is}}{\Delta t}\!=\!\text{\Is}\frac{\cos(\omega_\text{EDM}^0\Delta t)-1}{\Delta t} \sim -10^{-23}~\text{A/s}
\end{align}
which is entirely negligible. However, the way to measure this effect is clear: use a large \Is, a low temperature and suppressed self-inductances $L_\phi$ and $L_\text{z}$. This could be achieved, for example by inducing a current in a wire sample that encircles a  cylinder made of a superconducting material with a low free magnetization like Zn, and lower the temperature below its critical temperature.

\section{Fixed temperature standard deviations and signal discontinuities}\label{app:Tstddev}

\begin{figure*}
\resizebox{17 cm}{!}{
\begin{tabular}{ccc}
  \includegraphics*{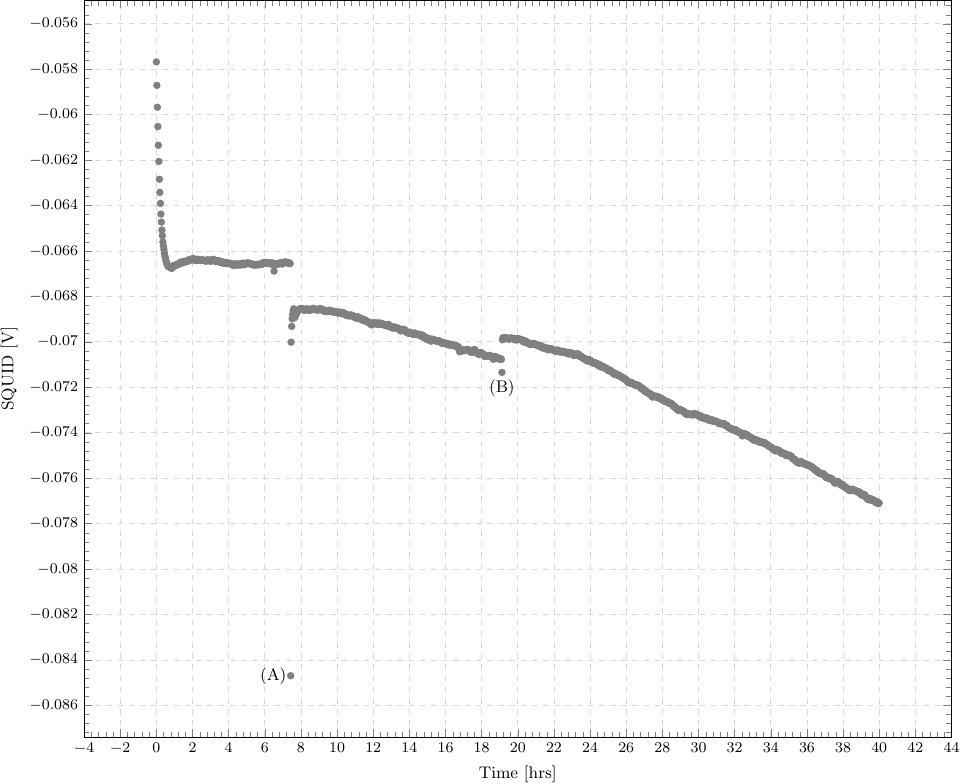}
&
  \includegraphics[width=15.8cm]{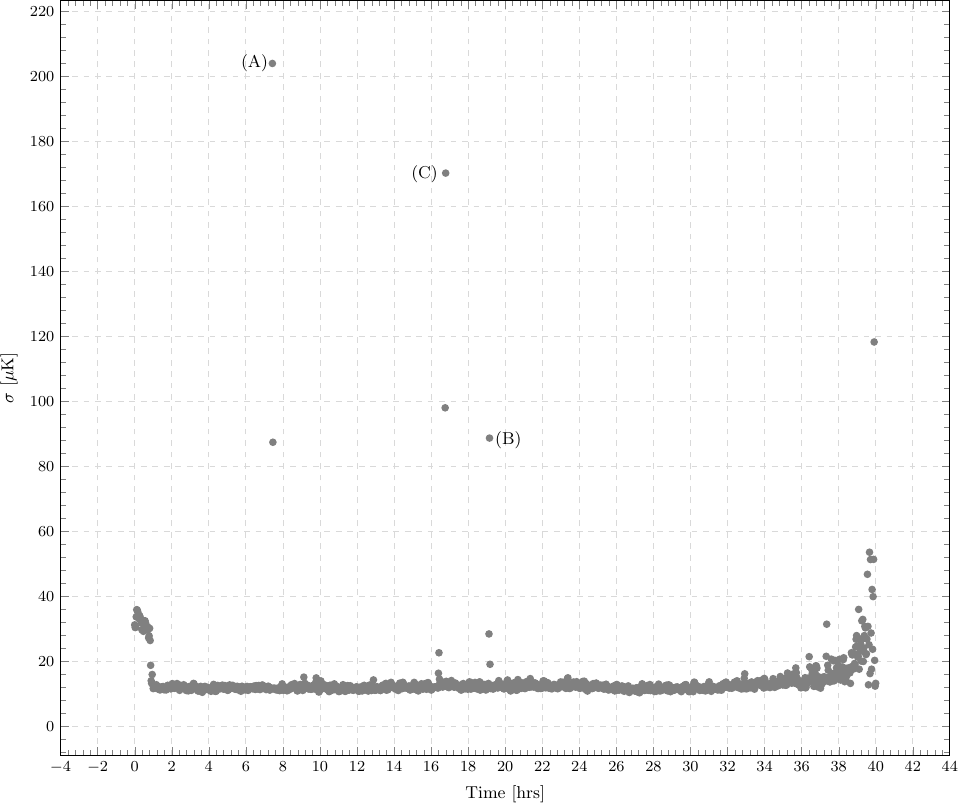}
&
 \includegraphics*{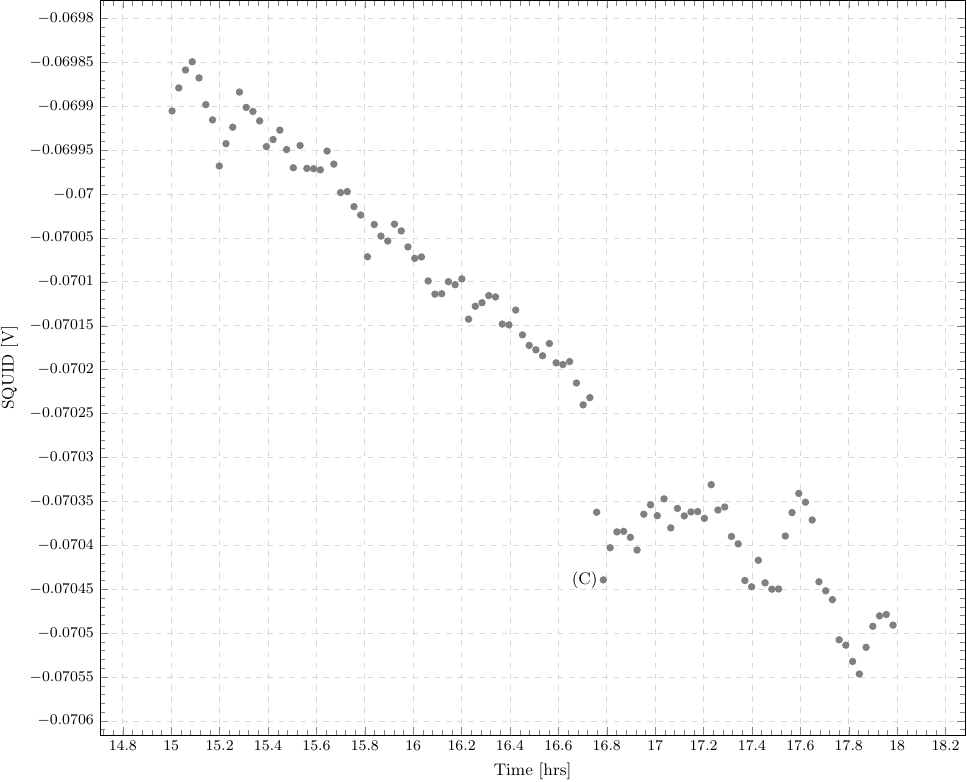}
\end{tabular}
}
\caption{Ta standard deviation plots for $T$=0.075~K, \Ia=0.53~A. The standard deviation was in the range $10\mu\text{K}<\sigma<15\mu$K (other than the jumps and the end of the run) for all the Ta fixed temperature data runs. }\label{SDplots}
\end{figure*}

\begin{figure}
\centering
\resizebox{7 cm}{!}{
 \includegraphics*{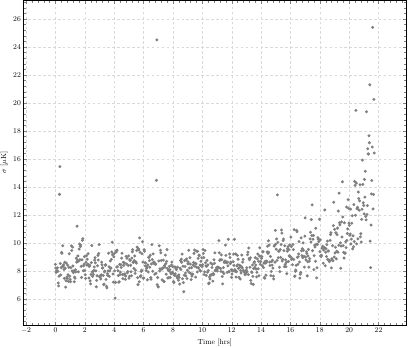}
}
\caption{Pb standard deviation plot for $T$=0.0525~K, \Ia=0.542~A. The impact of a better thermal contact during REGEN is evident in the general trend of the standard deviation $\sigma<10~\mu$K compared to Ta, as well as the much smaller jumps.}\label{TsetSDPb}
\end{figure}

\begin{center}
\begin{table}[t]
\begin{tabular}{|l|l|l|}
\hline
\stc{530}{52.5}{Ta}(1)    &\stc{265}{52.5}{Ta}(0)    &\stc{0}{52.5}{Ta}(0)     \\
\hline
\stc{530}{65}{Ta}(0)    &\stc{265}{65}{Ta}(0)    &\stc{0}{65}{Ta}(0)     \\
\hline
\stc{530}{75}{Ta}(2)   &\stc{265}{75}{Ta}(0)     &\stc{0}{75}{Ta}(1)    \\
\hline
\stc{530}{85}{Ta}(2)   &\stc{265}{85}{Ta}(0)    &\stc{0}{85}{Ta}(0)    \\
\hline
\stc{530}{105}{Ta}(0)    &\stc{265}{105}{Ta}(0)   &\stc{0}{105}{Ta}(0)     \\
\hline
\end{tabular}
\caption{Number of offsets in parentheses applied to each Ta dataset. None of the Pb datasets have offsets except for  \stc{134}{75}{Pb} with 1 offset. None of the zero-field datasets have offsets.}
\label{offsetsNum}
\end{table}
\end{center}

During a data collection run, discontinuities in the SQUID signal occasionally appeared. Most of the discontinuities were matched to abrupt temperature variations visible as sudden jumps in the temperature standard deviation; these discontinuities occurred only during fixed temperature runs. The remainder of the discontinuities were due to ``flux jumps'' of the SQUID~\cite{STARCRYO}.

For example, the SQUID dataset \stc{530}{75}{Ta} run had two, relatively small sudden jumps labeled (A) and (B), and they were both matched to jumps in the temperature standard deviation as shown in \Fig{SDplots}, also labeled (A) and (B),  plotted in the left and middle figures. These discontinuities induced by abrupt changes in the temperature feature a `tail' during which the system returns to equilibrium. Sometimes that tail is very pronounced as was the case for point (A) while at other times less so. This difference may be due to the fact that each data point is an average of raw data over 100~s. On occasion, a temperature jump like point (C) in the middle figure does not result in an obvious change in the SQUID signal. However, if one blows up the region in the SQUID signal that should have been affected by that jump, it is clearly there as shown in the rightmost plot in \Fig{SDplots} where the discontinuity and tail are small but visible. In order to obtain the corresponding plot in the third row, first column of \Fig{fixedTempCurr-T}, offsets of -0.002427304~V and 0.000939462~V were subtracted from points 269 and 688 $\in$ \stc{530}{75}{Ta}  respectively, and all subsequent points. Applying these offsets allowed the use of all the data points $\in$\stc{530}{75}{Ta} in a single linear regression, as opposed to applying the linear regression to three subsets of \stc{530}{75}{Ta} which increases the standard error.

The offsets were chosen to restore continuity  between the \stc{530}{75}{Ta} plot segment anterior to the jump and the \stc{530}{75}{Ta} segment posterior to the jump. For the signal jump (A), that meant setting \stc{530}{75}{Ta} data point 271 equal to  data point 267 while for signal jump (B), the offset was determined by making data point 690 equal to data point 687. Note that these jumps were very rare. For example, these two jumps occurred in a data set with 1438 data points. The number of offsets applied to each dataset is provided in \Tab{offsetsNum} and it is seen that a total of 6 offsets was applied to 4 fixed temperature datasets. None of the zero-field dataset have offsets. Furthermore, data point 268 in \stc{530}{75}{Ta} distorted the $R^2$ and was  set equal to data point 267; the modification of data point 268 resulted in an insignificant change of 0.6\% in the value of the linear regression slope of \dstc{265}{530}{75}{Ta}{m,n}, and an even more insignificant change of 0.04\% in the value of $_{265}\!\bar{\gamma}^{\text{Ta}}$.

As for the cause of these jumps, the ADR regulates the temperature through a feedback loop modification of the pill current \ips~ after each reading of the temperature. Hence, noise in the thermal circuit and parasitic heat load in the pill~\cite{Shirron2014} can sometimes result in overcompensation of \ips, which is why no such discontinuity was seen in the zero-field datasets. Although rare, it was found that the number of jumps could be further mitigated by improving the thermal contact between the heat switch and the pills during REGEN. This was achieved through the application of a torque to the heat switch as seen in \Fig{mylar} for the Pb experimental runs. Compare a typical fixed temperature standard deviation plot for Pb, \Fig{TsetSDPb}, with those shown for Ta, \Fig{SDplots}: the Pb temperature standard deviation is lower and more stable.

As for the flux jumps, they are discussed in the SQUID manufacturer's user guide~\cite{STARCRYO} in Fig.~3-2 on P-18. Those jumps are due the fact that the SQUID voltage signal in the LOCK mode is multivalued for the same flux. Hence, the SQUID will occasionally spontaneously `jump' to the value closest to 0~V by an amount which is an integer multiple of the calibration fluxon voltage of 0.743~V. These jumps are not necessarily due to an external physical signal, do not have a tail since the system  remains in equilibrium, and they were identified and subtracted out automatically by the software with a notification. These flux jumps were rare occurring only 5 times in over 1000 hours of Ta data.

\section{$L=1$, $S=1$ average $\mu_\text{e}$ }\label{app:avrMagMom}

For a Cooper pair with orbital angular momentum $L=1$ and spin $S=1$, the possible values for the total angular momentum are $J=0,1,2$. If the Cooper pair is in a uniform magnetic field $\mathbf{B}=B\hat{\mathbf{z}}$, the mean magnetization of the Cooper pair can be calculated from statistical mechanics. The average spin magnetic moment of a Cooper pair is
\begin{align}
&\bar{\mu}_\text{CP}\!=\nonumber \\
&\frac{   \sum\limits_{J\!=\!0}^{2}\sum\limits_{M_J\!=\!-J}^{J}\sum\limits_{m_L\!=\!-1}^{1}\sum\limits_{m_s\!=\!-1}^{1}\!\!\!\delta_{M_J,m_L\!+\!m_s}g_\text{e}m_s\mu_\text{B}e^{\beta(g_\text{e}m_s\!+\!g_Lm_L)\mu_\text{B}B}    }{  \sum\limits_{J\!=\!0}^{2}\sum\limits_{M_J\!=\!-J}^{J}\sum\limits_{m_L\!=\!-1}^{1}\sum\limits_{m_s\!=\!-1}^{1}\!\!\!\delta_{M_J,m_L\!+\!m_s}e^{\beta(g_\text{e}m_s\!+\!g_Lm_L)\mu_\text{B}B}   }
\end{align}
where $\beta\equiv1/(k_\text{B}T)$, $g_\text{e}$ is the electron gyromagnetic factor, $g_L=1$ and $\mu_\text{B}$ is the Bohr magneton. Using the Kronecker delta to perform the sum over $m_L=M_J-m_s$ leads to
\begin{align}
&\bar{\mu}_\text{CP}=g_\text{e}\mu_\text{B}\frac{   \sum\limits_{J=0}^{2}\sum\limits_{M_J=-J}^{J}e^{\beta M_J\mu_\text{B}B}\sum\limits_{m_s=-1}^{1}m_se^{m_s\eta}    }{  \sum\limits_{J=0}^{2}\sum\limits_{M_J=-J}^{J}e^{\beta M_J\mu_\text{B}B}\sum\limits_{m_s=-1}^{1}e^{m_s\eta}   } \\
&\eta\equiv \beta(g_\text{e}-1)\mu_\text{B}B
\end{align}
The sums over $J$ and $M_J$ cancel in the ratio and we are left with
\begin{align}
&\bar{\mu}_\text{CP}=g_\text{e}\mu_\text{B}\frac{\sinh\left(\eta\right)\sinh\left(\frac{1}{2}\eta\right)}{\sinh\left(\frac{3}{2}\eta\right)} \\
&M_\text{CP}=N_\text{CP}\bar{\mu}_\text{CP}
\end{align}
where $M_\text{CP}$ is the Cooper pair magnetization and $N_\text{CP}$ is the number density of Cooper pairs.

\end{appendices}

%\nocite{*}
\bibliographystyle{aapmrev4-2}
\bibliography{NEDM-refs}
%\printbibliography

\end{document}